\documentclass[12pt]{article}       
\usepackage{ee}               
\usepackage{amsmath}
\usepackage{amsfonts}
\usepackage{amssymb}
\usepackage{graphicx}

\newtheorem{theorem}{Theorem}
\newtheorem{example}[theorem]{Example}

\begin{document}

\title{Frequentist and Bayesian confidence intervals}

\author{G\"{u}nter Zech}

\institute{Universit\"{a}t Siegen, D-57068 Siegen\\
\texttt{zech@physik.uni-siegen.de}}

\maketitle

\begin{abstract}
Frequentist (classical) and Bayesian approaches to the
construction of confidence limits are compared. Various examples
which illustrate specific problems are presented. The Likelihood
Principle and the Stopping Rule Paradox are discussed. The
performance of the different methods is investigated relative to
the properties coherence, precision, bias, universality,
simplicity. A proposal on how to define error limits in various
cases are derived from the comparison. They are based on the
likelihood function only and follow in most cases the general
practice in high energy physics. Classical methods are not
recommended because they violate the Likelihood Principle, they
can produce inconsistent results, suffer from lack of precision
and generality. Also the extreme Bayesian approach with arbitrary
choice of the prior probability density or priors deduced from
scaling laws is rejected.
\end{abstract}

\tableofcontents


\section{Introduction}

\subsection{Scope of this article}

The progress of experimental sciences to a large extent is due to the
assignment of uncertainties to experimental results. A measurement is
incomplete and more or less useless, unless an error interval is attributed to
it. The precision of measurements has to be known i) to combine data from
different experiments, ii) to deduce secondary parameters from it and iii) to
test predictions of theories. Different statistical methods have to be judged
on their ability to fulfill these tasks.

In the language of statistics a measurement and its error are an estimate of a
parameter and an estimate of a parameter interval - the confidence interval.
Both are to be inferred from a data sample or a single element drawn from a
statistical distribution which depends on that parameter. The statistical data
sample collected in an experiment is called an \emph{observation} which we
distinguish from a \emph{measurement}.\footnote{For example, an observation of
ten decay times of a certain particle species constitutes a sample drawn from
an exponential distribution with unknown slope parameter. The estimate of this
parameter provides the measurement of the mean life of the particle. When we
perform a least square fit of some function containing unknown parameters
(which we want to measure) to experimental data points (the observation), our
probability distribution is a $\chi^{2}$-distribution provided that the
individual deviations of the data points follow Gaussians. The observation, -
think of the reading of a meter, a drift time, a mass distribution, a number
of observed events - has no error assigned to it. The statistical uncertainty
is embedded in the probability distribution function describing it. Often the
observation and the measurement are numerically identical. In other cases, an
observation summarizes many experimental numbers and the measurement is the
outcome of a sophisticated data analysis.}

Parameter inference is a relatively non-controversial subject, but there is
still no consensus on how to define confidence intervals for the parameters
among the different schools of statistics, represented by frequentists and
Bayesians. One of the reasons for a continuing debate between these two
parties is that statistics is partially an experimental science, as stressed
by Jaynes \cite{jayn84}, and partially a mathematical discipline as expressed
by Fisher: The first sentence in his famous book on statistics \cite{fish90}
is ``The science of statistics is essentially a branch of Applied
Mathematics''. Thus one expects from statistical methods not only to handle
all kind of practical problems but also to be deducible from few axioms and to
provide correct solutions for sophisticated exotic examples, requirements
which are not even fulfilled by old and reputed sciences like physics.

Corresponding to the two main lines of statistical thought, we are confronted
with different kinds of error interval definitions, the classical (or
frequentist) one and some more or less Bayesian inspired ones. The majority of
particle physicists intellectually favor to the first but in practice use the
second. Both methods are mathematically consistent. In most cases their
results are very similar but there are also situations where they differ
considerably. These cases exhibit either low event numbers, large measurement
errors or parameters restricted by physical limits, like positive mass,
$|\cos|\leq1$, positive rates etc..

Some standard is badly needed. For example, there exist at present at least
eight different methods to the single problem to compute \ an upper limit for
Poisson distributed events.

The purpose of this article is not to repeat all the philosophical arguments
in favor of the Bayesian or the classical school. They can be found in many
text books for example in Refs. \cite{kend73, good83, jeff61} and more or less
profound articles and reports \cite{efro86, cous95}. Further references are
given in an article by Cousins \cite{cous95}. I am convinced that methods from
both schools are valid and partially complementary. Pattern recognition, noise
suppression, analysis of time series are fields where Bayesian methods
dominate, goodness-of-fit techniques are based on classical statistics.

We restrict our discussion to the evaluation of parameters of an otherwise
completely defined theory and of the confidence intervals of these parameters.
They are deduced from a comparison of the theoretical predictions to a clean,
unbiassed data sample. In this context, goodness-of-fit tests are not relevant.

The situation in physics is different from that in social, medical or economic
sciences where usually crude models have to be used which cannot be
parametrized in a unique way and thus forbid the use of a likelihood function.

In this report, the emphasis is mainly put on performance and less on the
mathematical and statistical foundation. An exception is a discussion of the
Likelihood Principle which is fundamental for modern statistics. The intention
is to apply the procedures to problems to be solved in physics and to judge
them on the basis of their usefulness. Even though the challenge is in real
physics cases it is in simple examples that we gain clarity and insight. Thus
simple examples are selected which illustrate the essential problems.

To judge the different definitions of confidence intervals, we introduce the
following set of criteria:

\begin{enumerate}
\item  Error intervals and one-sided limits have to measure the precision of
an experiment.

\item  They should be selective (exclude wrong parameter values, powerful in
classical notation \cite{neym37b}).

\item  They have to be unique and consistent: Equally precise measurements
have equal errors intervals. More precise measurements provide smaller
intervals than less precise measurements.

\item  Subjective input has to be avoided.

\item  The procedure should allow us to combine results from different
experiments with minimum loss of information.

\item  The interval should provide a firm basis for decisions, like excluding
a theory or stopping data taking in an experiment.

\item  The method should be as general as possible. Ad hoc solutions for
special cases should be avoided.

\item  Last not least, we emphasize simplicity and transparency.
\end{enumerate}

Most of these points have acquired little attention in the ongoing debate in
the physics community, and the professional statistical literature which is
dominated by applications in economics, medicine, biology and sociology hardly
touches our problems.

The present discussion focuses on upper limit determinations and here
especially on the Poisson case relevant in rare or exotic particle searches.
However, upper limits should not be regarded isolated from the general problem
of error assignment.

In the following section we will confront the classical method
with examples which demonstrate its main difficulties and
limitations. Section~3 deals with the unified approach proposed by
Feldman and Cousins \cite{feld98}. In Sect.~4 we investigate
methods based on the likelihood function and discuss related
problems. Section~5 is devoted to the Likelihood Principle and
Sect.~6 contains a systematic comparison of the methods with
respect to the issues mentioned above. Section~7, finally,
concludes with some recommendations. Part of the content of this
report has been presented in Ref. \cite{zech98}.

We emphasize low statistics experiments. For simplicity, we will usually
assume that a likelihood function of the parameters of interest is available
and that it has at most one significant maximum. For some applications
(averaging of results) not only an interval has to be estimated but also a
parameter point. Normally, the maximum likelihood estimate is chosen. In
classical approaches, the intervals do not necessarily contain the likelihood
estimate and special prescriptions for the combination of results are necessary.

In the first part of this report we will use undefined statistical terms like
\emph{precision} and \emph{inconsitency} and hope that they will become clear
from the context in which they are applied. We will come back to them in
Sect.~5.

\subsection{A first glance at the problem}

Before we start to discuss details, let us look in a very qualitative way at
the main difference between a frequentist approach (respecting the coverage
principle, see Sect.~2) and methods based on the likelihood function
(respecting the Likelihood Principle, see Sect.~5).

As a simple example we imagine an observation\footnote{We use capital letters
for variates and small letters for observations of the variate.} $x$ of a
variate (random variable\footnote{For the definition of statistical terms see
Ref. \cite{kend82}.}) $X$ and a probability distribution function\footnote{In
most cases we do not distinguish between discrete and continuous probability
distribution functions.} (pdf) $f(X|\theta)$ depending on an unknown parameter
$\theta$ which we estimate from $x$. How should we select the range of
parameters which we consider compatible with the data? In Fig.~1 we display
an observation $x$. The two pdfs correspond to two specific parameter values
$\theta_{1}$ and $\theta_{2}$.

A parameter value $\theta$ is supported by an observation $x$ if the
observation is located where its probability density $f(x|\theta)$ is high. If
we had to choose between $\theta_{1}$ and $\theta_{2}$, intuitively we would
prefer $\theta_{1}$, corresponding to the narrow peak because the probability
density $f(x|\theta_{1})$, i.e. the likelihood of $\theta_{1}$, is larger than
the likelihood $f(x|\theta_{2})$ and would include $\theta_{1}$ in a
confidence interval with higher priority than the competitor $\theta_{2}$.

Classical confidence limits (CCL) rely on tail probabilities. A parameter
value is accepted inside the confidence interval if the observation is not too
far in the tail of the corresponding pdf. Essentially, the integral over the
tail beyond $x$ determines whether $\theta$ is included. Classical methods
would preferentially accept the parameter $\theta_{2}$ corresponding to the
wide peak, the observation being less than one standard deviation off, and not
$\theta_{1}$. Thus, they may exclude parameter values that correspond to
higher likelihood than those which they include.

Which of the two approaches is the better one? Given $x$ with no additional
information we certainly would bet for $\theta_{1}$ with betting odds
corresponding to the likelihood ratio $f(x|\theta_{1})/f(x|\theta_{2})$ in
favor of $\theta_{1}$. However, we then clearly favor precise predictions over
crude ones. Assume $\theta_{2}$ applies. The chance to accept it inside a
certain likelihood interval is smaller than the corresponding chance for
$\theta_{1}$. The choice based on the likelihood function is unfair to
$\theta_{2}$.%
\begin{figure}
[ptb]
\begin{center}
\includegraphics*[width=.8\textwidth]%
{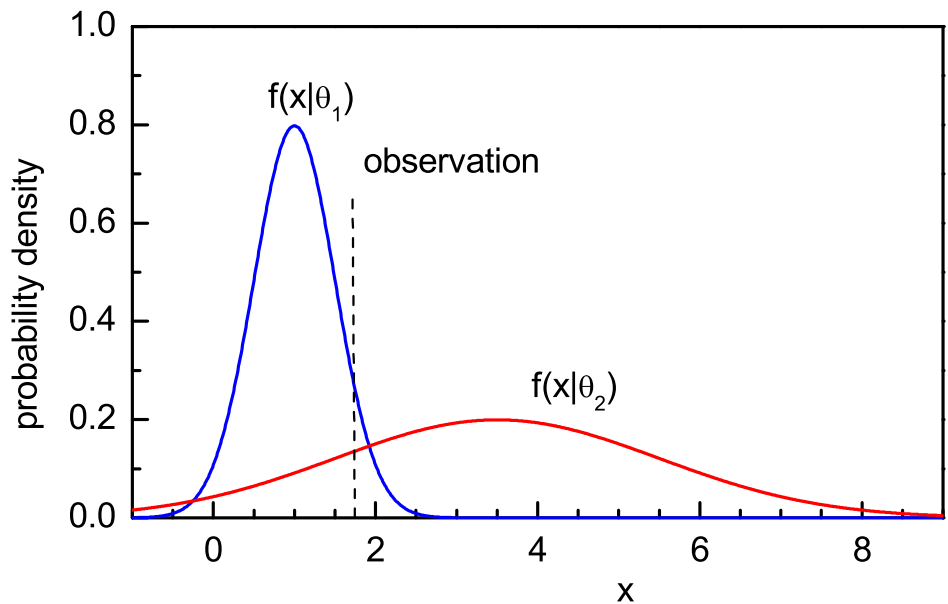}%
\caption{The likelihood is larger for parameter $\theta_{1}$, but
the observation is less then 1~st. dev. off $\theta_{2}$.
Classical approaches include $\theta_{2}$ and exclude $\theta_{1}$
within a 68.3\% confidence interval}
\end{center}
\end{figure}

The classical limits exhibit an integration in the sample space, thus
depending on the probability density of data that have not been observed.
Bayesians object to using such irrelevant information. (Why should we care
about the probability density $f(x^{\prime}|\theta)$ at $x^{\prime}$ when we
have observed $x$?) They rely on the likelihood function, transform it into a
probability density of the parameter and usually integrate the result to
compute probabilities or moments. Thus their conclusions depend on the
somewhat arbitrary choice of the parameter space\footnote{We consider only
uniform prior densities. There is no loss of generality, see Sect.~4.}. This
is not acceptable to frequentists.

The simplest and most common procedure is to avoid the integration and to
define intervals based solely on the likelihood function. It depends only on
the local probability density of the observed data and does not include
subjective or irrelevant elements. Admittedly, restricting interval estimation
to the information contained in the likelihood function - which is a mere
parametrization of the data - does not permit to deduce probabilities or
confidence levels in the probabilistic sense.

\section{Classical confidence limits}

The defining property of classical confidence limits \ (CCL) is
\emph{coverage}: If a large number $n$ of experiments perform measurements of
a parameter with confidence level\footnote{Throughout this article we use the
generic name ``confidence level'' which usually is reserved for to frequentist
statistics and there is equivalent to ``coverage'' or ``p-value''. Notations
attributed to Bayesian intervals are ``credibility'' or ``degree of belief''.
The confidence level $\alpha$ corresponds to the $1-\alpha$ in most of the
literature.} $\alpha$, in the limit $n\rightarrow\infty$, the fraction
$\alpha$ of the limits has to contain the true value of the parameter inside
the confidence limits.

In the following we show how confidence limits fulfilling the coverage
requirement can be constructed.

\subsection{Visualization}%

\begin{figure}[t]
\begin{center}
\includegraphics*[width=\textwidth]{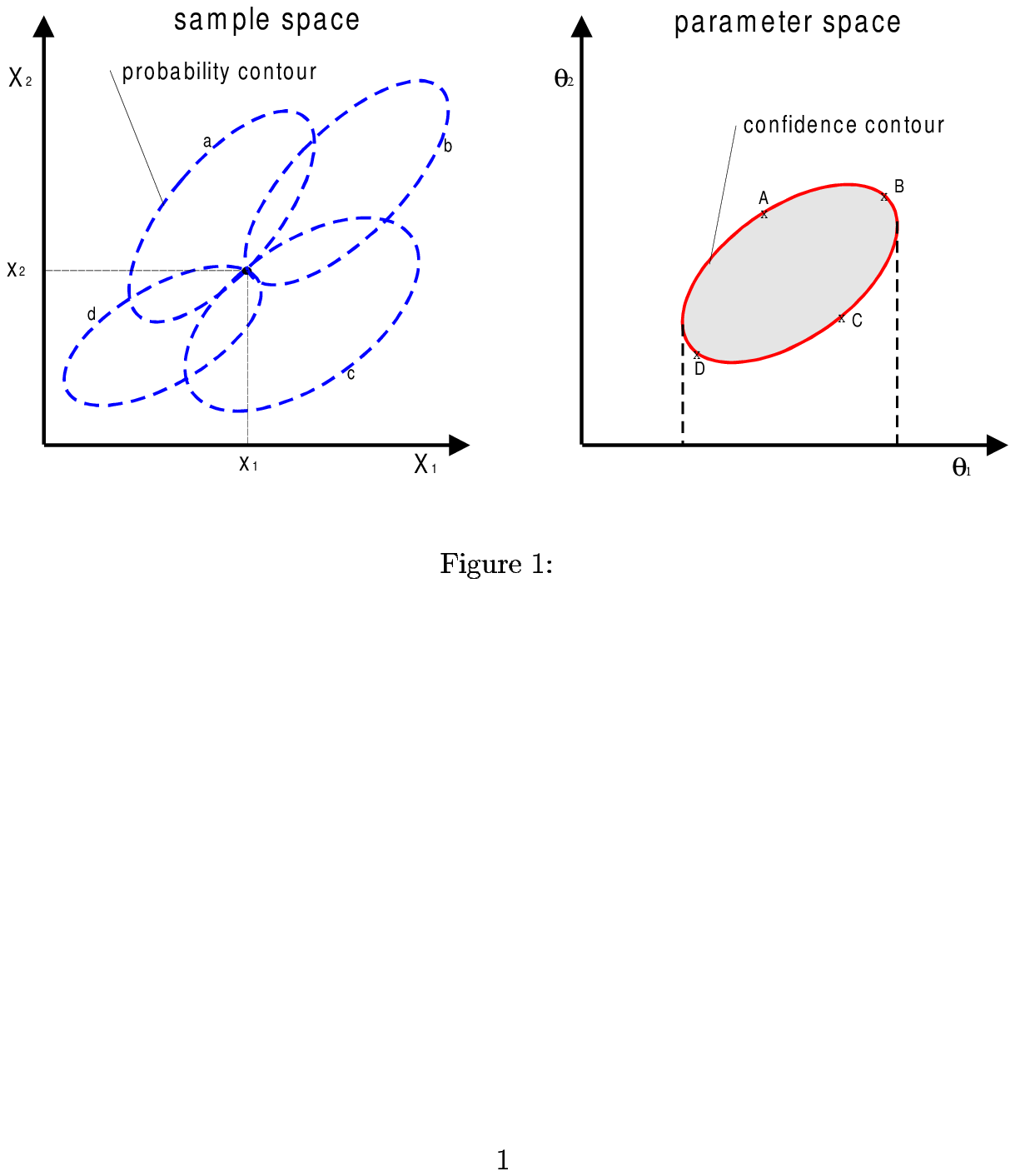}%
\caption{Two parameter classical confidence limit for an
observation $x_{1,}x_{2}$. The dashed contours labeled with small
letters in the sample space correspond to probability contours of
the parameter pairs labeled with capital letters in the parameter
space}
\end{center}
\end{figure}

\begin{figure}[t]
\begin{center}
\includegraphics*[width=.8\textwidth]{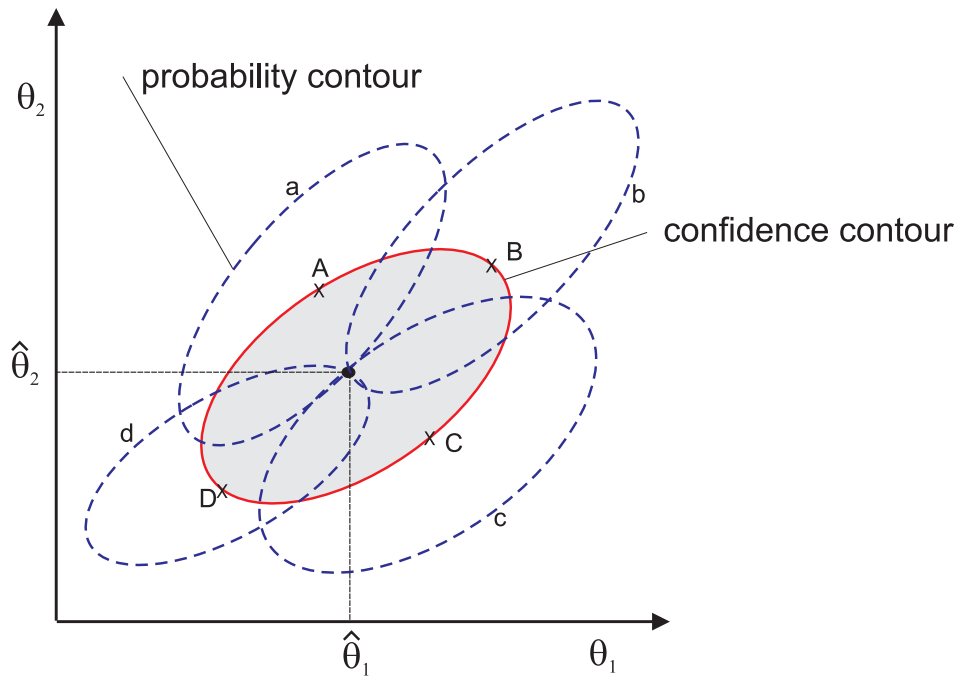}%
\caption{Two parameter classical confidence limit. The dashed
probability contours labeled with small letters contain an estimate of the
true value (capital letter) with probability $\alpha$}%
\end{center}
\end{figure}

We illustrate the concept of CCL for an observation consisting of a vector
($x_{1},x_{2}$) and a two-dimensional parameter space (see Fig.~2). In a
first step we associate to each point $\theta_{1},\theta_{2}$ in the parameter
space a closed \emph{probability contour} in the \emph{sample space}
containing an observation with probability $\alpha$. (The probability contour
is not necessarily a curve of constant probability. Given $\alpha$, there is
an infinite number of ways to define probability contours. Specific choices
will be discussed below.) For example, the probability contour labeled $a$ in
the sample space corresponds to the parameter values of point $A$ in the
parameter space. The curve (\emph{confidence contour}) connecting all points
in the \emph{parameter space} with probability contours in the sample space
passing through the actual observation $x_{1},x_{2}$ encloses the
\emph{confidence region} of confidence level (C.L.) or coverage $\alpha$. This
construction guarantees that all parameter values contained in the confidence
region contain the observation inside their probability contour. As a
consequence, whatever the values of the parameters realized in nature are,
measurements will produce with probability $\alpha$ a confidence contour which
contains these parameters.

Frequently, an observation is composed of many independent single observations
which are combined to an estimate $\hat{\theta}$ of the parameter. Then the
two plots of Fig.~2 can be combined to a single graph (see Fig.~3).

Figures 2 and 3 demonstrate some of the requirements necessary for the
construction of a simply connected, non-empty confidence region with coverage
exactly equal to $\alpha$:

\begin{enumerate}
\item  The sample space must be continuous. (Discrete distributions, thus all
digital measurements and in principle also Poisson processes are excluded.)

\item  The probability contours should enclose a simply connected region.

\item  The parameter space has to be continuous.

\item  The parameter space should be infinite.
\end{enumerate}

The restriction (1) usually is overcome by relaxing the requirement of exact
coverage. The probability contours are enlarged to contain at least the
fraction $\alpha$ of observations. In most cases, confidence limits with
minimum overcoverage are chosen. This makes sense only when the density of
\ points in the sample space is relatively large.\

A possibility to conserve exact coverage is to randomize the confidence belts.
The sample points are associated to the probability region of one or another
parameter according to an appropriate probability distribution. This method is
quite popular in some research fields and discussed extensively in the
statistical literature but will not be followed here. It introduces an
additional unnecessary uncertainty. A procedure cannot be optimum when
decisions depend on the outcome of coin tossing.

Discrete parameter spaces can lead to empty intervals if an observation is not
included in any of the probability contours.

Confidence intervals may be located in a region forbidden by physical
constraints. Mathematical constrains, like $\cos\leq1$, can produce empty
intervals in the standard classical approaches. This problem is absent in
unified approaches (see Sect.~3).

Frequentists usually require that confidence intervals do not contain regions
where the parameter is not defined. Then instead of unphysical intervals empty
intervals are obtained. Unphysical intervals, however, are more informative.
From the two statements ``$m=0$ with $68\%$ confidence`` and ``$-2eV<m<-1eV$
with $68\%$ confidence'', the second one is more valuable.

There is considerable freedom in the choice of the probability contours but to
ensure coverage \emph{their definition has to be independent of the result of
the experiment}. Usually, the contours are locations of constant probability density.

\subsection{Classical confidence limits in one dimension -- definitions}

For each possible value of the parameter $\theta$ we fix a probability
interval $[X_{1}(\theta)$, $X_{2}(\theta)]$ fulfilling
\[
P(X_{1}\leq X\leq X_{2}|\theta)=\int_{X_{1}}^{X_{2}}f(X|\theta)dX=\alpha
\]
where $f(X|\theta)$ is the pdf. Using the construction explained above, we
find for an observation $x$ the confidence limits $\theta_{low}$ and
$\theta_{high}$ from
\begin{align*}
x_{1}(\theta_{high})  & =x\\
x_{2}(\theta_{low})  & =x
\end{align*}

The definitions do not fix the limits completely. Additional constraints have
to be added. Some sensible choices are listed in Table 1. (We only consider
distributions with a single maximum.)%

\begin{table}[tbp]
\caption{Some choices for classical confidence intervals}%
\vskip5pt
\centering
\begin{tabular}
[c]{|l|l|}\hline
central interval & $P(X\leq X_{1}|\theta)=P(X\geq X_{2}|\theta)=(1-\alpha
)/2$\\
equal probability densities & $f(X_{1}|\theta)=f(X_{2}|\theta)$\\
minimum size & $\theta_{high}-\theta_{low}$ is minimum\\
symmetric & $\theta_{high}-\hat{\theta}=\hat{\theta}-\theta_{low}$\\
likelihood ratio ordering & $f(X_{1}|\theta)/f(X_{1}|\theta_{best}%
)=f(X_{2}|\theta)f(X_{2}|\theta_{best})$\\
one-sided & $\theta_{low}=-\infty$ or $\theta_{high}=\infty$\\\hline
\end{tabular}
\end{table}

\subsubsection{Central intervals.}

The standard choice is \emph{central intervals}. For a given parameter value
it is equally likely to fall into the lower tail and into the upper tail of
the distribution. The Particle Data Group (PDG) \cite{pdg98} advocates central
intervals, though the edition from the year 2000 also proposes intervals based
on the likelihood ratio ordering. Central intervals are invariant against
variable and parameter transformations. Their application is restricted to the
simple case with one variate and one parameter. When central intervals are
considered together with a point estimation (measurement, parameter fit), the
obvious parameter choice corresponds to the zero interval length limit
($\alpha=0$) which in most cases would not coincide with the maximum
likelihood estimate.

\subsubsection{Equal probability density intervals.}

\emph{Equal probability density intervals }are preferable because in the
majority of the cases they are shorter and less biassed than central intervals
and the concept is also applicable to the multidimensional case. They coincide
with central intervals for symmetric distributions. A disadvantage is the
non-invariance of the definition under variate transformations (see Sect.~6.4).

\subsubsection{Minimum size intervals.}

Of course we would like the confidence interval to be as short as possible
\cite{banc81}. The construction of \emph{minimum size intervals}, if possible
at all, is a difficult task \cite{kend73, banc81}. In Appendix B we illustrate
how pivotal quantities can be used to compute such intervals. Clearly, these
intervals depend by definition on the parameter choice. When we determine the
mean lifetime of a particle the minimum size limits for the lifetime will not
transform into limits of the decay constant with the same property.

\subsubsection{Symmetric intervals.}

Often it is reasonable to quote symmetric errors relative to an estimate
$\hat{\theta}$ of the parameter. Again this type of interval is difficult to
construct and not invariant under parameter transformations.

\subsubsection{Selective intervals.}

One could try to select confidence intervals which minimize the probability to
contain wrong parameter values \cite{neym37b}. These intervals are called
\emph{shortest} by Neyman and \emph{most selective} by Kendall and Stuart
\cite{ks2128}. Since this condition cannot be fulfilled for two-sided
intervals independent of the value of the true parameter, Neyman has proposed
the weaker condition that the coverage for all wrong parameter values has to
be always less than $\alpha$. It defines the \emph{shortest unbiassed} or
\emph{most selective unbiassed} (MSU) intervals\footnote{These intervals are
closely related to uniformly most powerful and uniformly most powerful
unbiased tests \cite{lehm59a}}.

\emph{Most selective unbiassed} intervals can be constructed \cite{ks2240}
with the likelihood ratio ordering (not to be mixed up with likelihood ratio
intervals). Here the probability contour (in the sample space) of a parameter
corresponds to constant $R$, defined by
\begin{equation}
R(X|\theta)=\frac{f(X|\theta)}{f(X|\theta_{best})}%
\end{equation}
where $\theta_{best}$ is the maximum likelihood estimate for a fictitious
observation $X$. Qualitatively this means that preferentially those values of
$X$ are added to the probability interval of $\theta$ where competitive values
of the parameter have a low likelihood. Selective intervals have the
attractive property that they are invariant under transformations of variables
\emph{and} parameters independent of their dimension. Usually the limits are
close to likelihood ratio intervals (see Sect.~4.2) and shorter than central
intervals\footnote{The likelihood ratio ordering minimizes the probability of
\ errors of the second kind for one sided intervals. For two-sided intervals
this probability can only be evaluated using relative prior probabilities of
the parameters located at the two sides of the confidence interval.
\cite{neym37b,kend73,feld00}.}. The construction of the limits is quite
tedious unless a simple sufficient statistic can be found. In the general
case, where sufficiency requires a full data sample of say 1000 events, one
has to find likelihood ratio contours in a 1000-dimensional space. For this
reason, MSU intervals have not become popular in the past and were assumed to
be useful only for the so-called exponential family \cite{lehm59a} of pdfs
where a reduction of the sample space by means of sufficient statistics is
admitted. Nowadays, the computation has become easily feasible\ on
PCs.\ However, the programming effort is not negligible.

The likelihood ratio ordering is applied in the unified approach \cite{feld98}
which will be discussed in Sect.~3.

\subsubsection{One-sided intervals.}

\emph{One-sided intervals} define upper and lower limits. Here, obviously the
limit should be a function of a sufficient statistic or equivalently of the likelihood.

\subsubsection{Which definition is the best?}

No general answer can be given. The choice may be different when we are
interested in the uncertainty of a measurement of a particle track or in the
verification of a theory. The former case one may prefer criteria based on the
mean squared deviation of the limits from the true parameter value
\cite{hart64}. If probability arguments dominate, Neyman's \emph{MSU}
intervals are most attractive. Their only disadvantage is the complicated
numerical evaluation of the limits.

To simplify the discussion, in the following section (except for the first
example) we do not apply the Neyman construction but follow the more popular
line using central or equal probability intervals. The MSU prescription will
be treated separately in Sect.~3 dealing with the unified approach.%

\begin{figure}[t]
\begin{center}
\includegraphics*[width=3.6902in]{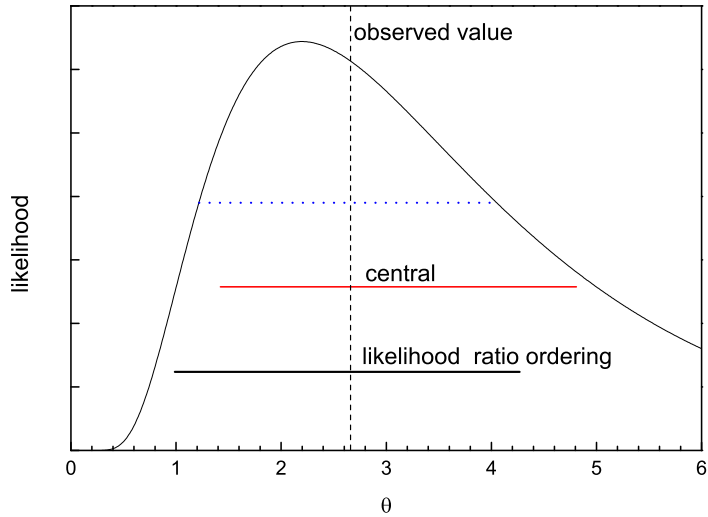}%
\caption{Position measurement from drift time. The error is due to
diffusion. Classical confidence intervals are shown together with
the likelihood function}
\end{center}
\end{figure}

\begin{figure}[t]
\begin{center}
\includegraphics*[width=3.3676in]{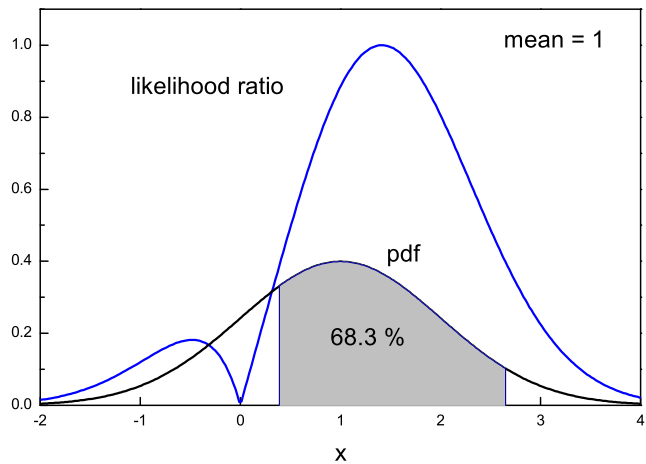}%
\caption{Likelihood ratio ordering. The likelihood ratios are equal at
the limits of the shaded probability interval}%
\end{center}
\end{figure}

\subsection{Two simple examples}

\begin{example}
To illustrate the concept of \ classical confidence levels in one dimension we
consider an electron moving in a gaseous detector. The observed drift time is
converted to a distance $x$. The uncertainty is due to diffusion. The
probability density is
\[
f(X)=\frac1{\sqrt{2\pi c\theta}}\exp\left[  -\frac{\left(  X-\theta\right)
^{2}}{2c\theta}\right]
\]
where $c$ is a known constant. Fig.~4 shows an observation, the likelihood
function and the confidence limits for a central interval and for a MSU
interval. The likelihood ratio ordering is indicated in Fig.~5 for
$\theta=1$. The 68.3 \% probability interval covers the region of $x$ where
the likelihood ratio is largest. The dip at $X=0$ is due to the fact that
$f(0|\theta_{best})=f(0|\theta=0)=\infty$. The MSU (likelihood ratio ordering)
interval is shorter than the central interval and very close to the likelihood
ratio interval which is bound by parameter values of equal likelihood (see
Sect.~4.2).
\end{example}

We now turn to problematic situations.

\begin{example}
\textbf{\ }We extract the slope of a linear distribution
\begin{equation}
f(X|\theta)=\frac12(1+\theta X)\label{eqslope}%
\end{equation}
with $-1<X<1$. Our data sample consists of two events observed at symmetric
locations, $x_{1},x_{2}$. From the observed sample mean $s=(x_{1}+x_{2})/2$ we
construct the central limits
\[
\int_{-1}^{s}g(s^{\prime}|\theta_{high})ds^{\prime}=\int_{s}^{1}g(s^{\prime
}|\theta_{low})ds^{\prime}=\frac{1-\alpha}2
\]
where $g(s|\theta)$ is the distribution of the sample mean of two observations
each following $f(X|\theta)$. For the specific result a) $x_{1}=-0.5$,
$x_{2}=0.5,$ we find $s=0$ and $\theta_{low}=-0.44$, $\theta_{high}=0.44$ with
$C.L.=0.683$. If instead the result of the observation had been b) $x_{1}=0$,
$x_{2}=0$ or c) $x_{1}=-1$, $x_{2}=1$ the same confidence limits would have
been obtained as in case a).
\end{example}

We realize that experimental data which are non-informative as in case b) can
provide stringent limits. This is compensated by relatively loose limits for
informative observations as in case c) and a well defined coverage is obtained.

The reason for the poor behavior is that the first moment of the linear
distribution is not a sufficient statistic. The limits do not exhaust the
experimental information.

In the quoted example the behavior can be improved also in classical
statistics by working in the two-dimensional space of $x_{1},x_{2}$.

For a sample of size $n$ in the above example, probability contours have to be
defined in a $n$-dimensional space because no simple sufficient statistic is
available. The computation of the limits becomes complex but is easily
feasible with present PCs with non-negligible programming effort.

\subsection{Digital measurements}

\begin{example}
\textbf{\ }A particle track is passing at the unknown position $\mu$ through a
proportional wire chamber. The registered coordinate $x$ is set equal to the
wire location\ $x_{w}$. The probability density
\[
f(X|\mu)=\delta(X-x_{w})
\]
is independent of the true location $\mu$. Thus it is impossible to define a
classical confidence interval, except a trivial one with full overcoverage.
This difficulty is common to all digital measurements because they violate
condition 1 of Sect.~2.1. Thus a large class of measurements cannot be
handled in classical statistics.
\end{example}

Even frequentists use the obvious Bayesian solution (see Sect.~4) to this
problem. True values near the wire are always covered, those near the border never.

\subsection{External constraints}

One of the main objections to classical confidence limits is related to
situations where the parameter space is restricted by external constraints. To
illustrate the problem, we take up an often quoted example:%

\begin{figure}[tp]
\begin{center}
\includegraphics*[width=.8\textwidth]{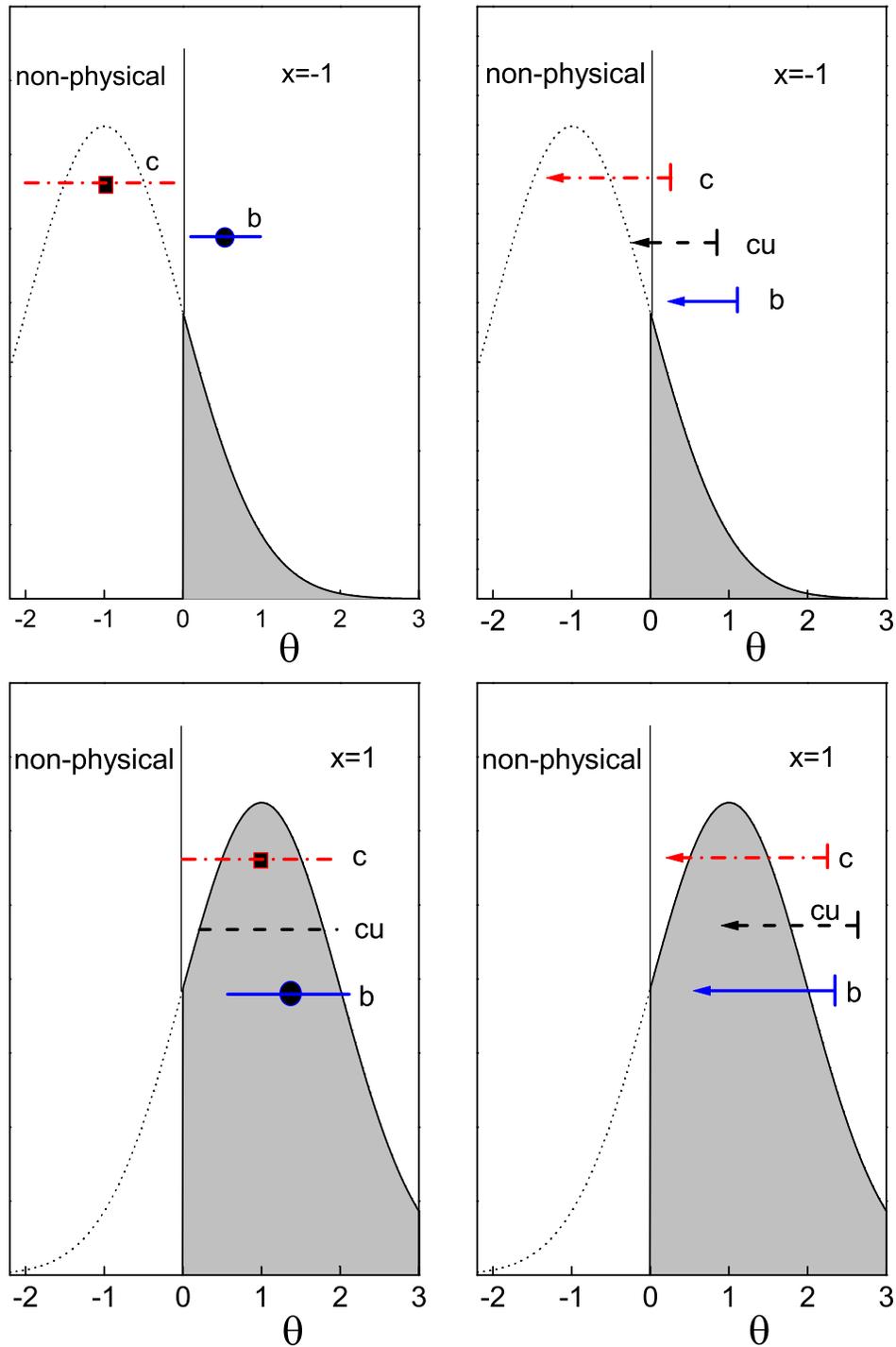}%
\caption{Confidence limits for a measurement with Gaussian errors
and a physical boundary. The left side shows 68.3\% confidence
intervals, the right side 90\% upper limits. The observed values
of $\theta$ are $-1$ (top) and $+1$ (bottom). The labels refer to
classical (c), unified classical (cu) and Bayesian (b)}%
\end{center}
\end{figure}

\begin{example}
A physical quantity like the mass of a particle with observations following
normal distributions is constrained to positive values. Figure~6 shows typical
central confidence bounds which extend into the unphysical region. In extreme
cases an observation which due to a statistical fluctuation is located in the
unphysical region may produce a 90 \% confidence interval which does not cover
physical values at all\footnote{In classical statistics the situation is
usually described by a zero-length interval. I prefer unphysical intervals
which retain more of the experimental information.}.
\end{example}

An similar situation for the standard classical method is also encountered in
the following example corresponding to the distribution of Example~2.%

\begin{figure}[t]
\begin{center}
\includegraphics*[width=.8\textwidth]{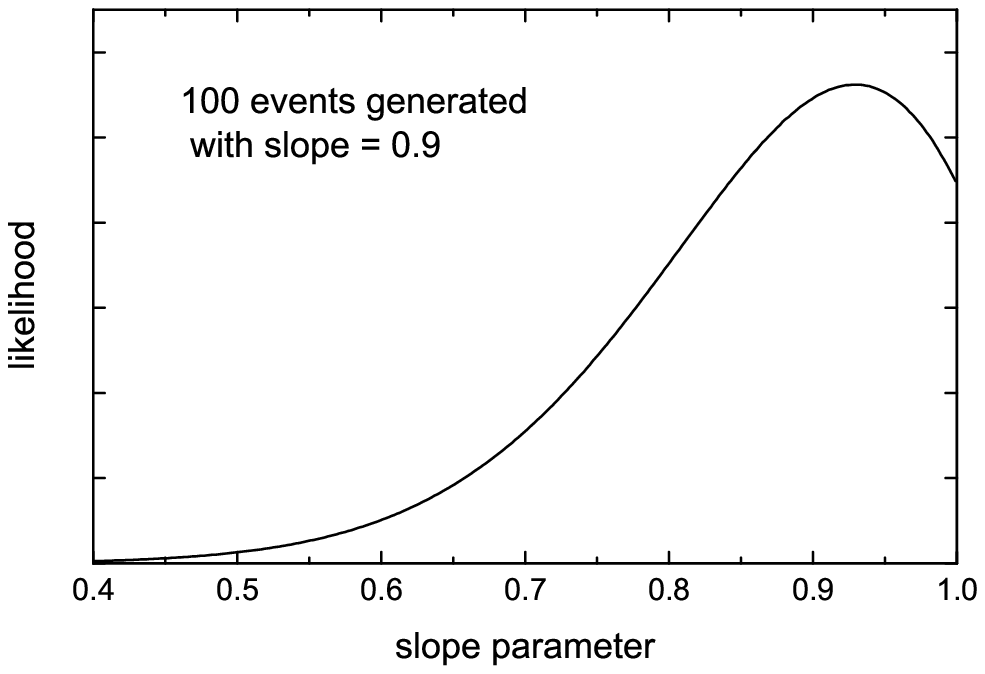}%
\caption{Likelihood of the slope parameter of a linear distribution.
Central classical 68.3\% confidence limits cannot be defined because the
slope is undefined outside the physical interval}%
\end{center}
\end{figure}

\begin{example}
For a sample of 100 events following the distribution of Eq.~(2), a likelihood
analysis gives a best value for the slope parameter of $\hat{\theta}=0.92$
(see Fig.~7). When we try to compute central classical $68.3\%$. confidence
bounds we get $\theta_{low}=0.82$ and find no upper bound inside the allowed
range $-1\leq\theta\leq1$. We cannot compute a limit above $1$ either because
slopes $\theta>1$ are not acceptable since they would produce negative
probabilities for small values of $X$.
\end{example}

Frequentists may solve this problem by choosing asymmetric probability
intervals (excluding central and equal probability intervals) which then
should be defined before the measurement is done.

Another difficulty arises for parameters which are restricted from both sides:

\begin{example}
A particle passes through a small scintillator and another
position sensitive detector with Gaussian resolution. Both
boundaries of the classical error interval are in the region
forbidden by the scintillator signal (see Fig.~8). The classical
interval is twice as large as the r.m.s. width. One would restrict
it to the allowed region, but the error interval is meaningless.
\end{example}%

\begin{figure}[ptb]
\begin{center}
\includegraphics*[width=3.7023in]{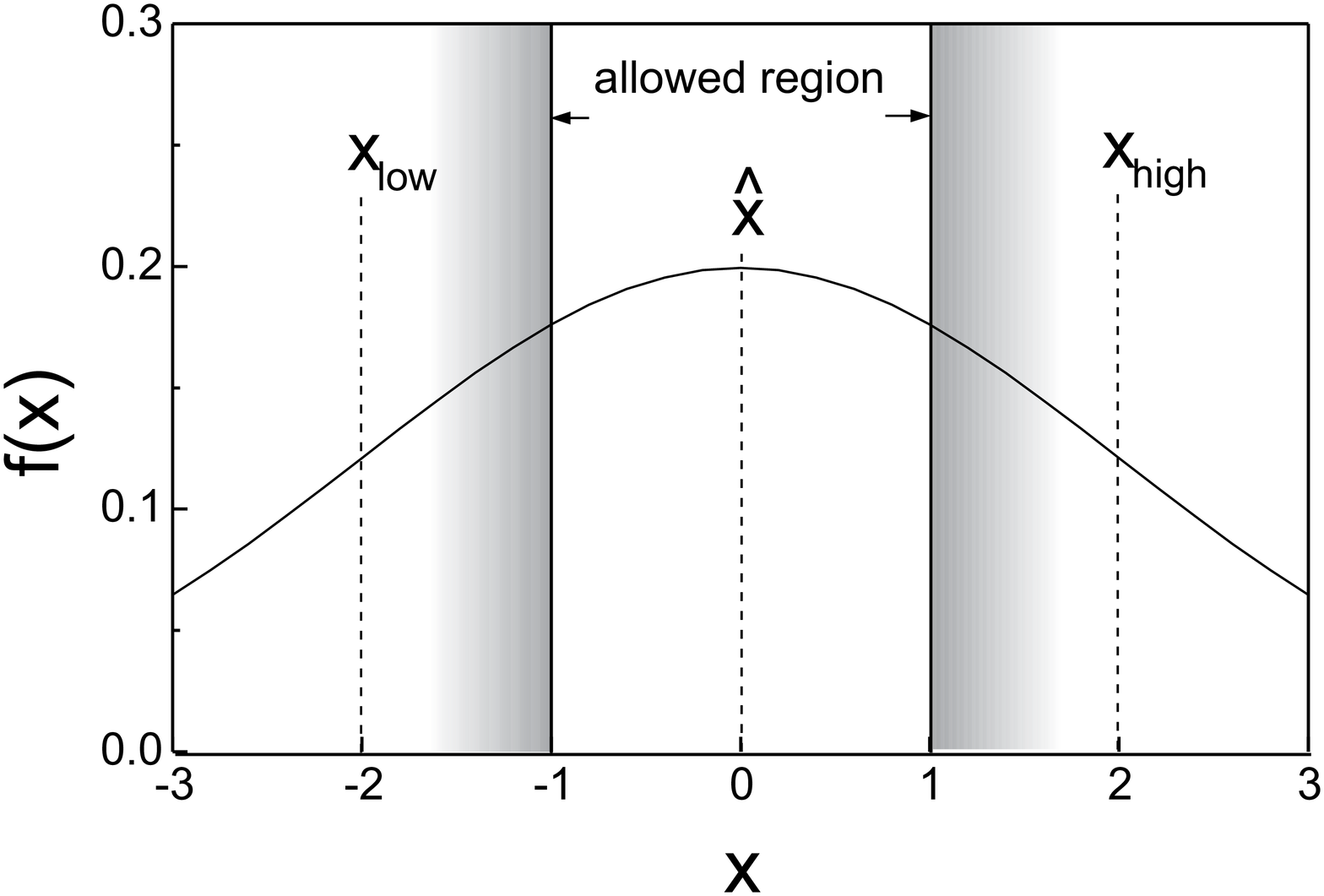}%
\caption{Classical confidence bounds cannot be applied to a parameter
space bound from both sides}%
\end{center}
\end{figure}

\subsection{Classical confidence limits with several parameters}

In case several parameters have to be determined, the notion of central
intervals no longer makes sense and the construction of intervals of minimal
size is exceedingly complex. It is advisable to use curves of equal
probability or likelihood ratio boundaries to define the probability region.

Very disturbing confidence intervals may be obtained in several dimensions
when one tries to minimize the size of the interval. Here, a standard example
is the case of the simple normal distribution in more than two dimension where
occasionally ridiculously tiny intervals are obtained \cite{bergp9}.

\subsection{Nuisance parameters}

In most experimental situations we have to estimate several parameters but are
interested in only one of them. The other parameters, the nuisance parameters,
have to be eliminated. A large number of research articles have been published
on the subject. A detailed discussion and references can be found in Basu's
work \cite{basu114}.

A formal way to eliminate the nuisance parameter without violating the exact
coverage principle is the following: For a given pdf $f(X|\theta,\nu)$ where
$\theta$ is the parameter of interest and $\nu$ the nuisance parameter. we
search for a statistic $Y_{\nu}$ which is ancillary in $\nu$ (contains no
information on $\nu$) and use the pdf $f(Y_{\nu}|\theta)$ to determine the
confidence interval. Usually $Y_{\nu}$ will not be sufficient for $\theta$ and
the approach is not optimum since part of the experimental information has to
be neglected. The implementation of this approach is not straight forward. Of
practical interest are the following methods.

\subsubsection{Factorization and restructuring.}

It is easy to eliminate the nuisance parameter if the pdf factorizes in a
function $f_{\nu}$ of the nuisance parameter $\nu$ and a function $f_{\theta}$
of the parameter of interest $\theta$.
\begin{equation}
f(X|\theta,\nu)=f_{\theta}(X|\theta)f_{\nu}(X|\nu)
\end{equation}
Then the limits for $\theta$ are independent of a specific choice of $\nu$. If
this condition is not realized, one may try to restructure the problem looking
for another parameter $\eta(\theta,\nu)$ which fulfills the condition.
\[
f(X|\theta,\eta)=f_{\theta}(X|\theta)f_{\eta}(X|\eta)
\]
Then again fixing the new parameter $\eta$ to an arbitrary value has no
influence on the confidence limits for the parameter of interest and we may
just forget about it.

The following example is taken from Edwards \cite{edwa92}.

\begin{example}
The absorption of a piece of material is determined from the two rates
$r_{1},r_{2}$ registered with and without the absorber. The rates follow
Poisson distributions with mean values $\theta_{1},\theta_{2}$:
\[
f(R_{1},R_{2})=\frac{e^{-(\theta_{1}+\theta_{2})}\theta_{1}^{R_{1}}\theta
_{2}^{R_{2}}}{R_{1}!R_{2}!}
\]
We are interested in the ratio $\phi=\theta_{1}/\theta_{2}$ which is directly
related to the absorption parameter. By a clever choice of the nuisance
parameter $\nu$
\begin{align*}
\phi=\frac{\theta_{1}}{\theta_{2}}\\
\nu=\theta_{1}+\theta_{2}%
\end{align*}
we get
\begin{align*}
f=e^{-\nu}\nu^{R_{1}+R_{2}}\frac{(1+1/\phi)^{R_{1}}(1+\phi)^{R_{2}}}%
{R_{1}!R_{2}!}\\
=f_{\nu}(R_{1}+R_{2}|\nu)f_{\phi}(R_{1},R_{2}|\phi)
\end{align*}
where the dependence of $f$ on the interesting parameter $\phi$ and the
function of the nuisance parameter $\nu$ factorize. The sum $R_{1}+R_{2}$
which is proportional to the measuring time, obviously contains no information
about the absorption parameter. It is ancillary and we can condition on
$R_{1}+R_{2}$. It is interesting to notice that $f_{\phi}$
\[
f_{\phi}=\frac{(1+1/\phi)^{R_{1}}(1+\phi)^{R_{2}}}{R_{1}!R_{2}!}
\]
is not a function of the absorption ratio $R_{1}/R_{2}$ only.\footnote{An
interesting application is the determination of $\varepsilon^{\prime}$ from a
double ratio of CP violating kaon decays.}\
\end{example}

A lot of effort has been invested to solve the factorization problem
\cite{kalb70}. The solutions usually yield the same results as integrating out
the nuisance parameter.

That restructuring is not always possible is seen in the following example.

\begin{example}
The mean decay rate $\gamma$ is to be determined from a sample of 20 events
which contain an unknown number of background events with decay rate
$\gamma_{b}=0.2$. Likelihood contours for the two parameters $\gamma$ and the
number of signal events are shown in Fig.~9. There is no obvious way to
disentangle the parameters.
\end{example}%

\begin{figure}[ptb]
\begin{center}
\includegraphics*[width=4.3569in]{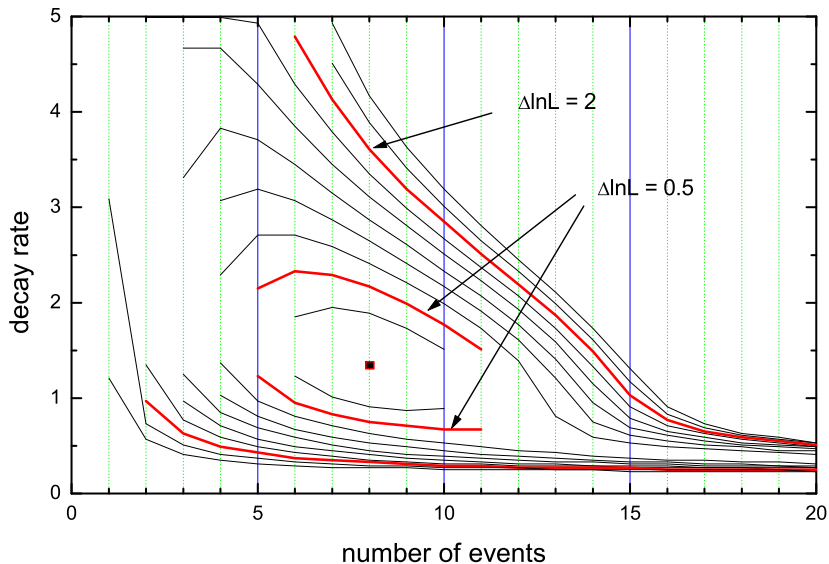}%
\caption{Likelihood contours for Example 8. For better visualization
the discrete values of the nuisance parameter ``number of events'' are
connected}%
\end{center}
\end{figure}

\subsubsection{Global limits.}

A possible but not satisfactory solution to eliminate a nuisance parameter is
to compute a conservative limit. How this can be done is indicated by the
dashed lines in the right hand graph of Fig.~2 for the two parameter case: A
global confidence interval is computed for all parameters including the
nuisance parameters. The projection of the contour onto the parameter of
interest provides a conservative confidence interval for this parameter.

Looking again at Fig.~2, we realize that by squeezing the confidence ellipse
in the direction of the parameter of interest and by stretching it in the
other direction, the confidence level probably can be maintained but a
narrower projected interval could be obtained. To perform this deformation in
a consistent way such that \ minimum overcoverage is guaranteed is certainly
not easy. An example is discussed in Sect.~2.9.

\subsubsection{Other methods.}

Another popular method used by classical statisticians \cite{cous00, feldcl}
to eliminate the nuisance parameter is to estimate it away, to replace it by
the best estimate. This proposal does not only violate the coverage principle
but is clearly incorrect when the parameters are correlated. The stronger the
correlation between the parameters is, the smaller the confidence interval
will become. In the limit of full correlation it will approach zero!

\subsection{Upper and lower limits}

Frequently, we want to give upper or lower limits for a physical parameter.
Then, of course, we use one sided bounds.

\begin{example}
In a measurement of the neutrino mass the result is $\hat{m}=(-2\pm2)\,eV$
with Gaussian errors independent of the true value. A 90\% confidence upper
limit $m_{u}$ is defined classically by
\[
0.9=\int_{\hat{m}}^{\infty}\eta(X|m_{u},2)dX
\]
where $\eta$ denotes the normal distribution centered at $m_{u}$ with width $2
$. The upper 90\% confidence limit is $m_{u}<0.6eV$. For a measurement
$\hat{m}=(-4\pm2)\,eV$ the limit would be $m_{u}<-1.4\,eV$ in the unphysical
region. This is a frequently discussed problem. The fact that we confirm with
90\% confidence something which is obviously wrong does not contradict the
concept of classical confidence limits. The coverage is guaranteed for an
ensemble of experiments and the unphysical interval occur only in a small
fraction of them. Whether such confidence statements are of any use is a
different story\footnote{Savage et al. \cite{sava62}: ``The only use I know
for a confidence interval is to have confidence in it.''}.
\end{example}

\subsection{Upper limits for Poisson distributed signals}

In particle physics, by far the most frequent case is the calculation of upper
limits for the mean of Poisson distributed numbers, $P(k|\mu)=e^{-\mu}\mu
^{k}/k!$ (To conform to the notations used in physics, we do not consistently
apply the convention to use capital letters for variates.) As stated above,
for discrete data the frequentist approach has to accept overcoverage.
(Remember, we do not consider randomization.) In the approximation with
minimum overcoverage the upper limit $\mu$ with confidence $\alpha$ for $n$
observed events is given by:
\begin{align*}
\alpha & =\sum_{i=n+1}^{\infty}P(i|\mu)\\
1-\alpha & =\sum_{i=0}^{n}P(i|\mu)
\end{align*}

In words, this means: If the limit corresponded to the true parameter value
$\mu$, the probability to observe $n$ events or less were equal to $1-\alpha$.%

\begin{figure}[ptb]
\begin{center}
\includegraphics*[width=4.1053in]{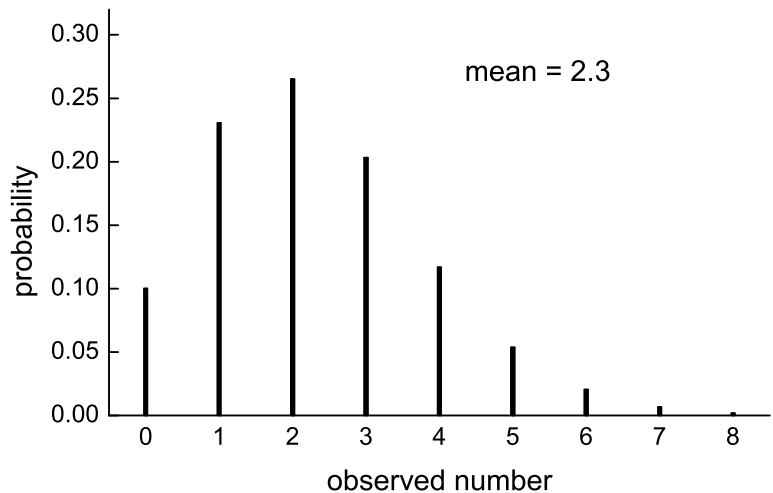}%
\caption{Construction of the classical 90\% confidence upper limit for
zero events observed. The Poisson distribution with mean 2.3 predicts zero
events in 10\% of the cases}%
\end{center}
\end{figure}

In the majority of search experiments no event ($n=0$) is found. The classical
90\% upper confidence limit is then $\mu=2.3$ as shown in Fig.~10.

Now assume the true value is $\mu=0$. Obviously, no event will be found in
repeated experiments and the coverage is 100\% independent of the nominal
confidence value. It is impossible to avoid the complete overcoverage.
Published exotic particle searches are much more often right than indicated by
the given confidence level.%

\begin{figure}[t]
\begin{center}
\includegraphics*[width=\textwidth]{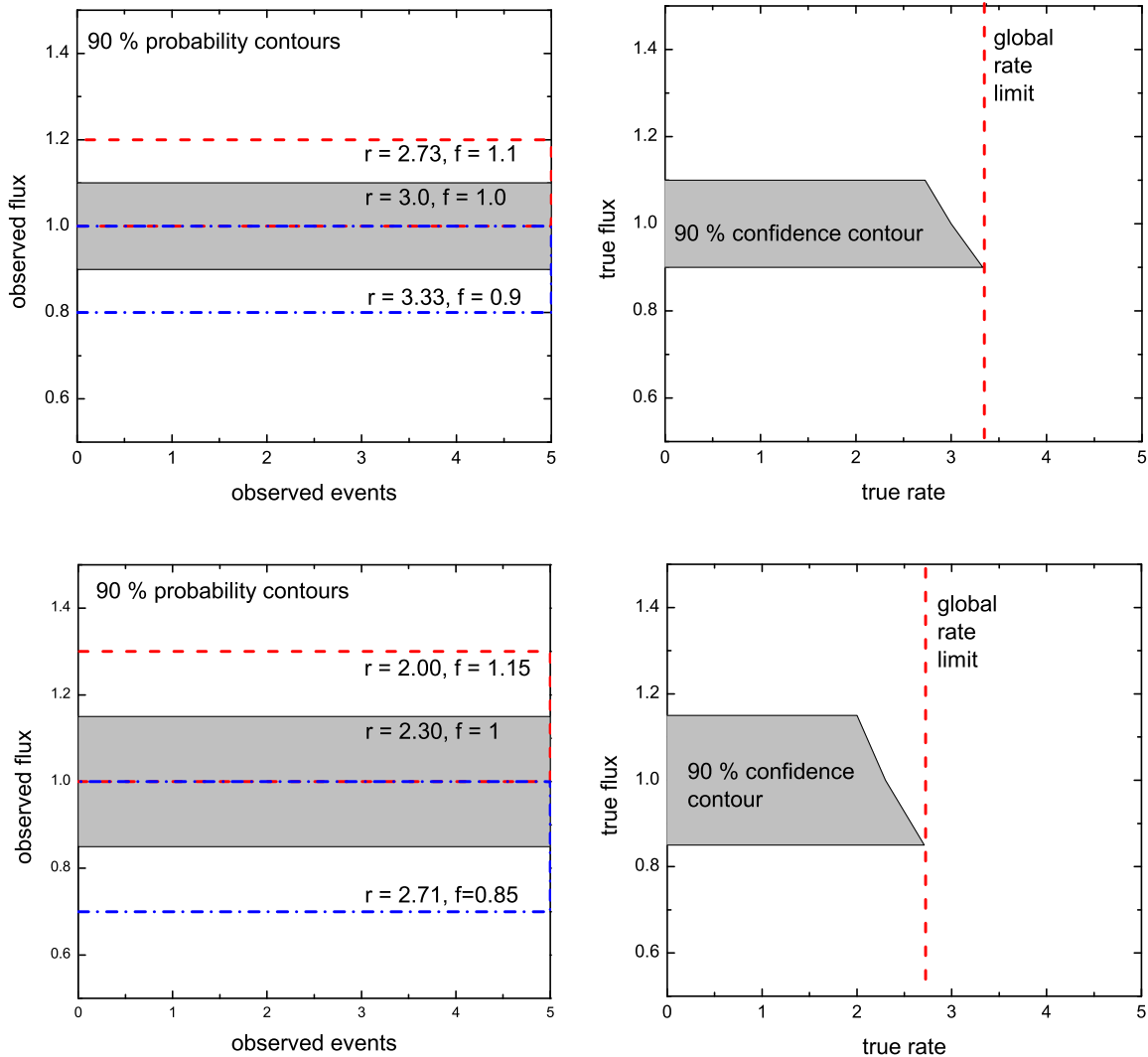}%
\caption{Probability contours (left) and confidence regions (right)
for a Poisson rate with flux uncertainty. The conservative 90\% upper limit of
the rate is indicated by the dashed line. The lower plots with 3~st. dev.
limits in the flux provide more restrictive 90\% limits than the upper 2~st.
dev. flux limits}%
\end{center}
\end{figure}

\subsubsection{Rate limits with uncertainty in the luminosity.}

A rate is obtained by dividing the number of observed events by the luminosity
or flux which usually is not exactly known. We are only interested in the
ratio, the flux is a nuisance parameter which is to be eliminated. In Refs.
\cite{cous95, cous00} it was claimed, that the flux uncertainty improves the
frequentist limit. This would be an inconsistent result. We apply the
procedure outlined above in Sect.~2.72 and obtain a conservative global limit.

\begin{example}
We observe zero events $n=0$ of a certain type and measure the flux
$f=1\pm0.05$ and want to compute an upper limit for the rate. To simplify the
problem for the present purpose, we assume that the observed flux follows a
Gaussian distribution with width independent of the mean value. The event
number is Poisson distributed.\ As usual, we first have to construct
probability contours in the sample space. Figure~11 left shows \ 90\%
probability contours for three different combinations of true flux and true
rate. Each combination is chosen such that the observation $n=0,f=1$ is
located at the border of the probability region. In the upper plot the range
for the observed flux was selected smaller than in the lower plot. The
corresponding confidence contours are displayed at the right hand side. The
global limits for the rate are indicated by a dashed line. There is an
infinite number of ways to construct the 90\% probability region. Each of them
will lead to a different global limit. The construction belonging to the lower
plots obviously is superior to the first because it gives a more restrictive
limit. An optimization has not been done. In any case the limit is
considerably worse than in an experiment without uncertainty in the flux in
contradiction to Refs. \cite{cous95, cous00}.
\end{example}

\subsubsection{Poisson limits with background.}

The situation becomes complex when the experimental data contain background
indistinguishable from signal. Assuming the background expectation $b$ is
precisely known the probability to find $k$ events (background plus signal)
is
\begin{align}
W(k)  & =\sum_{i=0}^{k}\sum_{j=1}^{k}P(i|\mu)Q(j|b)\delta_{i+j,k}\\
& =\sum_{i=0}^{k}P(i|\mu)Q(k-i|b)
\end{align}
with $Q(j|b)$ the background distribution. (We sum over all combinations of
background $j$ and signal $i$ which add up to $k$ events.) Usually the
background also follows a Poisson distribution.
\begin{align}
W(k)  & =\sum_{i=0}^{k}P(i|\mu)P(k-i|b)\\
& =P(k|\mu+b)
\end{align}
Then the probability to find less than or equal to $n$ events is
\begin{align}
1-\alpha & =\sum_{k=0}^{n}W(k)\nonumber\\
& =\sum_{k=0}^{n}P(k|\mu+b)
\end{align}

Solving the last equation for $\mu$, we get the upper limit with confidence
$\alpha$. Apparently, for $n$ given, the limit becomes more restrictive the
larger is $b$.

In experiments with large background, occasionally, due to background
fluctuations, the numerical evaluation produces even negative limits or zero
length limits. The limits then do not represent the precision of the
experiment which certainly is not infinite.

It is instructive to study the case where no event is observed but where
background is expected:

\begin{example}
In a garden there are apple and pear trees near together. Usually during night
some pears fall from the trees. One morning looking from his window, the
proprietor who is interested in apples find that no fruit is lying in the
grass. If there were any, since it is still quite dark, he would be unable to
distinguish apples from pears. He concludes that the mean rate of falling
apples per night is less the 2.3 with 90\% confidence level. His wife who is a
classical statistician tells him that his rate limit is too high because he
has forgotten to subtract the expected pears background. He argues, ``there
are no pears'', but she insists and explains him that if he ignores the pears
that could have been there but weren't, he would violate the coverage
requirement. In the meantime it has become bright outside and pears and apples
- which both are not there - would now be distinguishable. Even though the
evidence has not changed, the classical limit has.
\end{example}

The 90\% confidence limits for zero events observed and background expectation
$b=0$ is $\mu=2.3$. For $b=2$ it is $\mu^{\prime}=0.3$ much lower.
\emph{Classical confidence limits are different for two experiments with
exactly the same experimental evidence relative to the signal (no signal event
seen)}. This conclusion is absolutely unacceptable. The classical procedure is
inconsistent in this case: The two experimental results $n=0,b=0$ and
$n=0,b=0$ warrant the same conclusion for the rate $\mu$ but lead to different
interval sizes.

Frequentists argue that in the classical approach the long-term properties of
many experiments are well defined. This is not in contradiction to the fact
that there are severe problems in the evaluation of single experiments with
low statistics. We will come back to this discussion in Sect.~6.

Feldman and Cousins consider the objections to the classical result as ``based
on a misplaced Bayesian interpretation of classical intervals'' \cite{feld98}.
It is hard to detect a Bayesian origin in a generally accepted principle in
science, namely, two observations containing the same information should give
identical results. The criticism here is not that CCLs are inherently wrong
but that their application\textbf{\ }to the computation of upper limits when
background is expected does not make sense, i.e. these limits do not measure
the precision of the experiment which is the only porpose of error intervals.
This is also illustrated in the following example which is a more scientific
replicate of Example 11:

\begin{example}
\textbf{\ }An experiment is undertaken to search for events predicted by some
exotic theory. In a pre-defined kinematic region \emph{no event} is found. A
search in a corresponding control region predicts $b$ background events. An
upper limit is computed. After the limit has been published a student
discovers a new kinematical cut which completely eliminates all background.
The improved analysis produces a much less stringent classical limit than the
original one!
\end{example}

The 90\% upper limits for some special cases are collected in Table~2. The
upper rate limit for no event found but three background events expected is negative.%

\begin{table}[tbp]
\caption{90 percent confidence limits in frequentist and Bayesian approaches for
n observed events and background expectation~b}%
\vskip7pt
\centering
\small
\begin{tabular}
[c]{|l|l|l|l|l|l|}\hline
& n=0, b=0 & n=0, b=1 & n=0, b=2 & n=0, b=3 & n=2, b=2\\\hline
standard classical & 2.30 & 1.30 & 0.30 & -0.70 & 3.32\\\hline
unified classical & 2.44 & 1.61 & 1.26 & 1.08 & 3.91\\\hline
uniform Bayesian & 2.30 & 2.30 & 2.30 & 2.30 & 3.88\\\hline
\end{tabular}
\end{table}

\subsubsection{Limits with re-normalized background.}

To avoid the unacceptable situation, I had proposed \cite{zech89} a modified
frequentist approach to the calculation of the Poissonian limits. It takes
into account that the background $k$ has to be less or equal to the number $n
$ of observed events. For example, if $n=1$, we know for sure that the
background is either $0$ or $1$. Then only the probability ratio $Q(1)/Q(0)$
is relevant, $Q(k>1)$ is irrelevant for the inference of the signal from the
available data. The a priori background distribution for the condition ($k\leq
n$) is then:
\[
Q^{\prime}(k|b)=\frac{Q(k|b)}{\sum_{i=1}^{n}Q(i|b)}
\]
We replace $Q$ by $Q^{\prime}$ in Eq.~(5) and obtain for the Poisson case:
\begin{equation}
1-\alpha=\frac{\sum_{k=0}^{n}P(k|\mu+b)}{\sum_{k=0}^{n}P(k|b)}%
\end{equation}
The interpretation is: The probability $1-\alpha$ to observe less or equal $n$
events (signal + background) for a signal mean equal $\mu$, a background mean
equal $b$ with the restriction that the background does not exceed the
observed number is given by Eq.~(9). The resulting limits respect the
Likelihood Principle (see Sect.~5) and thus are consistent. The standard
classical limits depend on the background distribution for background larger
than the observed event number. This information which clearly is irrelevant
for estimating the signal is ignored in the modified approach (Equ. 9).

The formula 9 accidentally coincides with that of the uniform Bayesian method.
Interesting applications of the method with some variations are found in Refs.
\cite{read97, gan98, read00, con00}.

Formula 9 has been criticized by Highland \cite{high97} because the method
does not respect the coverage requirement. This is correct, but coverage had
not be claimed in my paper. A reply is given in Ref. \cite{zech97} and a
further discussion can be found in Ref. \cite{cous00a}.

\ In view of the unavoidable complete overcoverage for $\mu=0$ of all
classical methods, the moderate overcoverage of the Relation 9 which avoids
inconsistencies seems to be acceptable to many pragmatic frequentists.

\subsubsection{Uncertainty in the background prediction.}

Often the background expectation $b$ is not known precisely. We have to
distinguish two cases.

a) The pdf $g(b)$ of $b$ is known. Then we can integrate over $b$ and obtain
for the conventional classical expression
\[
1-\alpha=\int g(b)\sum_{k=0}^{n}P(k|\mu+b)db
\]
and the modified frequentist Formula 9 becomes%

\[
1-\alpha=\frac{\int g(b)\sum_{k=0}^{n}P(k|\mu+b)db}{\int g(b)\sum_{k=0}%
^{n}P(k|b)db}
\]

b) We know only the likelihood function of $b$ for example if it is estimated
from side bands or from other measurements with limited statistics. Then $b$
is a nuisance parameter. Thus the methods outlined in Sect.~2.7 have to be
applied. Again, I would not support the proposal of Cousins \cite{cous00},
``replace the nuisance parameter by the best estimate'' since the result often
violates the coverage principle, the coverage is not necessarily greater than
or equal to $\alpha$ for all values of the signal. The ``global'' method
(Sect.~2.72) should be used to eliminate the nuisance parameter.

\subsection{Discrete parameters}

For discrete parameters usually it does not make much sense to give error
limits but we would like to define relative confidence values for the parameters.

\begin{example}
\textbf{\ }Two theories H$_{1}$, H$_{2}$ predict the time of an earthquake
with Gaussian resolution:
\begin{align*}
\text{H}_{1}\text{{}}:t_{1}=(\text{ }7.50\pm2.25)\text{ h}\\
\text{H}_{2}\text{{}}:\text{ }t_{2}=(50\pm100)\text{ h}%
\end{align*}
The event actually takes place at time $t_{m}=10$ h. The predictions together
with the observation are displayed in Fig.~12 top. The prediction of H$_{2}$
is rather vague but includes the observation within half a standard deviation.
H$_{1}$ is rather precise but misses the observation within the errors. There
is no obvious way to associate classical confidence levels to the two possible
solutions. A rational extension of the frequentist methods applied to
continuous parameters would be to compare the tail probabilities of the two
hypotheses. Tail probabilities, however, are quite misleading as will be
illustrated below in Example 14. They would support H$_{2}$ contrary to our
intuition which is clearly in favor of H$_{1}$. For this reason classical
statisticians prefer to apply the Neyman-Pearson test based on the likelihood
ratio, in our case equal to 26 in favor of H$_{1}$.
\end{example}

A similar example is analyzed in detail in an article by Jeffreys and Berger
\cite{jeff92}.%

\begin{figure}[pt]
\begin{center}
\includegraphics*[width=.75\textwidth]{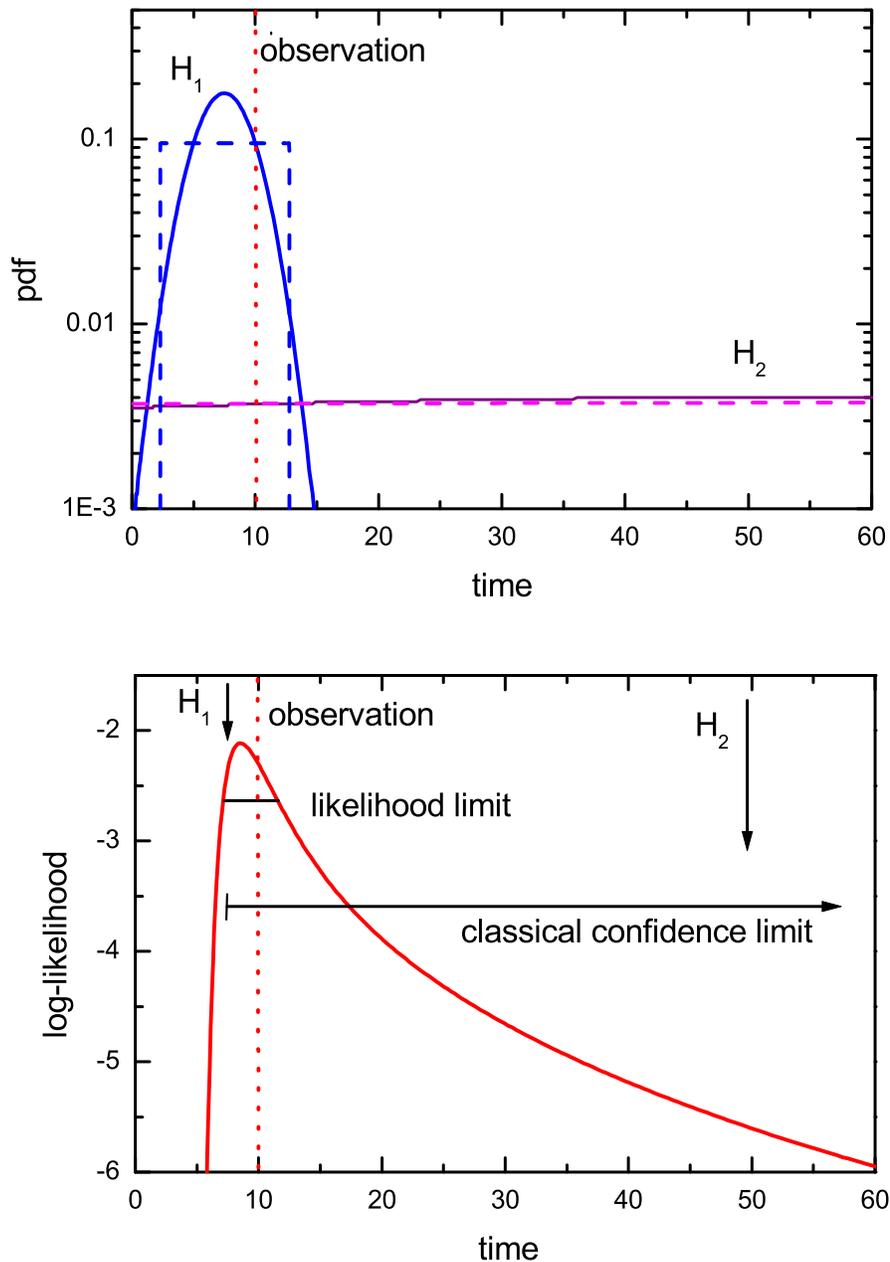}%
\caption{Predictions from two discrete hypothesis H$_{1}$,H$_{2}$
and observation (top) and log-likelihood for H$_{3}$, a
parametrization of H$_{1}$ and H$_{2}$ (bottom). The likelihood
ratio strongly favors H$_{1}$ which
is excluded by the classical confidence limits}%
\end{center}
\end{figure}

It is interesting to consider the modified example:

\begin{example}
Another theory, H$_{3}(t_{3}),$ depending on the unknown parameter $t_{3}$
predicts the Gaussian probability density
\[
f(T)=\frac{25}{\sqrt{2\pi}t_{3}{}^{2}}\exp\left(  -\frac{625(T-t_{3})^{2}%
}{2t_{3}^{4}}\right)
\]
for the time $T$. The classical confidence limits for the same
observation as above, $t_{m}=10$ h, are $7.66$ h $<t_{3}<\infty$,
is slightly excluding $t_{1} $ but in perfect agreement with
$t_{2}$. Fig.~12 bottom shows the corresponding likelihood
function and the classical confidence limit.
\end{example}

The probability density of our example has been constructed such that it
includes H$_{1}$ and H$_{2}$ for the specific values of $t_{3}$ equal
$t_{1},t_{2}.$ Thus the likelihood ratio $f(t_{1})/f(t_{2})$ is identical to
that of the previous example. Let us assume that the alternative theories
H$_{1}$ and H$_{2}$ (which are both compatible with H$_{3}$) were developed
after H$_{3}$. H$_{1}$ which is by far more likely than H$_{2}$ could have
been excluded on the basis of the observation and the classical confidence limits.

A decisions in favor of one of two alternative hypothesis, according to the
Neyman-Pearson Lemma should be based on the likelihood ratio only. Here,
frequentists usually do not consider the full sample space and do not compute
the coverage for the two hypotheses.

What happens, when we add to the first two discrete parameter values a third,
forth, fifth parameter and so on? We may construct a transition from the
discrete case to the continuous one by adding more and more hypotheses. At a
certain point, frequentists would switch from characterizing the situation by
likelihood ratios to using coverage intervals.

The two classical concepts, confidence limit and Neyman-Pearson test, lack a
common basis. Frequentists would argue that there are two different methods
for two different goals. The question is then: What are there two different
goals exactly and why are we interested in the likelihood ratio in one
situation and in coverage in the other?

\section{Unified approaches}

Feldman and Cousins \cite{feld98} have proposed a new approach to the
computation of \emph{classical confidence bounds} which avoids the occurrence
of non-null (or alternatively non-physical) confidence regions, one of the
most problematic features of the conventional classical confidence limits. In
addition it unifies the two procedures ``\emph{computation of confidence
intervals}'' and ``\emph{computation of one-sided confidence limits}''. The
unified treatment has already been adopted by several experiments and is
recommended by the Particle Data Group \cite{pdg98}. However, as shown below,
it has serious drawbacks.

\subsection{Basic ideas of the unified approach}

The unified approach has two basic ingredients:

1) It unifies the two procedures ``computation of a two-sided interval'' and
``computation of an upper limit''. The scientist fixes the confidence level
before he looks at the data and then the data provide either an error interval
or \ a lower or upper limit, depending on the possibility to obtain an error
interval within the allowed physical region. The method thus avoids a
violation of the coverage principle which is obvious in the commonly used
procedure where the selection of one or two sided bounds is based on the data
and personal prejudice.

2) It uses the \emph{likelihood ratio ordering} principle (see Sect.~2.2)
which has the attractive property to be invariant against transformations of
the sample space variables. In addition, unphysical intervals or limits are
avoided. Here the trick is to require that the quantity $\theta_{best}$ of
relation (1) is inside the range allowed by the laws of physics. As discussed
in Sect.~2, the likelihood ordering corresponds to MSU intervals and is an
old concept. What is new is the application to cases where the parameter range
is restricted by external bounds.

In practice, the main impact of this method is on the computation of upper
limits. Experiments have ``improved'' their upper limits by switching to the
unified approach \cite{karm99,baud99}.

The new approach has attractive properties, however, all those problems like
the treatment of nuisance parameters which are intrinsic to the philosophy of
classical statistics remain unsolved.

Kendall and Stuart \cite{kend235} write in connection with the likelihood
ratio ordering (notation slightly modified) ``For the likelihood ratio method
to be useful ... the distribution of $R(X|\theta)$ has to be free of nuisance parameters.''

Additional complications are introduced by the requirement that $\theta
_{best}$ has to be inside the allowed parameter space \cite{zech98}.
Pathological cases have also been presented by Punzi \cite{punz99} and Bouchet
\cite{bouc00}.

\subsection{Difficulties with two-sided constraints}

One of the advantages of the unified approach is the improved handling of
physical bounds. This is shown in Fig.~6 for a Gaussian with external
bounds. The unified intervals avoid unphysical values of the parameter but
unfortunately this problem persists when a parameter is bounded from both sides.

\begin{example}
We resume Example 6, Fig.~8. A particle track is measured by a combination
of a proportional wire chamber and a position detector with Gaussian
resolution. Let us assume an observation $\hat{x}=0$ of a parameter $\mu$ with
a physical bound $-1<\mu<1$ and a Gaussian resolution of $\sigma=1.1$. The
Bayesian r.m.s. error computed by integrating the likelihood function is 0.54.
Since there are two boundaries the procedure applied in Example 4 no longer
works. Requiring a 68.3\% confidence level produces at the same time an upper
and a lower limit. It is impossible to fulfill the coverage requirement,
except if the complete range of $x$ is taken (complete coverage).
\end{example}

A similar example is discussed in the classical book by Kendall and Stuart
\cite{ken167}: ``It may be true, but would be absurd to assert $-1\leq\mu
\leq+2$ if we know already that $0\leq\mu\leq1$. Of course we could truncate
our interval to accord with the prior information. In our example, we could
assert $0\leq\mu\leq1$: the observation would have added nothing to our knowledge.''

\subsection{External constraint and distributions with tails}

\textbf{\ }Difficulties occur also for location parameters with one-sided
physical bounds when the resolution function has tails. In the unified
approach, unphysical confidence intervals are avoided by adding preferentially
those parts of the sample space to the probability region which are located in
the unphysical region. However, the corresponding likelihood ratio ordering
may produce disconnected probability regions and consequently also
disconnected confidence intervals if the pdf decreases slowly in the
unphysical part of the sample space.\ Disconnected confidence intervals do not
make sense for measurements following smooth probability distributions with a
single maximum.%

\begin{figure}[ptb]
\begin{center}
\includegraphics*[width=.75\textwidth]{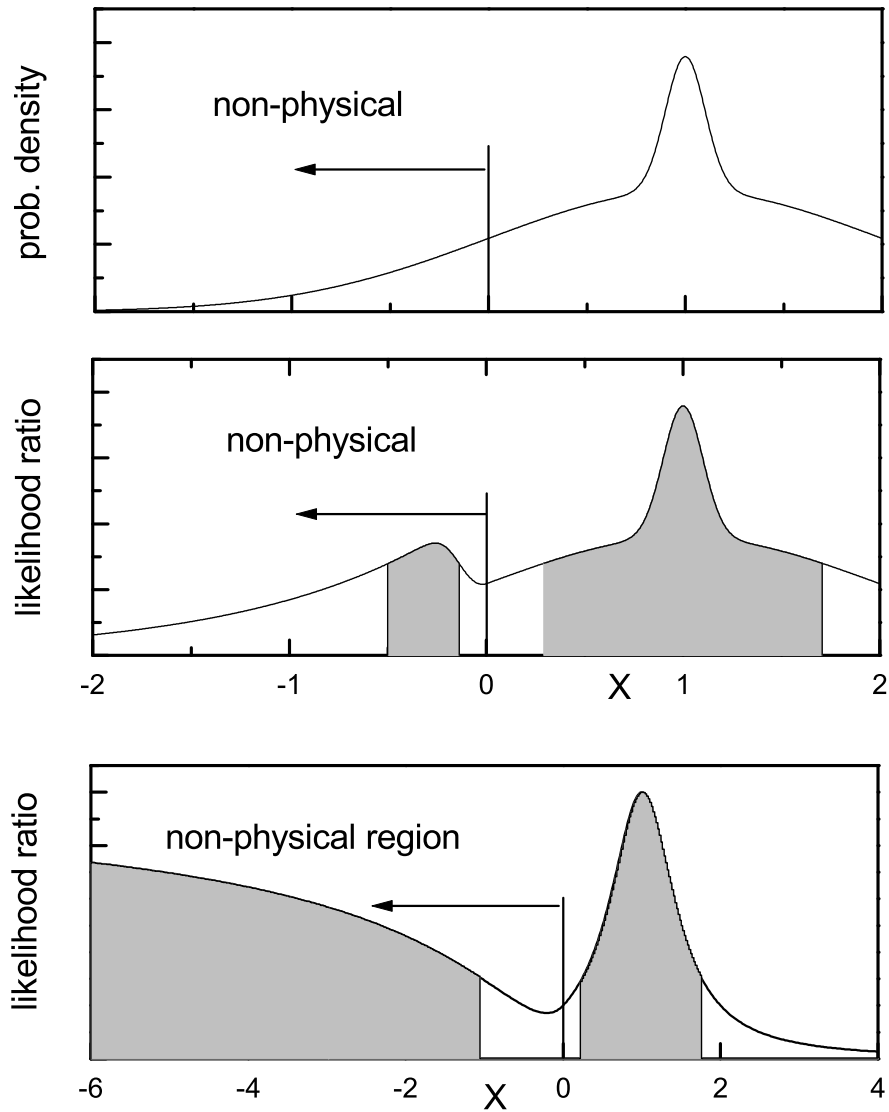}%
\caption{Probability distribution (top) and corresponding likelihood
ratio (center) for the superposition of two Gaussians in the unified approach.
Since regions are added to the probability interval using the likelihood ratio
as an ordering scheme, disconnected intervals (shaded) are obtained. A
Breit-Wigner pdf shows a similar behavior (bottom)}%
\end{center}
\end{figure}

\begin{example}
We consider the superposition of a narrow and a wide Gaussian. (It is quite
common that distributions have non-Gaussian tails.)
\begin{equation}
f(X|\mu)=\frac1{\sqrt{2\pi}}\left\{  0.9\exp\left(  -(X-\mu)^{2}/2\right)
+\exp\left(  -(X-\mu)^{2}/0.02\right)  \right\}
\end{equation}
with the additional requirement of positive parameter values $\mu$. This pdf
is shown in Fig.~13 top for $\mu=1$. The likelihood ratio for an observation
$x=1$ is displayed in the center of Fig.~13. Adding $X$-values to the
probability interval according to the likelihood ratio produces disconnected
probability interval regions which do not make sense. It is impossible to
construct simply connected confidence intervals.
\end{example}

The same difficulty arises for the Breit-Wigner distribution, where again the
likelihood ratio is shown in Fig.~13 bottom for an observation $x=1$.

In fact, it is only a very special class of distributions for which we can
guarantee that we do not get disconnected confidence intervals. For pdfs
depending on $|X-\theta|$ like those we just discussed, disconnected
probability intervals are avoided if the likelihood ratio $R=f(X|\theta
)/f(X|\theta_{best})$ decreases monotonically from the center of the
distribution towards the tails. The requirement $dR/dX<0$ for $X>\theta$ is
equivalent to
\begin{align*}
\frac d{dX}\ln f(X|\theta_{best})  & >\frac d{dX}\ln f(X|\theta)\\
\frac d{dX}\ln f(X|\theta+\Delta)  & >\frac d{dX}\ln f(X|\theta)\\
\frac d{dX}\ln f(X-\Delta|\theta)  & >\frac d{dX}\ln f(X|\theta)\\
\frac d{dX}\frac{\ln f(X|\theta)-\ln f(X-\Delta|\theta)}\Delta & <0
\end{align*}
where we have set $\Delta=\theta_{best}-\theta$. All values of $\Delta$ can
occur. In the limit $\Delta\rightarrow0$ we get
\[
\frac{d^{2}\ln f}{dX^{2}}<0
\]

Thus, the problem is absent for pdfs with convex logarithms. Integrating the
last relation twice, we find that near physical boundaries the unified
approach in the present form is essentially restricted to Gaussian like pdfs.

\subsection{Artificial correlations of independent parameters}

Since the confidence intervals are defined through coverage, one has to be
careful in the interpretation of error bounds near physical boundaries:
Artificial correlations between parameters are introduced.

\begin{example}
The uncorrelated variates $X,Y$ follows a two dimensional
Gaussian. The two variances are equal and known, the center
$\mu_{x},\mu_{y}$ is unknown. Figure~14 shows schematically the
circular probability contour for the conventional classical
approach and the modified contour of the unified approach which
has shrunk in $X$ due to the boundary in $Y$. An observation
$x,y$ near the boundary will lead to confidence contours which
reflect this modified shape. The error for $\mu_{x}$ shrinks due
to the unphysical $\mu_{y}$ region.
\end{example}%

\begin{figure}[ptb]
\begin{center}
\includegraphics*[width=3.9712in]{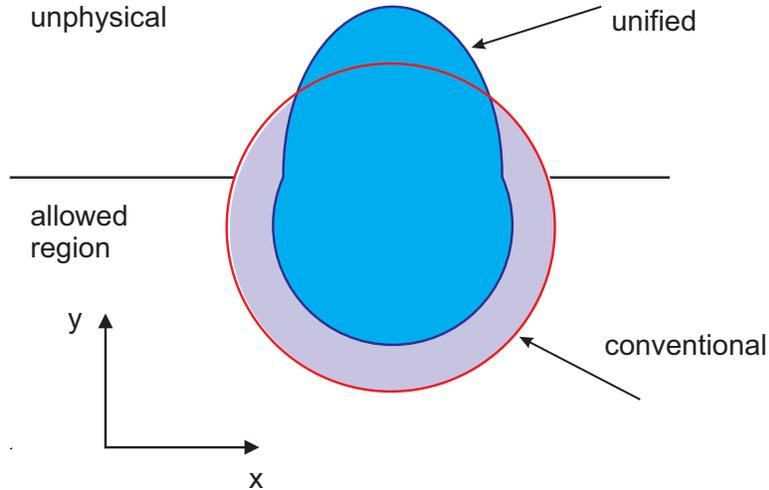}%
\caption{Probability contours (schematic) for a two-dimensional
Gaussian near a boundary in the unified approach}
\end{center}
\end{figure}

Probably one would try to avoid \ the artificially introduced correlation and
treat the two coordinates independently but even a slight real correlation
would inhibit this possibility.

\subsection{Several bounded parameters}

The prescription to restrict the value $\theta_{best}$ to the physically
allowed region is easily extended to the situation where we have to determine
confidence limits for several bounded parameters. However, this case has not
been studied and very strange results might be obtained in some cases. Feldman
and Cousins \cite{feld98} apply their method to a two-dimensional toy model
and show that the method works technically. A thorough study is still missing.

\subsection{Upper Poisson limits}%

\begin{figure}[ptb]
\begin{center}
\includegraphics*[width=.8\textwidth]{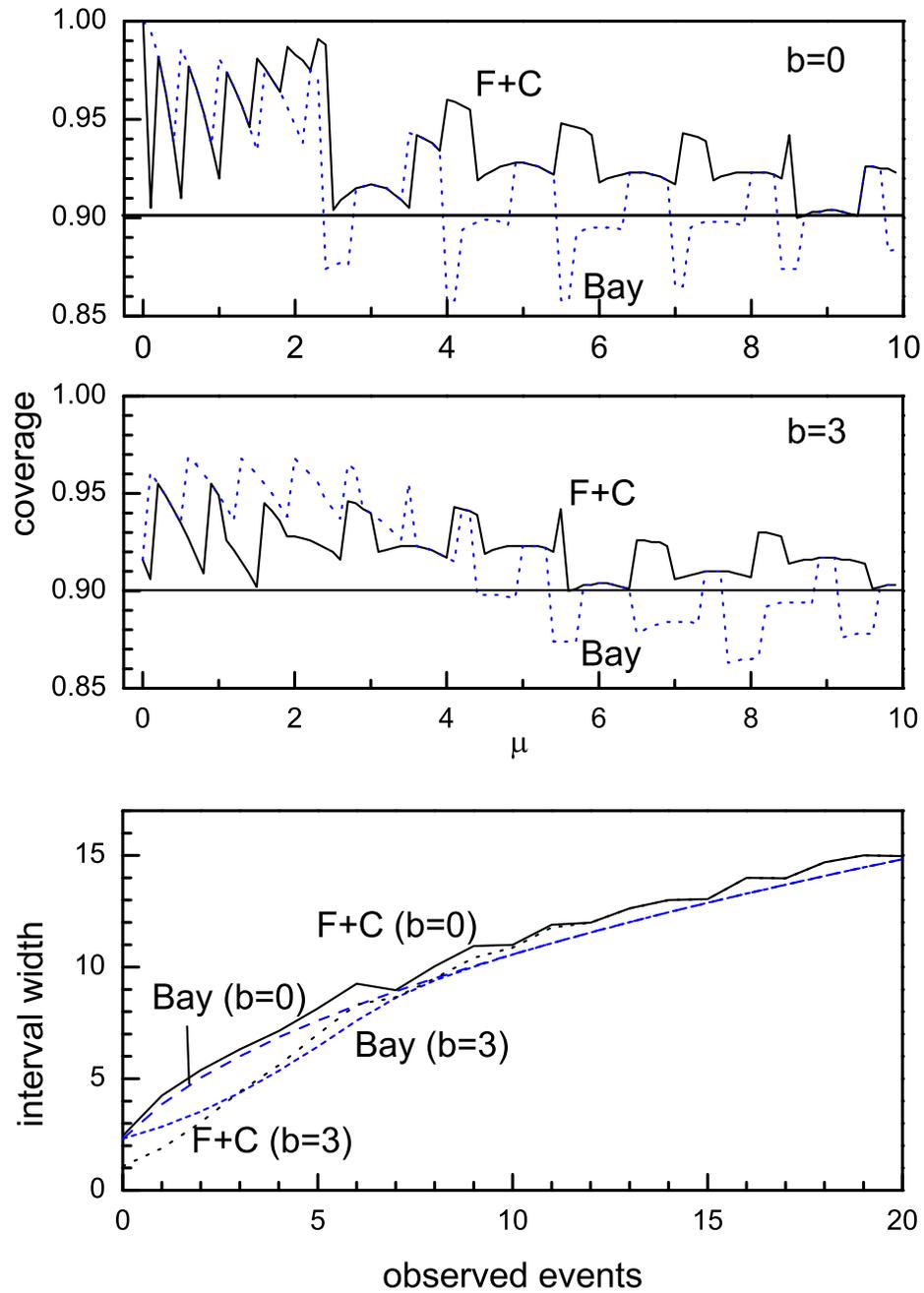}%
\caption{Coverage and confidence interval width for the unified
approach and the Bayesian case. The Bayesian curves are computed
according to the unification prescription}
\end{center}
\end{figure}

Table 2 contains $90\%$ C.L. upper limits for Poisson distributed signals with
background. For the case $n=0,b=3$ the uniform approach avoids the unphysical
limit of the conventional classical method but finds a limit which is more
restrictive than that of a much more sensitive experiment with no background
expected and twice the flux! Compared to the conventional approach, the
situation has improved - from a Bayesian point of view - but the basic problem
is not solved and the inconsistencies discussed in Sect.~2.9 persist.%

\begin{figure}[ptb]
\begin{center}
\includegraphics*[width=.9\textwidth]{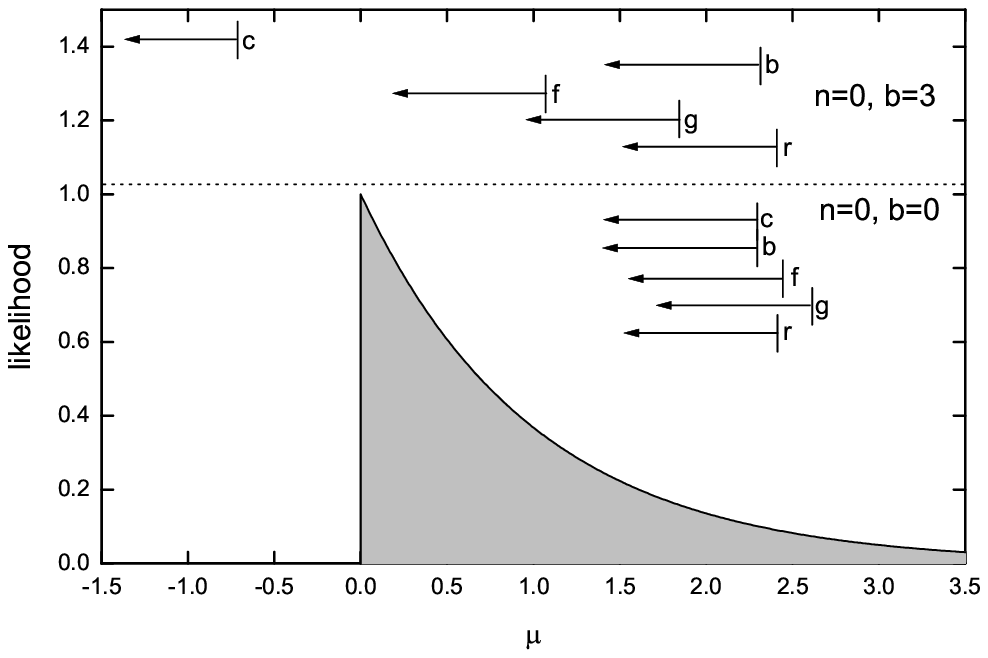}%
\caption{Likelihood function for zero observed events and 90\%
confidence upper limits with and without background expectation.
The labels refer to \cite{feld98} (f), Bayesian (b), \cite{giun99}
(g) and \cite{roe99} (r)}
\end{center}
\end{figure}

Figure 15 compares the coverage and the interval lengths of the unified method
with the Bayesian one with uniform prior (to be discussed below). The nominal
coverage is 90\%. Unavoidably for $\mu=0$, $b=0$ there is maximum
over-coverage and in the range $0\leq\mu\leq2.2$ the average coverage is 96\%.

\subsection{Restriction due to unification}

Let us assume that in a search for a Susy particle a positive result is found
which however is compatible with background within two standard deviations.
Certainly, one would prefer to publish an upper limit to a measurement,
contrary to the prescription of the unified method. The authors defend their
method arguing that a measurement with a given error interval always can be
interpreted as a limit - which is true.

\subsection{Alternative unified methods}

Recently modified versions of the unified treatment of the Poisson case have
been published \cite{giun99, roe99, punz99, roe00, mand00}.

The results of Giunti \cite{giun99} are nearer to the uniform Bayesian ones,
but wider than those of Feldman and Cousins.

Roe and Woodroofe \cite{roe99} re-invented the method proposed by myself ten
years earlier \cite{zech89} and added the unification principle. The authors
introduced the concept of \ ``ancillary variable'' for the background and
condition on its upper limit given by $n$. Since this number obviously is not
an ancillary statistic, their concept fails to provide the required coverage
properties. In their second paper \cite{roe00} they present a Bayesian method
which is claimed to have good coverage properties, but this is also true for
the standard Bayesian approach.

Punzi \cite{punz99} modifies the definition of confidence intervals such that
they fulfill the Likelihood Principle. This approach is in some sense
attractive but the intervals become quite wide and it is difficult to
interpret them as standard error bounds.

Ciampolillo \cite{ciam98} uses the \ likelihood statistic which maximizes the
likelihood inside the physically allowed region as estimator. In most cases it
coincides with the parameter or parameter set which maximizes the likelihood
function. Usually, the maximum likelihood estimator is not a sufficient
statistic and thus will not provide optimum precision (see Example 2).
Similarly, Mandelkern and Schulz \cite{mand00} use an estimator confined to
the physical domain. Both approaches lead to intervals which are independent
of the location of the observation within the unphysical region. I find it
quite unsatisfactory that two observations following Gaussian pdfs
$x_{1}=-0.1$ and $x_{2}=-2$ with the bound $\mu>0$ and width $\sigma=1$ yield
the same confidence interval. Certainly, our betting odds would be different
in the two cases.

In summary, from all proposed unified methods only the Feldman/Cousins
approach has a simple statistical foundation. Zech/Roe/Woodroofe's method is
not correct from a classical frequentist point of view, Punzi's method is
theoretically interesting but not suited for error definitions, and the other
prescriptions represent unsatisfactory ad-hoc solutions.%

\begin{figure}[t]
\begin{center}
\includegraphics*[width=.9\textwidth]{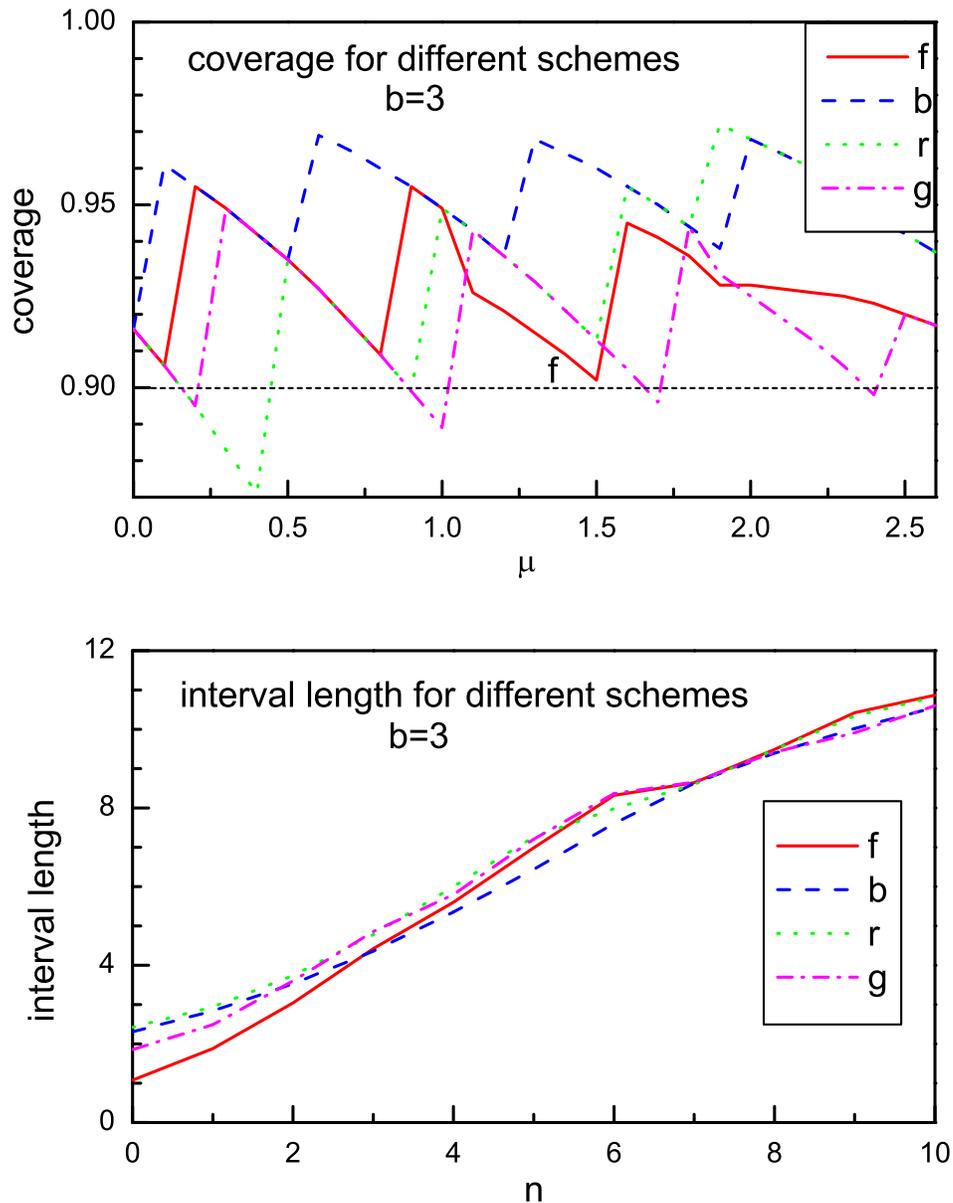}%
\caption{Top: Coverage as a function of the Poisson rate for
expected background $b=3$. The nominal coverage is 90\%. The
labels refer to \cite{feld98} (f), Bayesian (b), \cite{giun99} (g)
and \cite{roe99} (r). Bottom: Interval length as a function of the
observed number of events}
\end{center}
\end{figure}

The Figs.~16 and 17 compare upper limits, coverage and interval lengths for
some approaches for $b=0$ and $b=3$. (Part of the data has been taken from
Ref. \cite{roe99}.) The likelihood functions for $n=0,$ $b=0$ and $n=0,$ $b=3$
are of course identical up to an irrelevant constant factor. Apparently,
Feldman and Cousins avoid under-coverage while the other approaches try to
come closer to the nominal confidence level but in some cases are below. The
interval widths are similar for all methods.

\section{Likelihood ratio limits and Bayesian confidence intervals}

\subsection{Inverse probability}

Bayesians treat parameters as random variables (see also Appendix A). The
combined probability density $f(X,\theta)$ of the variate $X$ and the
parameter $\theta$ can be conditioned on the outcome of one of the two
variates using Bayes theorem:%

\begin{align}
f(X,\theta)  & =f_{x}(X|\theta)\pi_{\theta}(\theta)=f_{\theta}(\theta
|X)\pi_{x}(X)\nonumber\\
f_{\theta}(\theta|X)  & =\frac{f_{x}(X|\theta)\pi_{\theta}(\theta)}{\pi
_{x}(X)}%
\end{align}
where the functions $\pi$ are the marginal densities. The density $\pi
_{\theta}(\theta)$ in this context\textbf{\ }usually is called prior density
of the parameter and gives the probability density for $\theta$ prior to the
observation $x$ of $X$. For a given observation $x$ the conditional density
$f_{x}$ can be identified with the likelihood function. The marginal
distribution $\pi_{x}$ is just a multiplicative factor independent of $\theta$
and is eliminated by the normalization requirement.
\begin{align}
f_{\theta}(\theta|x)  & \propto L(x,\theta)\pi_{\theta}(\theta)\nonumber\\
f_{\theta}(\theta|x)  & =\frac{L(x,\theta)\pi_{\theta}(\theta)}{\int_{-\infty
}^{\infty}L(x,\theta)\pi_{\theta}(\theta)d\theta}%
\end{align}

The prior density $\pi_{\theta}$ has to guarantee that the normalization
integral is finite.

In the literature $f_{\theta}$ often is called \emph{inverse probability} to
emphasize the change of role between $X$ and $\theta$.

The relation (12) contains one parameter and one observation, but $x$ can also
be interpreted as a ``vector'' ($x_{1},x_{2},..)$, a set of individual
observations of independent and identically distributed (i.i.d.) \ variates,
following the same distribution $f_{0}(X|\theta)$
\[
f(X|\theta)=\prod_{i}f_{0}(X_{i}|\theta)
\]
or any kind of statistic and $\theta$ may be a set of parameters.

\subsection{Interval definition}%

\begin{figure}[ptb]
\begin{center}
\includegraphics*[width=.9\textwidth]{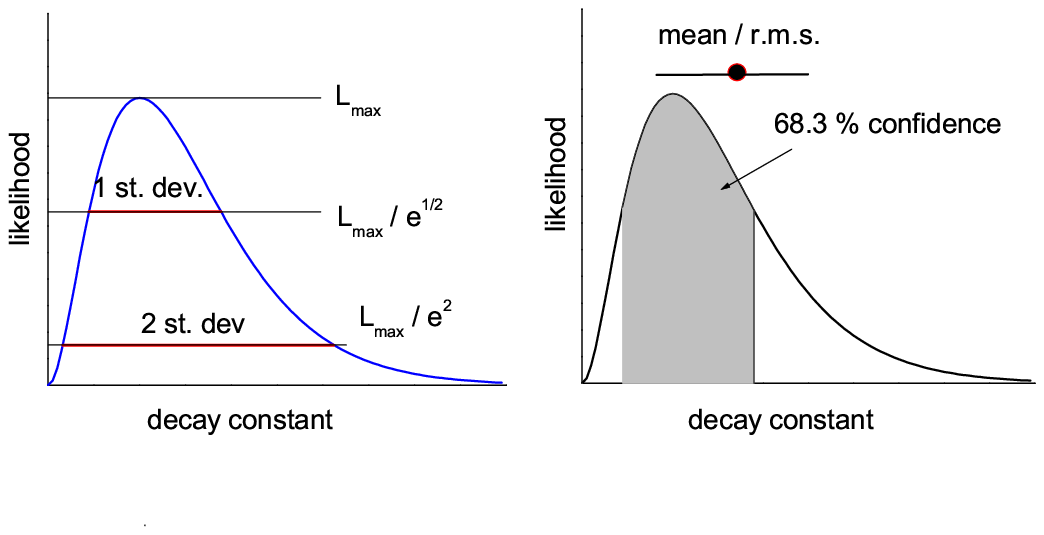}%
\caption{Likelihood ratio limits (left) and Bayesian limits
(right)}
\end{center}
\end{figure}

\subsubsection{Bayesian intervals.}

Having the probability density of the parameter in hand, it is easy to compute
mean values, r.m.s. errors or probabilities for upper or lower bounds (see
Fig.~18).

There is again some freedom in defining the Bayesian limits. One possibility
would be to quote the mean and the variance of the parameter, another
possibility is to compute intervals of a given probability\footnote{In the
literature the expression \emph{degree of belief} is used instead of
probability to emphasize the dependence on a prior density. For simplicity we
will stick to the expression confidence level.}. The first possibility
emphasizes error propagation, the second is better adapted to hypothesis
testing. In the latter case, the interval boundaries correspond to equal
probability density of the parameter\footnote{Central intervals are wider and
restricted to one dimension.} (equal likelihood for a uniform prior). Again a
lack of standardization is apparent for the choice of the interval.

\subsubsection{Likelihood ratio intervals.}

Since the likelihood function itself represents the information of the data
relative to the parameter of interest and is independent of the problematic
choice of a prior it makes sense to publish this function or to parametrize
it. Usually, the log-likelihood, in this context called support function
\cite{jeff36} or support-by-data an expression introduced by Hacking
\cite{hack65}, is mathematically more convenient than the likelihood function itself.

For continuous parameters, usually likelihood limits are given. For a single
parameter, the two equivalent conditions
\begin{align}
L_{\max}/L(\theta_{low})  & =L_{\max}/L(\theta_{high})=e^{\Delta
}\label{elikelim}\\
\ln L(\theta_{low})  & =\ln L_{\max}-\Delta=\ln L(\theta_{high})
\end{align}
where $L_{\max}$ is the maximum of the function \emph{in the allowed parameter
range} $\theta_{\min}<\theta<\theta_{\max}$, fix a likelihood ratio interval
$\theta_{low}<$ $\theta<\theta_{high}$ (see Fig.~18). The values $\Delta=0.5$
and $\Delta=2$ define one and two standard deviation likelihood limits
(support intervals). In the one parameter case limit where $f(X)$ is a
Gaussian (the width being independent of the parameter) these bounds
correspond to classical.C.Ls. of $0.683$ and $0.954$ confidence.

The one-dimensional likelihood limits transform into boundaries in several
dimensions. For example for two parameters we get the confidence contour in
the $\theta_{1}-\theta_{2}$ plane.
\[
\ln L(x;\theta_{1},\theta_{2})=\ln L_{\max}-\Delta
\]
The value of $\Delta L$ is again equal to 0.5 (2) for the 1st. dev. (2 st.
dev.) case. The contour is invariant under transformations of the parameter space.

Upper Poisson limits are usually computed from Bayesian probability intervals.
D'Agostini \cite{dago00cr} emphasizes the likelihood ratio as a sensible
measure of upper limits, for example in Higgs searches. In this way the
dependence of the limit on the prior of the Bayesian methods is avoided.

\subsection{Problems with likelihood ratio intervals}

Many of the situations where the classical methods find difficulties are also
problematic for likelihood ratio intervals.

\begin{itemize}
\item  The elimination of nuisance parameters is as problematic as in the
frequentist methods.

\item  Digital measurements have constant likelihood functions and cannot be handled.

\item  The error limits for functions with long tails (like the Breit-Wigner
pdf) are misleading.

\item  When the likelihood function has its mathematical maximum outside the
physical region ($L_{\max}$ is then at the edge of the physical region), the
resulting one-sided likelihood ratio interval for $\Delta=0.5$ may be
unreasonably short.

\item  The same situation occurs when the maximum is near the border of the
allowed sample space.
\end{itemize}

A frequently discussed example is:

\begin{example}
The width $\theta$ of a uniform distribution
\[
f(X|\theta)=\frac1\theta;\;0<X<\theta
\]
is estimated from $n$ observations $x_{i}$. The likelihood function is
\[
L=\theta^{-n};\;\theta\geq x_{\max}
\]
with a $1/\sqrt{e}$ likelihood ratio interval of $x_{\max}<\theta<x_{\max
}e^{1/(2n)}$ which for $n=10$ is only about half of the classical and the
Bayesian widths. (This example is relevant for the determination of the time
zero $t_{0}$ from a sample of registered drift times.)
\end{example}

We will come back to likelihood ratio intervals in sections 6 and 7. In the
following we concentrate on the Bayesian method with uniform prior but free
choice of parameter.

\subsection{The prior parameter density and the parameter choice}

The Bayesian method would be ideal if we knew the prior of the parameter.
Clearly the problem is in the prior parameter density and some statisticians
completely reject the whole concept of a prior density. The following example
demonstrates that at least in some cases the Bayesian way is plausible.

\begin{example}
An unstable particle with known mean life $\tau$ has decayed at time $\theta$.
We are interested in $\theta$. An experiment with Gaussian resolution $s$,
$\eta(T|\theta,s)$ finds the value $t$. The prior density $\pi(\theta)$ for
$\theta$ is proportional to $\exp(-\theta/\tau).$ Using (12) we get
\[
f_{\theta}(\theta)=\frac{\eta(t|\theta,s)e^{-\theta/\tau}}{\int_{0}^{\infty
}\eta(t|\theta,s)e^{-\theta/\tau}d\theta}
\]
\end{example}

However, there is no obvious way to fix the prior density in the following example:

\begin{example}
We find a significant peak in a mass spectrum and associate it to a so far
unknown particle. To compute the probability density of the mass $M$ we need
to know the a priory mass density $\pi(M)$.
\end{example}

There are all kind of intermediate cases:

\begin{example}
When we reconstruct particle tracks from wire chamber data, we always assume a
flat track density (uniform prior) between two adjacent wires. Is this
assumption which is deduced from experience justified also for very rare event
types? Is it valid independent of the wire spacing?
\end{example}

In absence of quantitative information on the prior parameter density, there
is no rigorous way to fix it. Some Bayesians use scaling laws to select a
specific prior \cite{jeff61} or propose to express our complete ignorance
about a parameter by the choice of a uniform prior distribution (Bayes'
Principle). An example for the application of scaling is presented in Appendix
C. Some scientists invoke the Principle of Maximum Entropy to fix the prior
density \cite{jayn68}.

I cannot find convincing these arguments, but nevertheless I consider it very
sensible to choose a flat prior as is common practice. This attitude is shared
by most physicists \cite{dagos98}\footnote{A different point of view is
expressed in Refs. \cite{cous95, pros88}.}. I do not know of any important
experimental result in particle physics analyzed with a non-uniform prior. In
Example 21 it is quite legitimate to assume a uniform track density at the
scale of the wire spacing.

Similarly for a fit of the Z$^{0}$-mass there are no reasons to prefer a
priori a certain mass within the small range allowed by the measurement. This
fact translates also into the quasi independence of the result of the
parameter selection. Assuming a flat prior for the mass squared would not
noticeably alter the result.

The constant prior density is what physicists use in practice. The probability
density is then obtained by normalizing the likelihood
function\footnote{Normalizing the likelihood function means technically
$L(\theta)/\int_{-\infty}^{\infty}L(\theta)d\theta$ and is the Bayesian pdf
for $\theta$ obtained from a uniform prior $\pi(\theta)=const$.}. The obvious
objection to this recipe is that it is not invariant against parameter
transformations. A flat prior of $\theta_{1}$ is incompatible with a flat
prior of $\theta_{2}$ unless the relation between the two parameters is linear
since we have to fulfill
\[
\pi_{1}(\theta_{1})d\theta_{1}=\pi_{2}(\theta_{2})d\theta_{2}
\]

The formal contradiction often is unimportant in practice unless the intervals
are so large that a linear relation between $\theta_{1}$ and $\theta_{2}$ is a
bad approximation.

One should also note that fixing the prior density to be uniform does not
really restrict the Bayesian choice: There is the additional freedom to select
the parameter. A parameter $\theta_{1}$ with arbitrary prior can be
transformed into a parameter $\theta_{2}(\theta_{1})$ with a constant prior
density. Thus, for example, a physicist who would like to choose a prior
density $\pi_{\tau}\propto1/\tau^{2}$ for the mean life $\tau$ is advised to
use instead the decay parameter $\gamma=1/\tau$ with a constant prior density.
In the following we will stick to the convention of a uniform prior but allow
for a free choice of the parameter.

The convention to use uniform prior pdfs is convenient because it simplifies
the presentation of results. It is enough to state the parameter and the
interval. The scientist has to select a parameter space for which he has no
strong a priori preference for specific parameter values.

An example where the parameter choice matters is the following:

\begin{example}
The decay time of a particle is recorded. The single observation $t$ is used
to estimate its mean life $\tau$. Using a flat prior we find the posterior
parameter density
\[
f(\tau)\sim\frac{1}{\tau}e^{-t/\tau}
\]
which is not normalizable.
\end{example}

Thus a flat prior density is not always a sensible choice for an arbitrarily
selected parameter. In the preceding example the reason is clear: A flat prior
density would predict the same probability for a mean life of an unknown
particle in the two intervals $0<\tau<1ns$ and $1s<\tau<1s+1ns$ which
obviously is a fairly exotic choice. When we choose instead of the mean life
the decay constant $\gamma$ with a flat prior, we obtain
\[
f(\gamma)=\frac{\gamma e^{-\gamma t}}{\int_{0}^{\infty}\gamma e^{-\gamma
t}d\gamma}=t^{2}\gamma e^{-\gamma t}
\]

The second choice is also supported by the more Gaussian shape of the
likelihood function of $\gamma$ as illustrated in Fig.~19 for the
observation of two decays. The figure gives also the distributions of the
likelihood estimators to indicate how frequent they are.%

\begin{figure}[ptb]
\begin{center}
\includegraphics*[width=.8\textwidth]{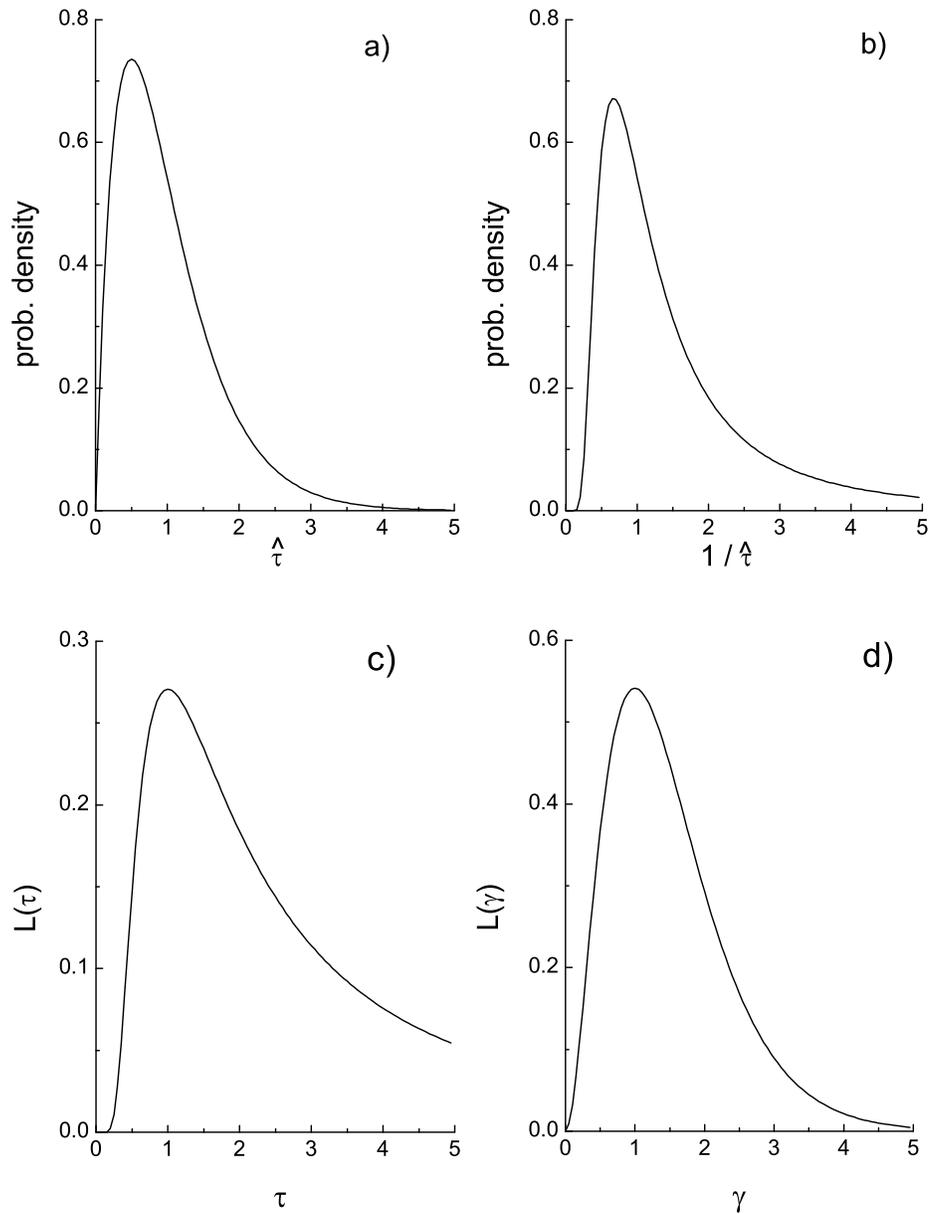}%
\caption{Distribution of the max. likelihood estimates for the
mean life (a) and the decay constant (b) for two decays. The
corresponding likelihood functions are shown in (c) and (d) for
estimates equal to one}
\end{center}
\end{figure}

One possible criterion for a sensible parameter choice is the shape of the
likelihood function. One could try to find a parameter with approximately
Gaussian shaped likelihood function and use a constant prior.

We resume the digital position measurement by a proportional wire chamber
(Example 21). The likelihood is constant inside the interval allowed by the
measuring device. It does not make sense to transform the parameter to obtain
a Gaussian shape. Here it is common practice to use a constant prior
probability \ and to associate to the central value the r.m.s. error interval
$\pm2mm/\sqrt{12}$ of a flat probability distribution.

\subsection{External constraints}

Bayesians account for external constraints by using vanishing prior functions
in the unphysical region which is equivalent to normalizing the parameter
density to the relevant parameter interval as is illustrated in Fig.~20. The
same procedure is applied to the Gaussian of Example~4 in Fig.~6.%

\begin{figure}[ptb]
\begin{center}
\includegraphics*[width=3.6781in]{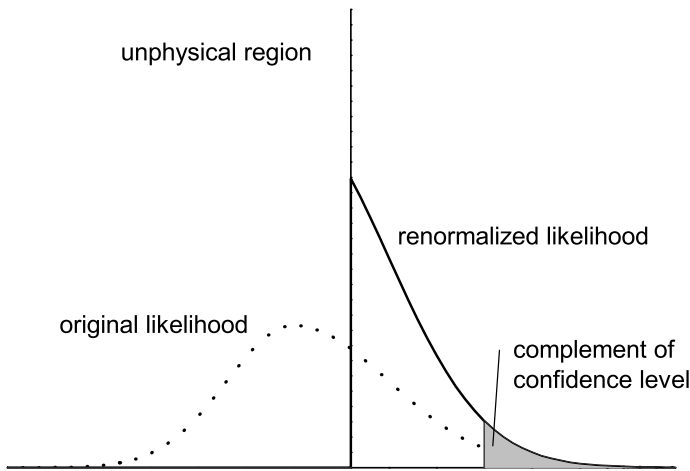}%
\caption{Likelihood function renormalized to the physically
allowed region. The construction of the upper limit is indicated}
\end{center}
\end{figure}

\subsection{Several parameters and nuisance parameters}

The extension of the Bayesian scheme to several parameters is straight
forward. Nuisance parameters can be integrated out. To do so, in principle, we
have to choose the full prior density of all parameters.%

\[
f(\theta|x)\sim\int L(\theta,\phi,x)\pi(\theta,\phi)d\phi
\]
Even in case the likelihood function of the parameters factorizes we cannot
separate the parameters without making the additional assumption that also the
prior density can be written as a product of independent priors.

Since we have decided to accept only uniform prior densities the difficulty is absent.

\subsection{Upper and lower limits}

Bayesians compute the upper limits $\mu$ from the tail of the inverse
probability density which for a constant prior is the normalized likelihood function.%
\begin{equation}
\alpha=\int_{\mu}^{\infty}L(\mu^{\prime})d\mu^{\prime}{\Large /}\int_{-\infty
}^{\infty}L(\mu^{\prime})d\mu^{\prime}%
\end{equation}

For the Poisson process with Poisson distributed background we get the limit
$\mu$ from%
\[
\alpha=\int_{\mu}^{\infty}P(n|\mu^{\prime}+b)d\mu^{\prime}=1-\frac{\sum
_{i=0}^{n}P(i|\mu+b)}{\sum_{i=0}^{n}P(i|b)}
\]
which is identical to the frequentist formula with normalization to possible
background \cite{zech89}, and, for zero background expectation to the
classical limit. Figure~21 compares the classical 90\% confidence limits to
the Bayesian ones for 4 expected background events as a function of the number
$n$ of observed events. The width of the horizontal bins indicates the
frequency of the occurrence of a certain number $n$ of observed events for the
specific case where $\mu=0$. The unphysical limits of the classical approach
for $n=0,1$ occur in about 9\% of the cases.%

\begin{figure}[ptb]
\begin{center}
\includegraphics*[width=.9\textwidth]{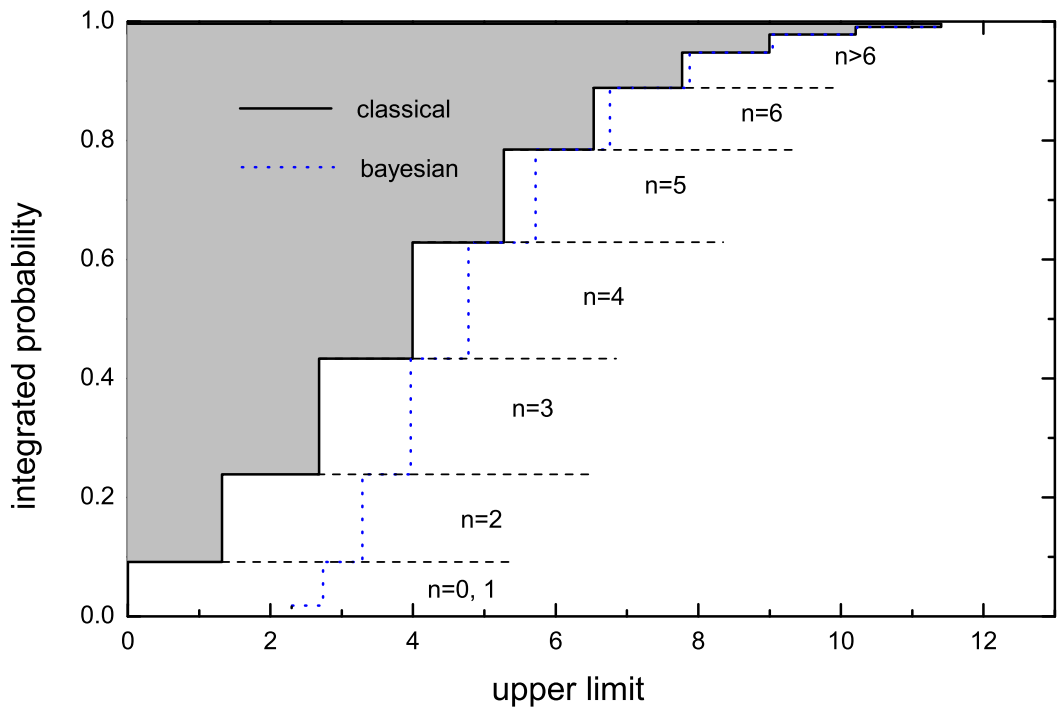}%
\caption{90\% c.l. upper limits in the Poisson case for the
classical and the Bayesian approach. The expected background is 4,
the parameter $n$ is the number of observed events. The unphysical
limits of the classical approach for $n<2$ are suppressed. The
integrated probability corresponds to the true value $\mu=0$}
\end{center}
\end{figure}

It is easy to generalize the Bayesian procedure for arbitrarily distributed
background with the probability $Q(m|b)$ to obtain $m$ background events for a
given expectation $b$.%
\[
L(\mu)=\sum_{i=0}^{n}P(i|\mu)Q(n-i|b)
\]
The likelihood function has to be inserted in Equ. 15. The background
parameter $b$ may follow another distribution $g(b)$%
\[
L(\mu)=\sum_{i=0}^{n}P(i|\mu)\int_{-\infty}^{\infty}Q(n-i|b)g(b)db
\]

The latter expression is useful if the background mean has an uncertainty for
example when it is deduced from the sidebands of an experimental histogram.

\subsection{Discrete parameters}

The discrete case is automatically included in the continuous one if we allow
for $\delta$-functions. The probability for hypothesis $i$ is given by
\[
P_{i}=\frac{L_{i}\pi_{i}}{\sum_{i}L_{i}\pi_{i}}
\]
where the likelihoods are multiplied by the prior probabilities. The
confidence level for an interval translates into the confidence attributed to
a hypothesis. Usually, one would set the priors all equal and obtain the
simple likelihood ratio used in Example 13. Another example (quoted in Ref.
\cite{cous95}) follows:

\begin{example}
The Mark II collaboration reported a measurement of the number of neutrinos
$N_{\nu}=2.8\pm0.6$ which is slightly lower than the known number of three
neutrino generations. They take advantage of the negative fluctuation of their
result and derive from the tail distribution the 95\% classical confidence
limit $N_{\nu}<3.9$. A Bayesian would quote the ratio of the likelihoods for
the hypothesis of three and four neutrino species, $L(3|2.8\pm0.6)/L(4|2.8\pm
0.6)=7.0$ (assuming Gaussian distributed errors) which in principle could be
multiplied by prior probabilities. The difference between these results is
considerable: The classical tail limit indicates a factor $20$ in favor of the
three neutrino hypothesis. The likelihood ratio certainly is better suited to
describe the experimental situation.
\end{example}

\section{The Likelihood Principle and information}

In this article I hope to convince the reader that all parameter and interval
inference should be based solely on the likelihood function. To this end we
start with a discussion of the Likelihood Principle which plays a key role in
modern statistics.

\subsection{The Likelihood Principle}

Many statisticians (Fisher, Barnard, Birnbaum, Hacking, Basu and others) have
contributed to the development of the Likelihood Principle (LP). References,
derivations and extensive discussions are found in a summary of Basu's
articles \cite{basu88} and the book by Berger and Wolpert \cite{berg84}.

We assume that our observation $x$ follows a given probability density
$f(X|\theta)$ with unknown parameter $\theta$.

The Likelihood Principle states:

\emph{The information contained in an observation }$x$\emph{\ with respect to
the parameter }$\theta$\emph{\ is summarized by the likelihood function
}$L(\theta|x)$.

In other words, $L(\theta|x)$ of the actual observation is all what matters
for the parameter inference.

A multiplicative constant $c(x)$ in $L$ (corresponding to an additive constant
not depending on $\theta$ in the log-likelihood) is irrelevant. Methods not
respecting the LP -- inference not based on $L$ alone -- do not use the full
information contained the data sample\footnote{Information as used here is not
to be mistaken with the technical meaning given to this word by Fisher
\cite{fish25b}.}.

The LP asserts that \emph{the information relative to a parameter is identical
in two experiments with proportional likelihood functions and identical
parameter space, even when the pdfs are different} (see Stein's example
below). Parameter inference should be based on the likelihood function alone.

The Poisson problem, with zero events observed but background expected, is a
nice example \ illustrating the LP. The likelihood function is independent of
the background expectation but the pdf depends on it. According to the LP, the
information relative to the signal is independent of the background
expectation. Classical confidence intervals are in conflict with the LP,
likelihood ratio limits and Bayesian intervals are in agreement with it.

The LP relies on the strict validity of the probability density which is used,
a condition usually fulfilled in exact sciences. It is less useful in social,
economical and medical applications where only approximative theoretical
descriptions are available. Methods based on the likelihood function often are
very sensitive to unknown biasses, background and losses.

The LP can be derived \cite{berg84, basu88} from the Sufficiency Principle:

\emph{If }$T^{\prime}=T(x_{1}^{\prime},x_{2}^{\prime},...x_{n}^{\prime}%
)$\emph{\ is a sufficient statistic\footnote{A statistic $t$ is sufficient for
a parameter $\theta$, if the distribution of a sample given $T$ does not
depend on $\theta$.} in an experiment and if }$T^{\prime}=T^{\prime\prime
}=T(x_{1}^{\prime\prime},x_{2}^{\prime\prime},...x_{n}^{\prime\prime}%
)$\emph{\ the same value is obtained in another experiment, both experiments
following }$f(X_{1},X_{2},...X_{n}|\theta)$\emph{\ the evidence for }$\theta
$\emph{\ is the same in both experiments}.

and the Weak Conditioning Principle (in sloppy notation):

\emph{Performing the mixed experiment where randomly one of the experiments
}$E_{1}$ \emph{and }$E_{2}$\emph{\ is selected is equivalent to performing
directly the selected experiment}.

We illustrate the Weak Conditioning Principle with a simple example: An
experiment $E_{1}$ has two position sensitive detectors D$_{1}$ with low
resolution and D$_{2}$ with high resolution. The efficiency of D$_{2}$ is low.
When D$_{2}$ has seen a\ particle, only the measurement of D$_{2}$ is used.
For this specific measurement the experiment with both detectors is equivalent
to the experiment $E_{2}$ with only one detector D$_{2}$. The Weak
Conditioning Principle requires that inference of the position should be the
same in both experiments. The presence of a detector which is not used should
not matter.

However, there exist reservations against the proof of the LP and some
prominent statisticians do not believe in it. Barnard and Birnbaum, originally
promoters of the LP, ultimately came close to rejecting it or to strongly
restrict its applicability. One of the reasons for rejecting the LP is its
incompatibility with classical concepts. We will come back to some of the most
disturbing examples. Other criticism is related to questioning the Sufficiency
Principle and to unacceptable consequences of the LP in specific cases
\cite{kalb75, barn62, barn82, birn62a}. Most statisticians, frequentists and
Bayesians, usually accept that the likelihood function summarizes the
experimental information but some statisticians like Barnard feel that
parameter inference should include system information like the pdf or, if
relevant, the condition applied to stop data taking (see below).

Independent of the formal proof, the LP can be justified by simple arguments.

The LP almost follows from the optimal performance of the likelihood ratio
test for a discrete hypothesis. It is hardly avoidable if the concept of a
prior probability density for the parameter is accepted. Then Bayes' theorem
can be used to deduce the probability density $g(\theta|x)\,$of the parameter
$\theta$ conditioned on the outcome $x$ of an experiment (see Eq.~(11))
\begin{equation}
g(\theta|x)\sim L(\theta|x)\pi(\theta)\label{ebayesth}%
\end{equation}

Bayes' theorem is \ a simple consequence of probability calculus and accepted
by all statisticians. For known prior information $\pi(\theta)$, the
distribution $g(\theta|x)$ contains all we know about $\theta$. We can compute
the expectation value for the mean, all other moments and probabilities for
arbitrary intervals\footnote{It is interesting to notice that the sample pdf
does not enter in Relation 16.}. Since $\pi$ does not depend on $x$ the full
sample information is contained in $L$. If we accept that $L(\theta|x)$
contains the full sample information about $\theta$ regardless of the known
prior, then we also have to accept that if $\pi(\theta)$ is unknown, all
information is contained in the likelihood function.

There are two possibilities to escape these conclusions based on Bayes' Theorem:

\begin{itemize}
\item  We can deny the existence of the prior density. In deed, at first
glance it is not obvious what meaning we should attribute for example to a
prior density of the muon neutrino mass but to derive LP it is not necessary
to assume that \ the prior $\pi$ is known, it is sufficient to assume that it
exists, that God has chosen this mass at random according to an arbitrary
distribution $\pi(\theta)$ which we do not know, or in other words that in the
big bang a random process has fixed the particle masses. Such an assumption
would certainly not affect our scientific analysis techniques.

\item  We can argue that $g(\theta|x)$ is not all we want to know about
$\theta$. In this case we have to demonstrate that the information missing in
$g$ is relevant to inference of $\theta$.
\end{itemize}

There is one point of caution: \emph{The integral }$\int L(\theta|x)\pi
(\theta)dxd\theta$\emph{\ has to be finite} and \emph{the prior parameter
density has to be normalizable}. Thus in some cases the assumption of an
improper uniform prior extending over an infinite parameter range can lead to
problematic results. In practice, this is not a problem. For example,
restricting the Higgs mass to positive values below the Planck mass would be
acceptable to every educated scientist.

The formal proof of the LP does not require the existence of a prior.

\subsection{Some objections to the Likelihood Principle}

\subsubsection{The likelihood function does not explore the full sample space.}

One of the objection to the validity of LP is that the likelihood function
cannot contain all relevant information because it is not possible to compute
frequentist confidence intervals from the likelihood function alone.

In some occasions, like in Example 19, we know the prior density and inference
on the parameter $\theta$ can be based on Bayes' Theorem. Bayes' Theorem
allows us, using Eq.~(11) or equivalently Eq.~(16) to compute the full
pdf $g(\theta)$ and the probabilities to contain $\theta$ for arbitrary
intervals. On the other hand the information $L(\theta|x),\pi(\theta)$
entering in Equ. 16 is not sufficient to compute the coverage of these
intervals. As a consequence, coverage intervals must rely on elements of
information which are irrelevant to $g(\theta)$. Full information relative to
$\theta$ does not imply full information relative to coverage. This conclusion
is independent on whether we know the prior or not. The fact that coverage
cannot be computed from the likelihood function is not a valid argument
against the LP.

Another argument against the frequentist objection is the following: Assuming
that information relevant for parameter inference is missing in the likelihood
function of small samples, we hardly can imagine that by combining the
likelihood functions of many small samples to a large sample the missing
information is recovered. Therefore also the likelihood function of large
samples should be insufficient to compute coverage intervals. However, in the
limit of large samples, the likelihood function usually approximates a
Gaussian and allows the derivation of frequentist intervals without requiring
the pdf.

The violation of the LP by classical methods does not imply that those are
wrong. The latter address a different question. Their goal is to provide
coverage while the application of the LP is restricted to parameter inference.

\subsubsection{Inference of Gaussian parameters from a single observation.}

Inferring the mean $\mu$ and the width $\sigma$ of a Gaussian from a single
observation $x$ we realize that the likelihood function is infinite for
$\mu=x,$ $\sigma=0$ which implies this solution with certainty if we apply the
LP. This was one of the examples that had cast doubt on Birnbaum's believe in
the LP \cite{barn82}. However, close inspection reveals that this situation
cannot occur. No measurement can be described by a Gaussian with zero variance.

\begin{figure}[ptb]
\begin{center}
\includegraphics*[width=\textwidth]{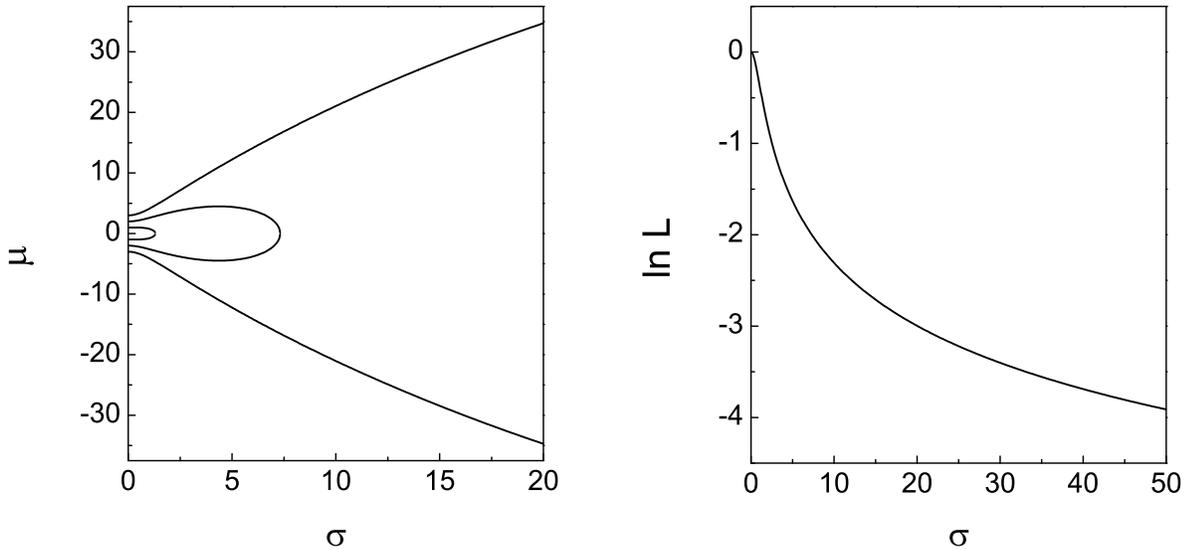}%
\caption{One, two and three st. dev. likelihood ratio
contours (left) and log-likelihood for $\mu=0$ as a function of
$\sigma$, Example~24}
\label{gausbirn}%
\end{center}
\end{figure}

\begin{example}
When we turn Birnbaum's Gaussian pdf into a more realistic pdf
\[
f(x|\theta,\sigma)=\frac{1}{\sqrt{2\pi(\sigma^{2}+s^{2})}}\exp\left[
-\frac{(x-\mu)^{2}}{2(\sigma^{2}+s^{2})}\right]
\]
where we interpret $s$ as the known experimental resolution, we still get a
maximum of the likelihood function at $\mu=x,\sigma=0$. \ For $s=1$ and an
observation at $x=0$ we find for the log-likelihood
\[
\ln L=-0.5\ln(\sigma^{2}+1)-\frac{\mu^{2}}{2(\sigma^{2}+1)}+const.
\]
which has a narrow peak but also a long tails in $\sigma$ and does not
contradict our intuition. Figure~\ref{gausbirn} shows the likelihood contours.
\end{example}%

\begin{example}
We may modify Birnbaum's example further, fix the parameter $\mu$ to $\mu=0 $
and assume $x=0$. This observation which hits exactly the prediction is likely
if the value of $\sigma$ is small. The log-likelihood function is shown in
Figure~\ref{gausbirn}. The 1, 2 and 3 st. dev. likelihood ratio limits are
$\sigma_{1}<1.31$, $\sigma_{2}<7.32$ and $\sigma_{3}<90$. Classical intervals
would include all values of $\sigma$ independent of the confidence level since
the observation is obviously compatible with any $\sigma$.
\end{example}

The LP does not contradict the frequentist result. It just states that all
information on $\sigma$ is in the likelihood function. Certainly, we learn
something about $\sigma$ when we observe a $x-\mu=0$, information which is
completely lost in frequentist intervals but partially conserved in likelihood
ratio intervals.

\subsection{The Stopping Rule Paradox}%

\begin{figure}[ptb]
\begin{center}
\includegraphics*[width=3.4921in]{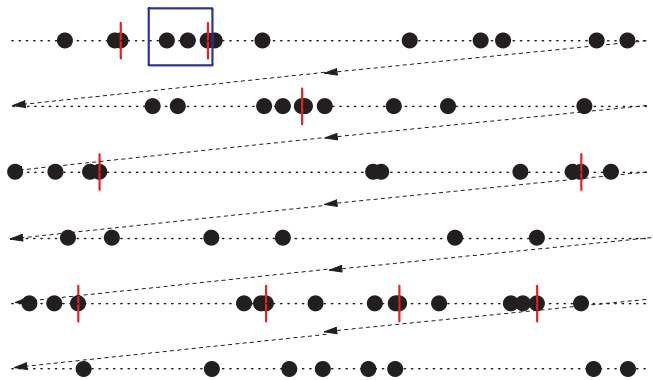}%
\caption{Rate measurement. The time sequence is stopped when 3
events are observed within 1 second (indicated by a bar) and a new
measurement is started. Since the combination of many conditioned
measurements is equivalent to a long unconditioned measurement,
this stopping rule cannot introduce a bias}
\end{center}
\end{figure}

The likelihood function is independent of the sequence of independent and
identically distributed observations. Thus, since according to the LP
parameter, inference should solely be based on the observed sample it should
be independent of any sequential sampling plan \cite{barn62}. For example,
stopping data taking in an experiment after a ``golden event'' has been
observed should not introduces a bias. This contradicts our naive intuition
and corresponds to the so-called \emph{Stopping Rule Paradox}. Similarly, some
people believe that one can bias the luck in a casino by playing until the
balance is positive and stopping then.

The Stopping Rule Principle is of considerable importance for experimental
techniques\footnote{Edwards, Lindman, Savage \cite{edwa63}: ``The irrelevance
of stopping rules to statistical inference restores a simplicity and freedom
to experimental design that had been lost by classical emphasis on
significance levels ... Many experimenters would like to feel free to collect
data until they have either conclusively proved their point, conclusively
disproved it, or run out of time, money, or patience.''}.

Frequentists claim that stopping rules introduce biasses and consequently deny
the LP. One has to admit that the Stopping Rule Principle is hard to digest.

It is easy to see that there is no bias in the example quoted above: Let us
imagine that we record experimental observations for infinitely long time
where many ``golden event'' are recorded. The experiment now is cut into many
sub-experiments, each stopping after a ``golden event'' and the
sub-experiments are analyzed separately. Since nothing has changed, the
sub-experiments cannot be biassed and since the sum of the log-likelihoods of
the subsamples is equal to the log-likelihood of the full sample it is also
guaranteed that combining the results of a large number of experiments
performed with an arbitrary stopping rule asymptotically give the correct
result. Of course the condition that the subsamples are unbiassed is necessary
but not sufficient for the validity of the Stopping Rule Principle. The effect
of the stopping rule ``stop when 3 events are found within 1 second'' is
presented in Fig.~23.

In the following example we explicitly study a specific stopping rule.%

\begin{figure}[ptb]
\begin{center}
\includegraphics*[width=\textwidth]{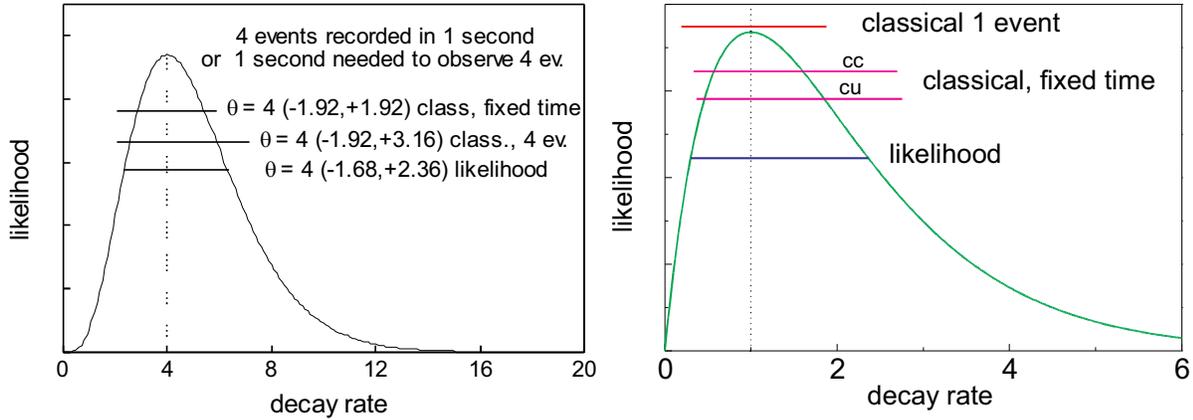}%
\caption{Likelihood for the decay rate. Left hand: Four events are
observed in one time unit. The classical error limits depend on the stopping
condition i.e. fixed time interval or fixed event number. Right hand: Same
evaluation for one event}
\end{center}
\end{figure}

\begin{example}
In a rate measurement 4 events have been observed within the time interval $t
$. Does the uncertainty on the rate $\theta$ depend on the two different
experimental conditions: a) The observation time had been fixed. b) The
experiment had been stopped after 4 events. In case a) the uncertainty is due
to the event number fluctuation and in b) to the time variation. The LP denies
any difference, only the data actually seen matter, while for a classical
treatment also the data that could have been observed are important. The
likelihood functions $L_{a},L_{b}$ derived from the Poisson distribution and
the decay time distribution, respectively, are proportional to each other and
depend only on the observation time and the number of events seen (see Fig.~24):
\begin{align*}
L_{a}(\theta,n)  & =P(n,\theta t)\\
& =\frac{e^{-\theta t}(\theta t)^{4}}{4!}\propto\theta^{4}e^{-\theta t}\\
L_{b}(\theta,t)  & =\frac{t^{3}\theta^{4}}{3!}e^{-\theta t}\propto\theta
^{4}e^{-\theta t}%
\end{align*}
\end{example}

This example is well suited to explain the essence of the LP. Followers of the
LP would bet the same amount of money that $\theta$ is located in a certain
arbitrarily selected interval in both experiments.

The stopping rule may be tested by a Monte Carlo simulation. In the previous
example we could select two arbitrary rate values $\theta_{1}$ and $\theta
_{2}$ and simulate a large number of the two kinds of experiments. For the
experiments of type a) only results with 4 events found are retained and the
ratio of successes $R_{a}$ for the two values of $\theta$ is formed. For the
experiments of type b) a small time slice around $t$ is fixed and $R_{b}$ is
formed from the number of cases where four events are recorded. The two ratios
should be equal if the information relative to the parameter is the same in
both variants. Actually, a simulation is not necessary. The quantities $R_{a}$
and $R_{b}$ are the likelihood ratios which agree because the likelihood
functions are the same up to a constant factor. The stopping rule does not
favor one or the other of the $\theta$ values. There is no bias.

A classical treatment produces inconsistent results for the two equivalent
experiments. The error limits depend on the way we are looking at the data
(Fig.~24).

The intuitive explanation for the independence of inference of a sequential
stopping rule indicates also its limitations. The stopping rule has to be
simple in the sense that the different pieces obtained by cutting an
infinitely long chain of observations cannot be classified differently.
Experiments stopped when the result is significant, when for a long time no
event has been recorded or when money has run out should be analyzed ignoring
the reason for stopping data taking.

The following example cited in the statistical literature is more subtle. Basu
presents a detailed discussion \cite{basup67}.

\begin{example}
We sample data $x_{i}$ following a Gaussian with unknown mean $\theta$ and
width equal to one. The mean $\bar{x}$ is formed after each observation. The
experiment is stopped after observation $n$ when one of the following
conditions is fulfilled:
\begin{equation}
|\bar{x}|>\frac{2}{\sqrt{n}}\text{, or }n=1000
\end{equation}
The idea behind this construction is to discriminate the value $\theta=0$
which will be off by 2 st. dev. in all cases where the first condition is
realized. (One may want to stop clinical trials when a 2 st. dev. significance
is reached for the effectiveness of a certain drug.) The second condition is
introduced to avoid extremely large samples for $\theta=0$ (non-performable
experiments). Let us assume that an experiment produces an event of size
$n=100$ and $|\bar{x}|=0.21>2/\sqrt{100}$. When we ignore the stopping
condition a frequentist would exclude the parameter value $\theta=0 $ with
$>95\%$ confidence (tail probability). Knowing the stopping rule, he would
realize that the first of the two conditions 17 has been fulfilled and
probably be willing to accept $\theta=0$ as a possible value. Wouldn't we
include the stopping condition into our considerations? Basu \cite{basup67}
and also Berger and Wolpert \cite{berg84} deny this and argue that our
intuition is misled. In fact, the likelihood at $\theta=0$ is not really
reduced by the stopping rule, it is the integrated tail which is small,
because the likelihood function becomes narrower with increasing $n$.
\end{example}

In the last example certainly is not easy to accept the likelihood solution
and it is less the problem to understand that $\theta=0$ is not excluded but
rather to accept that some value different from zero has a very large value of
the likelihood even when $\theta=0$ is correct.

The stopping rule changes the experiment and some rules are not very useful.
But this is not in contradiction to the fact, once the data are at hand,
knowing the stopping rule does not help to improve the inference of the
parameter of interest.

\subsection{Stein's example}

In 1960, L. J. Savage emphasized \cite{sava61}: ``If the LP were untenable,
clear-cut counter-examples by now would have come forward. But such examples
seem, rather, to illuminate, strengthen, and confirm the principle.'' One year
later C. Stein \cite{stei62} came up with his famous example which we will
discuss in this section. Since then, other more or less exotic
counter-examples have been invented to disprove the LP. The corresponding
distributions exhibit strange infinities and are far from what we can expect
in physics. Nevertheless, they are very instructive.%

\begin{figure}[ptb]
\begin{center}
\includegraphics*[width=3.6in]{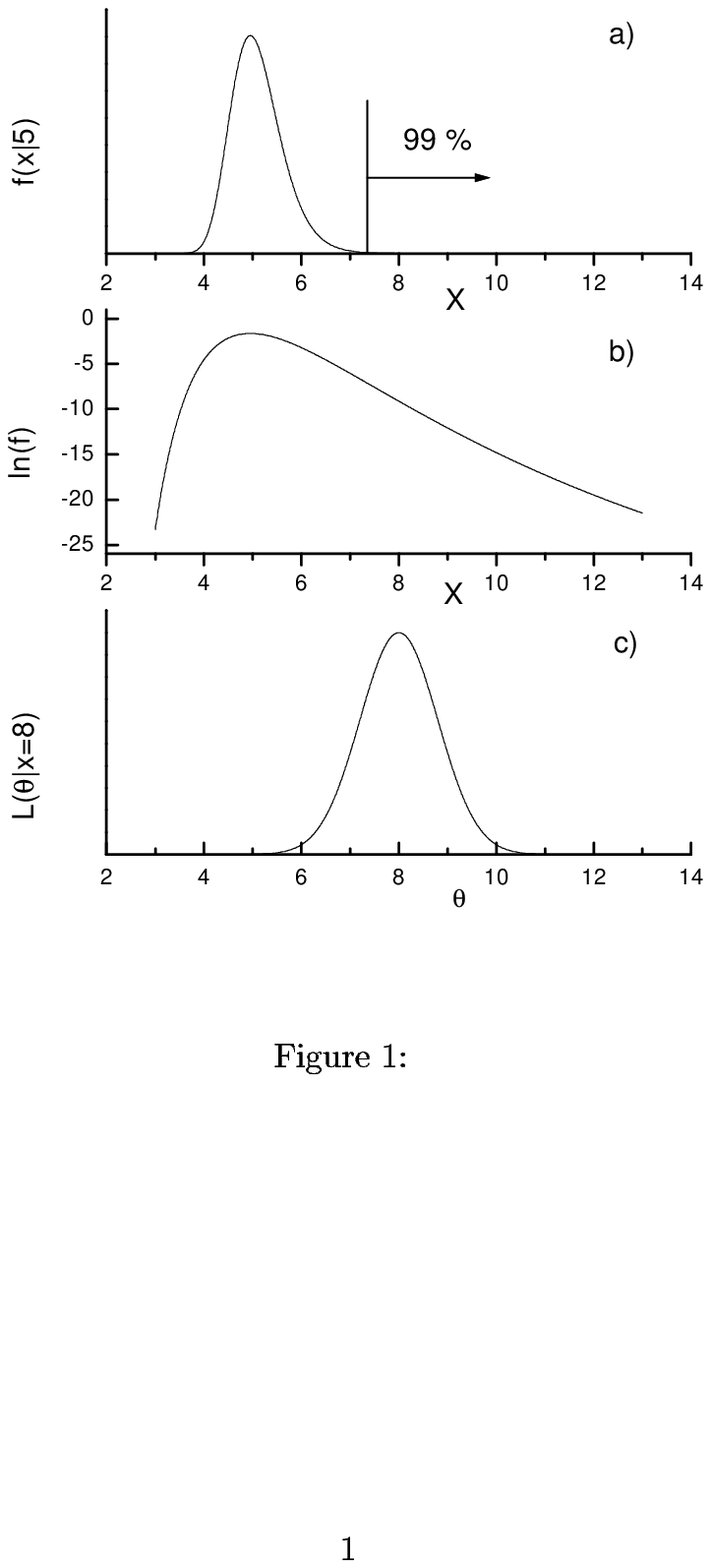}%
\caption{Stein's test function (a,b) with a long tail towards
large $x$. The maximum is close to the parameter value. The
Gaussian likelihood function for an observation $x=8$ is shown in~(c)}
\end{center}
\end{figure}

Stein has been introduced the probability density%
\[
f(X,\theta)=\frac{a}{X}\exp\left\{  -100\frac{(\theta-X)^{2}}{2X^{2}}\right\}
\text{, \hspace{1cm}for }0\leq X\leq b\theta
\]
where for simplicity we have fixed the slope parameter in the exponent.

The functional form resembles a Gaussian with width proportional to the mean
value as shown in Fig.~25a and 25b but it has a long tail towards large
values of $X$ which completely dominates the distribution. The range of $X$ is
restricted by the cutoff parameter $b$ to obtain a normalizable density. Since
the $1/X$ term almost insures convergence the upper limit of $X$ may be rather
large for not too large $a$. The two constants $a$, $b$ are fixed by the
normalization condition and the requirement that $X$\emph{\ exceeds }$\theta
$\emph{\ by at least a} \emph{factor of ten in 99\% of the cases}.
\begin{align}
\int_{0}^{b\theta}f(X,\theta)dX  & =1\nonumber\\
\int_{10\theta}^{b\theta}f(X,\theta)dX  & =0.99\label{stein2}%
\end{align}
The cutoff parameter is huge $b\theta>10^{(10^{22})}$ and $a\approx0.04$.
Figure 25a,b are plots of Stein's function with the specific choice
$\theta=5$. Only about $1\%$ of the $X$-values lie inside the bell shaped part
($X<50$) of the probability density, the remaining part is located in the long
right hand tail.

\subsubsection{The paradox and its solution.}

For a single observation $x$ the likelihood function for the Stein pdf is:
\[
L(\theta)\sim\exp\left\{  -100\frac{(\theta-x)^{2}}{2x^{2}}\right\}  \text{,
\hspace{1cm}for }x/b<\theta<\infty\text{ }
\]
It is proportional to the likelihood function of a Gaussian centered at $x$,
with width $s=x/10$ and restricted to positive $\theta$. It is displayed in
Fig.~25c for $x=8$. (There is a small difference in the validity range of
$\theta$ for the Gaussian and Stein's distribution which numerically is of no importance.)

If the likelihood function contains the full information relative to the
parameter, the inferences on $\theta$ should be the same for the very
different distributions, the Gaussian and the Stein function. However, one
might conclude from Eq.~(18) that the Stein parameter $\theta$ is 10 times
smaller than the observation $x$ in 99\% of the cases, while the Gaussian case
would favor a parameter in the vicinity of the observation \cite{stei62}. It
seems silly to select the same confidence interval in both cases.

The solution of Stein's paradox can be summarized as follows:

\begin{itemize}
\item  The LP does not relate different probability densities. It just states
that the likelihood function contains the full information relative to the
unknown parameter. Thus there is no contradiction in the fact that the same
likelihood belongs to two different probability densities.

\item  The assertion ``from $X$ $>10\theta$ in $99\%$ of the cases for given
$\theta$, independent of the value of $\theta$, follows that for given $x$ we
have $\theta$ $<x/10$ in $99\%$ of the case, independent of the value of $x$''
is not conclusive.

\item  Related to the last point: The same $\theta$ interval corresponds to
very different coverages in the two cases, Gaussian and Stein pdf. Coverage,
however, is a frequentist concept which is not computable from the likelihood
function. This is not specific to Stein's example and thus not relevant here.
\end{itemize}

It is easily seen that for any proper (normalizable) prior density the
inferences for the Gaussian and the Stein cases should be identical while the
classical intervals are rather strange in some cases.

Let us look at an example: We select a uniform prior function $\pi
(\theta)=0.1$ and $0<\theta<10$. In most cases one would get $x$ much larger
than $\theta$, say $x=10^{1000}$. Obviously, we cannot learn much from such an
observation, a fact which translates into a likelihood function which is
completely flat in the allowed $\theta$ range:%

\begin{align*}
L  & =\exp\left\{  -100\frac{(\theta-x)^{2}}{2x^{2}}\right\} \\
& \sim\exp\left\{  -100\theta/x\right\}  \approx1
\end{align*}
The same result would be obtained for a Gaussian pdf, except that the
occurrence of very large $x$ is rare.

Now assume $x=5$. Then
\[
L(\theta)\sim\exp\left\{  -\frac{(\theta-5)^{2}}{2\cdot0.5^{2}}\right\}
\text{, \hspace{1cm}for }5/b<\theta<10
\]
is a Gaussian of width $s=0.5$ and we guess that $\theta$ is near to $x$. In
both cases, for the Gaussian pdf and the Stein pdf the results derived from
the likelihood function are very reasonable. The difference between the two is
that in the former case likelihood functions centered around small $x$ are
frequent and in the latter those centered at large $x$, but for a given
observation this is irrelevant. The reader is invited to invent other prior
densities and to search for inconsistencies in the Bayesian treatment.

The argument of Stein relies on the following frequentist observation. Assume
$\theta$ values are selected in a range $\theta<\theta_{\max}$. Then in $99\%$
of all cases the likelihood function would favor $\hat{\theta}\approx
x>10\theta_{\max}$, an estimate which is considerably worse than the classical
one. The flaw of the argument is due to the inconsistency in selecting
specific test parameters $\theta$ and using a uniform prior of $\theta$
extending to some $10^{(10^{20})}\theta_{\max}$.

\subsubsection{Classical treatment.}

What would a classical treatment give for central 99\% confidence intervals?%

\begin{align*}
\int_{x}^{b\theta low}f(X,\theta_{low})dX  & =0.005\Longrightarrow\theta
_{low}\ll x\\
\int_{0}^{x}f(X,\theta_{high})dX  & =0.005\Longrightarrow\theta_{high}\ll x
\end{align*}

For any reasonably small $x$ both limits would be nearly zero. For very large
$x$ the classical procedure produces a rather wide $\theta$ interval.
Qualitatively the classical interval fails to contain the true value in the
1\% cases where the observation falls into the bell shaped part of the
distribution but produces a safe result in the remaining 99\% where $x$ is
located in the tail. Thus the correct coverage is guaranteed but in some cases
ridiculous error limits have to be quoted. This is the well known problem of
classical statistics: It can produce results that are known to be wrong. In
the remaining cases the interval will be enormously wide and extend to
extremely large values.

\begin{example}
A Susy Particle has a mass of $\theta=1$ eV. A physicist who does not know
this value designs an experiment to measure the mass in the range from $0.001
$ eV$<\theta<1$ TeV through a quantity $x$ depending on $\theta$ according to
the Stein distribution. (It is certainly not easy to determine a parameter
$\theta$ over 15 orders of magnitude from a quantity $x$ which may vary over
10$^{20}$ orders of magnitude.) \ A 99\% confidence level is envisaged. A
frequentist analysis of the data will provide wrong limits $0<\theta\ll0.1$ eV
in 1 \% of the cases and huge, useless intervals in the remaining ones (except
for a tiny fraction). In contrast, a Bayesian physicist will obtain a correct
and useful limit in 1\% of the cases and a result similar to the classical one
in the remaining cases.
\end{example}

It is impossible to invent a realistic experimental situation where the
Bayesian treatment of the Stein problem produces wrong results. The Stein
function plays with logarithmic infinities, which cannot occur in reality.
Limiting the uniform prior\footnote{Any other normalizable prior would produce
very similar results.} to a reasonable range or using other normalizable prior
densities avoids the paradoxical conclusion of Stein. Experimental conditions
always forbid improper prior densities.

\subsection{Stone's example}

Another nice example has been presented by M. Stone \cite{ston76} which is
presented here in simplified form. The Stone example is much closer to
problems arising in physics than the rather exotic Stein Paradox.%

\begin{figure}[ptb]
\begin{center}
\includegraphics*[width=2in]{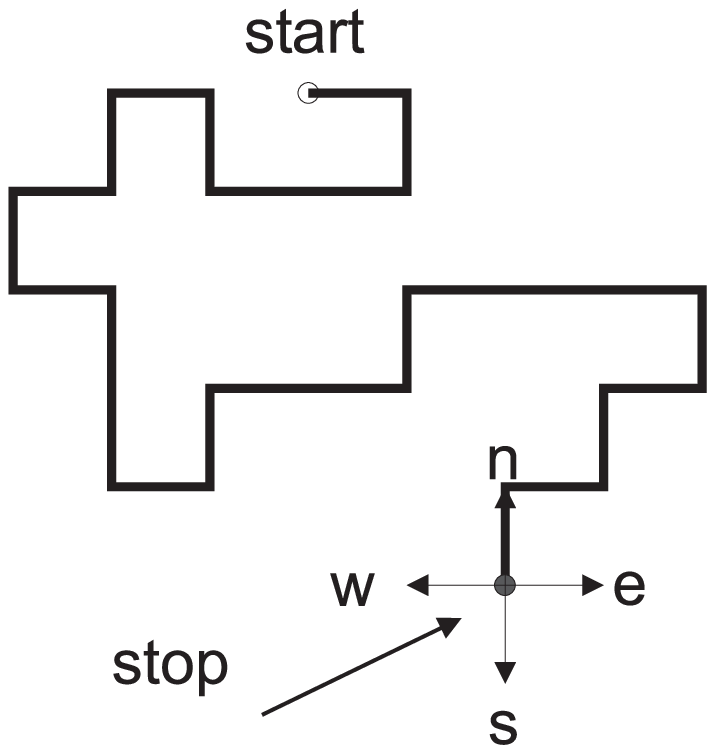}%
\caption{Modified Stone's example: The problem is to infer the
last step knowing start and stop positions}
\end{center}
\end{figure}

\begin{example}
A drunken sailer starts from a fixed intersection of streets and proceeds at
random through the square net of streets oriented along the directions N, W,
S, E (see Fig.~26)\footnote{The analogy to physics is not between the sailor
and drunken physicists but between his random walk and diffusion.}. At a
certain intersection $x$ he stops. The problem is to guess the last crossing
before the stop, i.e. the direction. We name the four possible origins
$n,w,s,e$. Starting in a random direction from $e$, the probability to arrive
at $x$ is 1/4. Thus the likelihood for $e$ is also 1/4. The same likelihoods
apply for the other three locations. With a ``non-informative'' prior for the
locations, one has to associate equal probabilities to all four possibilities.
This is in contradiction to the correct intuition that a position between the
starting point and the stop is more likely than 1/4.
\end{example}

The example does not directly involve the LP \cite{hill81} but it questions
inference based solely on the likelihood function. Here, the solution is
rather simple: There is information about the prior and it is clearly not
uniform. (In the original version of the example the prior has an even
stronger influence.) Including the prior information, and using the constant
likelihood we obtain a correct result \cite{hill81,berg84}.

It is not clear, what a classical solution of the problem would look like.

The fact, that so far no simple counter-example to the LP has been found
provides strong support to this principle.

\subsection{Goodness-of-fit tests}

Significance tests cannot be based on the LP since there is no parameter
differentiating between the hypothesis at hand and the rather diffuse
alternatives. For this reason they have been criticized\ especially by
Bayesians \cite{jeff61,good81} and some statisticians adhering to the LP
\cite{edwa92}. However, as long as no better alternatives are available,
classical methods should be used. Goodness-of-fit tests like the
Neyman-Pearson or Kolmogorov tests are very useful and cannot easily be
replaced by likelihood based recipes. Note that also here the subjectivity
problem is present. The binning has to be chosen according to the Bayesian
expectation of possible deviations. A rough binning neglects narrow deviations
and with fine binning in the Neyman-Pearson test the statistical fluctuations
hide systematic deviations \cite{zech95}. Also the fact that we use one-sided
tests has a Bayesian origin.

The approach to significance tests and classical confidence limits is very
similar, but significance tests are not related logically to interval estimation.

\subsection{Randomization}

Randomization is an important tool in many statistical applications like
clinical tests and survey studies. For example, we might select randomly fifty
CERN physicists and ask them their opinion on closing down the LEP machine. It
might turn out that by accident the majority of the selected physicists are
involved in LEP experiments. This post-experimental information has to be
ignored in a frequentist approach in contradiction to the LP but explains the outcome.

Strong criticism of randomization is expressed by Basu \cite{basu1978} and it
seems obvious that with a good unbiassed judgement one should be able to get
better performance than by random selection. Yet, there are cases where this
classical procedure is very useful - think of Monte Carlo simulation - and
especially for experiments with large statistics the Bayesian objections are
less convincing. This point of view is shared by prominent Bayesians like
Savage \cite{sava61}: ``The theory of personal probability ... does lead to
new insight into the role and limitations of randomization but does by no
means deprive randomization of its important function in statistics.''

In conclusion, there are useful frequentist concepts which are not respecting
the LP. We should neither exclude these methods nor use them as an argument
against the LP. Goodness-of-fit tests and randomization methods have no
bearing whatsoever on parameter interval estimation for a given pdf.

\subsection{Precision and consistency}

Following the LP we can fix an ordering principle for \emph{precision}.

Two measurements $E_{1},E_{2}$ of the same parameter are equally precise if
their likelihood functions $L_{1}(\theta)$ and $L_{2}(\theta)$ are
proportional to each other. $E_{1}$ is more precise than $E_{2}$ if they have
the maximum at the same parameter value $\theta_{\max}$ and if $L_{1}%
(\theta)<cL_{2}(\theta)$ for all other values of $\theta$ where the
normalization constant $c$ is fixed by $L_{1}(\theta_{\max})=cL_{2}%
(\theta_{\max})$. The narrower the likelihood function is, the more precise is
the measurement. This requirement should be acceptable to all statisticians.
We could extend our modest requirement to a general definition but then it
would single out a specific statistical method.

A statistical approach is inconsistent if it associates different error
intervals to equally precise measurements or if it produces smaller error
intervals for less precise measurements than for more precise ones.

\section{Comparison of the methods}

The quality of a confidence interval is measured

\begin{itemize}
\item  in the classical approach by coverage, the average number of intervals
containing the true value of a parameter,

\item  by the likelihood ratio in likelihood ratio intervals,

\item  by the probability obtained by integrating likelihood times uniform
prior over the interval.
\end{itemize}

Likelihood ratio and Bayesian intervals are related numerically. The standard
likelihood ratio intervals corresponds to 68.3 \% Bayesian probability in the
parameter space where the likelihood function is Gaussian and the prior is uniform.

\subsection{Difference in philosophy}

Before we compare the frequentist methods with the likelihood based approaches
in detail, let us come back to the example (see Fig.~1) presented in the
introduction which is repeated in more detail as Example 13/14.

Frequentists test the compatibility of data with a parameter and independent
of details of the theoretical prediction, they want to be right in a specified
fraction of cases. The confidence level is a kind of ``inclusion probability''
of the correct parameter value inside the interval and is fixed. The
``exclusion probability'' of wrong parameter values which depends on the width
of the probability distribution is at least partially ignored in the classical
approaches. In the language of discrete parameters one could say that the
``error of the second kind\footnote{Error of the second kind: Accepting a
hypothesis when it is false.}'' is not fully considered. The wider the
distribution is, the likelier it is to include a wrong parameter in the
confidence interval.

Since the confidence interval is based on an integration of the pdf it depends
on data that potentially could have been observed. The probability to observe
the actual value $x$ is not enough for classical reasoning. Intuitively, one
would find the two parameter values $\theta_{1}$ and $\theta_{2}$ which
satisfy $f(x|\theta_{1})=f(x|\theta_{2})$ equally acceptable, but the non
locality of the frequentist coverage introduces a difference.

On the other hand in concepts based on the likelihood function only the value
of the pdf \ at the observed data is important. The methods are local in the
sample space. To illustrate the idea behind it, it is simpler to argue with
discrete variables.

\begin{example}
Assume, we have a bag with many dice of seven different kinds, all differently
biassed, except for die type ``7'' which is unbiassed. We do not know the
number of dies of each kind. Each of the biassed dies favors a different
number with probability $1/2$. The remaining numbers have probabilities of
$1/10$. One of the dies of unknown type is selected and thrown. We observe the
number ``$3$'' and have to estimate the type. The following table contains the
probabilities $p(3|type)$ to observe ``$3$'' for the seven dies. Obviously, we
would select die ``3'' and we would not care about the probabilities
$p(1),p(2),p(4),p(5),p(6)$ for alternative outcomes. Thus we would base our
decision on the observation only and not consider the remaining sample space.
\end{example}

\begin{center}%
\begin{tabular}
[c]{|l|lllllll|}\hline
die type & 1 & 2 & 3 & 4 & 5 & 6 & 7\\
$p(3|type)$ & 0.1 & 0.1 & 0.5 & 0.1 & 0.1 & 0.1 & 1/6\\\hline
\end{tabular}
\end{center}

If we would repeat the experiment many times with the same die, we would
always select one of the first six die types and never die ``7''. Thus a
frequentist would argue that the procedure is inadmissible because the
procedure does not provide equal coverage to all hypotheses. A Bayesian would
not care about long term coverage. He wants to make the optimum decision in
the present case and select the die which gives the largest likelihood. Why is
long term coverage not an issue? The Bayesian updates his information by
multiplying the likelihoods of the individual trials and after some 20
experiments, he then, if ``7'' applies would probably find for die ``7'' a
higher likelihood than for the other dies.

Let us look at another very extreme example.

\begin{example}
An electronic code reader is able to identify the number $0.45301$. For any
other input number it will produce a random output $\theta$ between zero and
one. Both, the sample variable $X$ and the parameter $\theta$ are restricted
to the interval [0,1]. For the sake of the argument, let us assume that the
device has infinite resolution, and that numbers have an infinite number of
digits.
\[
f(X|\theta)=\left\{
\begin{array}
[c]{c}%
\delta(X-\theta)\ \ \ \ \ \ \text{for }\theta=0.45301\\
1\ \ \ \ \ \ \ \ \ \ \ \ \ \ \ \ \ \text{for }\theta\neq0.45301
\end{array}
\right\}
\]
\end{example}

The observation is $x=0.45301$. The classical confidence interval would be the
full interval $0\leq\theta\leq1$. The Bayesian result would be $\theta
=0.45301\pm0$ with 100 \% confidence. I believe, everybody would intuitively
support the Bayesian conclusion.

Classical confidence intervals are not well qualified for decision making.
Even advocates of the classical concept admit that for decision making
classical limits may not be optimum \cite{feld98}. The question is then, what
else should they be used for? Frequentists, however, argue that the intention
of inference is not necessarily to optimize decisions. Kendall and Stuart
\cite{ksp653} write: ``Some writers have gone so far as to argue that all
estimation and hypothesis-testing are, in fact, decision-making operations. We
emphatically disagree, both that all statistical inquiry emerges in decisions
and that the consequences of many decisions can be evaluated numerically.''

Narsky \cite{nars99} who compares several different approaches to the
estimation of upper Poisson limits, states: ``There is no such thing as the
best procedure for upper limit estimation. An experimentalist is free to
choose any procedure she/he likes, based on her/his belief and experience. The
only requirement is that the chosen procedure must have a strict mathematical
foundation.'' This opinion is typical for many papers on confidence limits.
However, ``the real test of the pudding is in its eating'' and not in the
beauty of the cooking recipe. We should state what we aim for and should not
forget that what we measure has practical implications, hopefully!

\subsection{Relevance, consistency and precision}

A. W. F. Edwards writes \cite{edwa92}: ``Relative support (He refers to a
hypothesis or a parameter) must be consistent in different applications, so
that we are content to react equally to equal values, and it must not be
affected by information judged intuitively to be irrelevant.'' We can check
whether we consider information as ``irrelevant'' when we consider betting
odds. When the amount of money we are ready to bet that $\theta$ is in a
certain interval is independent of \ part of the available information, then
this information is considered irrelevant.

The most obvious example for the violation of this principle in all classical
approaches is the Poisson case with background expected but where no event is
observed (Examples 11, 12). Other examples are related to the dependence of
classical results on the sampling plan (Example 26) For this reason alone,
classical limits are to be discarded as a useful description of uncertainty.

Formally, the inconsistency of classical limits, the fact that different
limits are obtained in equivalent situations manifests itself as a violation
of the LP which requires equal parameter inference for equal likelihood
functions. In the Poisson case \cite{zech97} for $n$ events found and $b$
background events expected, the likelihood function for the signal parameter
$\mu$ is
\[
L(\mu)\sim\sum_{b^{\prime}=0}^{n}P(b^{\prime}|b)P(n-b^{\prime}|\mu)
\]
and for $n=0$ it reduces to
\[
L(\mu)\propto e^{-\mu}
\]
independent of the background expectation. In Fig.~16 the likelihood
function is compared to the classical and Bayesian results.

Similarly, different stopping rules which produce identical likelihood
functions lead to different classical limits (Example 26). Another example is
presented by Berger and Berry \cite{berbery}.

In the previous section we discussed the notion of precision and established
an ordering principle.

This principle is violated in classical statistics in our standard Poisson
example: An experimental result of zero events $n=0$ found with background
expectation $b=0$ is clearly more exclusive than a result $n=1$ with $b=3$.
This obvious conclusion based on common sense is confirmed by the formal
argument that the measurement observing $n=0$ is the more precise because the
likelihood function $L(\theta|n=0,b=0)$ is steeper than $L(\theta|n=1,b=3)$
for all values of $\theta$. The corresponding 90 \% upper confidence limits
2.44 and 1.88 of the unified approach \cite{feld98} invert the sequence of precision.

All problems in the frequentist approach have the same origin: The limits
depend on the full sample pdf. However, there is no reason why we should care
about the probability of events that have not occurred.

These statements should not be misunderstood. The classical approach in itself
is not inconsistent but it does not provide a consistent measure of uncertainty.

The deficiency of classical upper limits is also recognized by some
frequentists. They accept that the classical confidence intervals are not
always useful measures of precision. But, what else should they be used for?
To cure the problem, Feldman and Cousins \cite{feld98} propose to supplement
the confidence limit by an additional quantity called
sensitivity\footnote{Oviously, frequentists realize that important information
is not contained in the frequentist limits interval.}. However, they do not
explain how to combine the two quantities to produce a sensitive measure for
the relative precision of different experimental results and how to use the
confidence interval to check the compatibility of the measurement with a
theoretical prediction.

Bayesian methods are in agreement with the LP and respect the consistency
property. Problems may occur for probability densities with multiple maxima
when the prior density is updated with increasing event number \cite{diac86}.
These exotic cases are not relevant in real physics problems and cannot be
handled by classical methods either. Complicated situations should be
represented by likelihood maps.

In the limit that the assumption of a uniform prior density is correct,
likelihood ratio intervals are optimum: For a given interval length the
probability to contain the true parameter value is maximum. In other words,
avoiding the notion of prior density, the likelihood interval contains the
parameter values best supported by the data.

\begin{sloppypar}
Classical methods hardly emphasize precision. Nevertheless they discuss
``minimum length'' and ``likelihood ratio'' intervals. Clearly, classical
intervals cannot be optimum when they violate the LP because then they ignore
available information. For example, it is easy to improve negative Poisson
limits or unphysical intervals. Qualitatively, the intervals from the unified
approach are expected to be superior to other classical intervals with respect
to precision.
\end{sloppypar}

Frequentists sometimes argue that the performance of different approaches
should not be compared looking at specific examples but that it is the average
performance which is relevant. On the other hand, also the average performance
is improved if more reasonable solutions are implemented in individual
disturbing cases. Why should it harm to modify intervals like zero length
intervals which with certainty do not include the true value?

\subsection{Coverage}

In the classical approach we have the attractive property that a well known
fraction of experimental results contain the true value of a measured
parameter within the quoted error limits. As explained in the introductory
example, coverage is a democratic \ principle which gives different true
parameter values the same chance to be included in a confidence interval.

Major criticism of the classical concepts is related to the requirement of
pre-experimental definition of the analysis methods necessary to guarantee
coverage. Conditioning on the actual observation is forbidden. However as
shown by Kiefer \cite{kief77}, conditioning and coverage are not always exclusive.

On the other hand for an individual experiment we cannot deduce from the
computed coverage a probability that the result is true and it is the latter
we need to make decisions.

\begin{example}
An experimenter tosses a coin to decide whether the Higgs mass is larger or
less than 100 GeV and finds that it is larger. The confidence level is 50\%.
It would be stupid to deduce from this ``observation'' a probability of 50\%
for the Higgs mass to be larger than the arbitrarily selected mass value of
100 GeV. Confidence is deduced from zero information.
\end{example}

The admiration of coverage by many physicists is due to an illusion: They
inadvertently mix coverage with the probability that the true value of the
parameter is located inside the confidence interval even if they
intellectually realize the difference. In reality, the only possible way to
derive such a probability is to invent a Bayesian prior to multiply it with
the likelihood function and to integrate the product over the confidence
interval. We cannot do any better. On the other hand, coverage is not a
useless quality as some Bayesians claim. Coverage intervals retain a large
fraction of the information contained in the likelihood function.

Coverage is a property of an ensemble of experiments and makes sense if an
experiment is repeated many times\footnote{Insurance companies may rely on
coverage.}. Let us assume that we have 10 experiments measuring with different
precision a particle mass and providing 90 \% confidence intervals each. It
will not be probable that more than one or two of the limits will not contain
the true value. This will certainly give us an idea where the true value may
be located, but a quantitative evaluation is difficult.

Although it is quite rare in physics applications that we have an ensemble of
similar experiments, let us assume that this is the case. What would we do
with our results? We would probably not be satisfied contemplating the
different confidence intervals but try to combine them\footnote{Insurance
companies cannot do this.}. If possible, we would go back to the original
data, combine them to a high statistics experiment and then forget about the
individual data sets. Thus, we have again a single experiment. There is no
obvious reason to consider ensembles. Otherwise we would subdivide high
statistics experiments in many low statistics measurements. In conclusion, the
problem of parameter inference can be discussed on the basis of single experiments.

The violation of coverage due to the separate treatment of upper limits and
intervals was the main motivation for the development of the unified approach.
In the years before the adoption of this new scheme, rare decay searches
usually presented Bayesian upper limits - not intervals - when the result was
compatible with background. It is hard to detect negative consequences from
the former procedure to the progress of physics.

The main problems with the coverage paradigm are the following:

\begin{itemize}
\item  Coverage often is only approximate. There is complete overcoverage in
Poisson distributed data with mean zero and in digital measurements.

\item  The usual treatment of nuisance parameters \cite{cous00} can lead to undercoverage.

\item  It is difficult to retain coverage when results are averaged.

\item  The principle is not universally applied. When the prior density is
known, like in Example 19 the coverage requirement is put aside.
\end{itemize}

Classical confidence cannot be attributed to intervals based on the likelihood
function. The PDG \cite{pdg98} gives the advice to determine the true coverage
by a Monte Carlo simulation. It is not possible to associate a single
classical confidence level to an arbitrary interval. Of course, coverage as a
function of the true value can be computed but it is not very economic to
first compress a measurement into an estimate and an error interval and the to
complement it by a function.

\subsection{Invariance under variate and parameter transformations}

We consider the following non-linear parameter transformation $\theta^{\prime
}=u(\theta)$ since linear transformations do not pose problems.
\begin{align}
\hat{\theta}^{\prime}  & =u(\hat{\theta})\\
\theta_{low}^{\prime}  & =u(\theta_{low})\nonumber\\
\theta_{high}^{\prime}  & =u(\theta_{high})\nonumber
\end{align}
Clearly, the probabilities $P(\theta_{low}<\theta<\theta_{high})$ and
$P(\theta_{low}^{\prime}<\theta^{\prime}<\theta_{high}^{\prime})$ are the same.

In addition, we may transform the sample space variables and consequently also
the probability densities $f(X)\longrightarrow g(X^{\prime})$.

Invariance means that i) the probability and ii) the interval limits are the
same independent of the initial choice of the parameter $\theta$ or
$\theta^{\prime}$ for which we compute the confidence interval and independent
of the sample variable to which the definition applies.

a) In the \emph{frequentist scheme} both conditions are satisfied with the
likelihood ratio ordering used in the unified approach and in the conventional
scheme when central intervals are chosen, a choice which is restricted to the
single variate case. Equal probability density intervals defined in $X$
usually differ from probability density intervals in $X^{\prime}$. Shortest
confidence intervals depend on the parameter choice.

General transformations of a multi-dimensional sample space may lead to
concave probability contours and ruin the whole concept.

Conceptually the parameter and sample variable dependence is acceptable: The
parameters and the sample variables have to be chosen pre-experimentally in a
reasonable way independent of the result of the experiment similarly to all
the other analysis procedures in the classical method. Nevertheless, it is
disturbing that a physicist's choice to determine a confidence interval from a
mass squared distribution instead from a simple mass distribution may produce
different limits\footnote{This happens, for example, if the probability
contours in the sample space are locations of constant pdf.}. Thus the
likelihood ratio ordering which is free of this problem is certainly to be
preferred to the other classical methods.

b) \emph{Likelihood ratio intervals} are strictly invariant under parameter
transformations. The support by data is independent of the choice of the
parameter. It is also invariant against transformations of the sample variable.

c) The \emph{Bayesian method} conserves probability if the prior is
transformed according to:
\[
\pi^{\prime}(\theta^{\prime})d\theta^{\prime}=\pi(\theta)d\theta
\]
Of course not both prior densities $\pi$ and $\pi^{\prime}$ can be uniform.
The invariance of the probability is guaranteed, but the size of the interval
depends on the primary parameter choice for which we required $\pi
(\theta)=const$. With our requirement of using uniform priors, the Bayesian
prescription is not invariant under transformations of the parameter.

The fact that the choice of a parameter space affects the result of an
analysis is quite common. Very useful parameters like root mean square errors
are not invariant. The success of inference procedures like pattern
recognition strongly relies on the adequate selection of the observation
space. \ We use Dalitz plots to detect resonances above a smooth background,
$r-\phi$ plots to find particle tracks in collider experiments, rapidity,
$q^{2}$, $x$ distributions etc.. Part of data analysis skills consists in
selecting the right variables and often results rely on an educated choice.

I find it rather natural that a scientist chooses not only the selection
criteria but also the parameter and sample space variables. In many cases
there is quite some common agreement on a reasonable choice (for example $1/p
$ for tracking, $m^{2}$ for neutrino mass limits, $\gamma$ for the decay rate).

\subsection{Error treatment}

Error propagation is simple when the distributions are Gaussian and the
functions are linear. All other cases are difficult to handle and very little
guidance can be found in the literature. A detailed discussion of the problem
is beyond the scope of this article.

Error propagation \ requires not only an error interval, but also a parameter
estimation. The usual choice for the latter is the maximum likelihood
estimate\footnote{In our context the least squares estimate is equivalent.}.
To classical methods other choices are better adapted. Central intervals are
well compatible with the estimate which we obtain when we shrink the
confidence interval to zero width. These estimates, however, usually are not
very efficient estimators.

The errors - better called uncertainties - of a measurement are usually not
accurately known and this is also not necessary. A precision of about 10\% is
good enough in most cases. Thus, we should be satisfied with reasonable approximations.

\subsubsection{Error definition.}

In the \emph{classical schemes} error intervals are typically 68.3 \%
confidence intervals. They are well defined, except for the arbitrariness of
the definition of the probability intervals.

Physicists usually publish the asymmetric \emph{likelihood ratio limits} which
provide approximate information of the shape of the likelihood function.

The \emph{Bayesian }probability distributions for parameters gives full
freedom (which is a disadvantage) in defining the error. An obvious choice is
to use mean value, variance, and, when necessary skewness and kurtosis of the
parameter probability density. This method has the advantage that tails in the
likelihood function are taken into account.%

\begin{figure}[ptb]
\begin{center}
\includegraphics*[width=3.9513in]{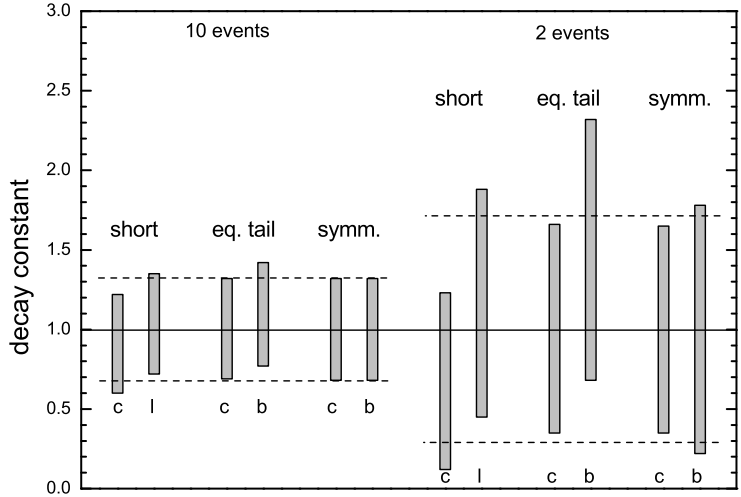}%
\caption{Lifetime measurements using 10 and 2 events. The error
bars from different methods are given for a measurement coinciding
with the true lifetime. The letters refer to classical (c),
likelihood (l) and Bayesian~(b)}
\end{center}
\end{figure}

A comparison of error bounds of lifetime measurements from the different
methods is presented in Fig.~27 where we show the error intervals for a
measurements of the decay parameter $\gamma=1$ obtained from the observations
of ten and two decays, respectively. In the figure \emph{equal tail} is
synonymous with \emph{central}, short indicates that the interval has minimum
length. There are considerable differences for small event numbers. In the
case of ten events the differences between various definitions are already negligible.

The Bureau International de Poids et Mesures (BIPM) and the International
Organization for Standardization (ISO) have issued some recommendations
\cite{bipm81} on how to define errors. After having distributed a
questionnaire among a large number of scientists BIPM and ISO have essentially
adopted a Bayesian inspired method. They recommend to quantify the uncertainty
by the variance and not by a classical confidence interval or a likelihood
ratio interval.

\subsubsection{Error propagation in general.}

To conserve the confidence level or the probability, all methods call for an
explicit transformation following generalized Eqs.~(19). The multivariate
problem $Y_{k}=Y_{k}(X_{1},X_{2},...X_{N})$ involves a variable transformation
plus the elimination of nuisance parameters. Here again, classical methods
fail. Only the Bayesian way permits a consistent solution of the problem.

Transformations conserving probability often produce strong biasses and very
asymmetric error bounds in the new variables. Following the ISO
recommendations one has to accept a certain violation of probability
conservation and work with the moments of the Bayesian distributions or the
asymmetric likelihood errors.

A reasonable approximation is obtained inserting the moments of the
probability densities of the variates $X_{i}$ into the Taylor expansions of
$Y_{k}$.

\subsubsection{Systematic errors.}

\begin{example}
Variations in the response of a calorimeter introduce a systematic uncertainty
in the measurement of some particle mass. Calibration runs are performed at
the beginning and the end of the experiment to obtain the energy scale. The
systematic r.m.s. uncertainty is estimated to be half the difference of the
observed variation. This estimate is rather crude and very little is known
about the corresponding probability density. We would certainly hesitate to
believe in a Gaussian three standard deviation probability derived from it. If
in our experiment calibration runs were performed with high frequency we could
get a rather precise knowledge of the fluctuations and derive a statistical
calibration error.
\end{example}

Usually, errors are called systematic if the are independent of the number of
observations, typically when they cannot be derived from Poisson or
multinomial distributions. However, the separation of statistical and
systematic errors often is rather artificial and often systematic errors can
be reduced with higher statistics. A wider definition includes all kind of
errors where the pdf is badly known.

A crudely known systematic error can be modelled by a pdf with long tails.

The combination of many different systematic error distributions with similar
variance, according to the Central Limit Theorem, can be approximated by a
Gaussian with variance equal to the sum of the individual variances. Then
non-Gaussian tails can be neglected.

It is useful to state statistical and systematic errors separately. For
example, when we observe a five standard deviation effect, we would like to
know whether the error is dominated by systematics or whether it is purely
statistical. Sometimes we have to combine the two kind of errors, when results
from different experiments are summarized or when a hypothesis is to be
tested. Then we should add the errors in quadrature. When a single systematic
error has similar size as the statistical error, we better use a conservative
estimate of the systematic error before combining the two.

It is not clear how to incorporate the frequentist notion of coverage in the
definition of systematic errors.

\subsection{Combining data}

One of the simpler aspects of error propagation is the computation of an
average from individual measurements of the same quantity.

Increasing precision by combining data from different experiments contrib\-utes
to a continuous progress of knowledge in science.

Following the LP we conclude:

\emph{The only way to combine measurements without loss of information is to
add their log-likelihood functions.}

This step does not require a definition of the error.

\emph{Frequentist methods} cannot compute confidence limits from the
likelihood function. Thus they have to invent other prescriptions to combine
data.\ No generally accepted method is known to me. Especially in unified
approaches additional information to the error limits is necessary to perform
sensible combinations. Complicated procedures have been invented to combine
upper limits \cite{alep97}.

The usual averaging procedure, weighting the individual measurement with
$\delta^{-2}$, the inverse of the error squared, assumes that the errors are
proportional to the standard deviation of a probability density of the
measured quantity and that the measurement itself corresponds to its
expectation value. Coverage properties are not considered.

The PDG \cite{pdg00} uses the following recipe for the combination of
measurements with asymmetric errors: The mean value is computed iteratively,
weighting the individual measurements with the inverse of the error squared.
The error is a function of the resulting mean value. When the mean value is
located inside the error interval a linear interpolation between the lower and
upper side errors is used. When the mean value is outside the interval, the
corresponding lower or upper error is chosen.

In principle, it would be more consistent to apply the linear error function
also outside the interval. Let us call this method \emph{PDG extended}. On the
other hand the pragmatic procedure of the PDG has the advantage to be less
sensitive to tails in the distributions.

Applying the usual prescriptions to add measurements with error intervals as
described in the PDG book i) certainly will not conserve coverage and ii) the
result will depend on the chosen scheme of defining the bounds.

For asymmetric likelihood ratio errors, the standard averaging procedure where
the inverse of the full width is used as a weighting factor is only
approximately correct. Better results are obtained with the PDG prescription.
Alternatively, one could construct approximate log-likelihood functions of
parabolic shape from the maximum likelihood estimate and the asymmetric errors
and then add the approximated log-likelihoods.

\begin{example}
We assume that an infinite number of identical experiments determine the mean
lifetime of a particle with true mean lifetime $\tau_{0}$ from the observation
of two decay times. The time distribution is exponential. The results
$\tau_{i}$ of the estimates from the individual experiments are averaged.
Table 3 summarize the results of different averaging procedures. The first
column presents the results when the procedure is applied to the lifetime
estimates $\tau_{i}$, the second column contains the results when the
averaging starts from the estimates of the decay constants $\gamma_{i}%
=1/\tau_{i}$. There is no unique prescription for averaging classical
intervals. To compute the classical value given in the table, the maximum
likelihood estimate and central intervals were used.
\end{example}%

\begin{table}[tbp]
\caption{Average of an infinite number of equivalent
lifetime measurements using different weighting procedures, see Example 34}%
\vskip7pt
\centering
\small
\begin{tabular}[c]{|lll|}
\hline
\multicolumn{1}{|l|}{method} & $<\tau/\tau_{0}>$ & $<\gamma/\gamma_{0}%
>$\\\hline
\multicolumn{1}{|l|}{adding log-likelihood functions} & $1$ & $1$\\
\multicolumn{1}{|l|}{classical, weight: $\delta^{-2}$} & $0.0$ & $0.67$\\
\multicolumn{1}{|l|}{likelihood, weight: PDG} & $0.26$ & $0.80$\\
\multicolumn{1}{|l|}{likelihood, weight: PDG extended} & $0.69$ & $0.85$\\
\multicolumn{1}{|l|}{Bayesian mean, uniform prior, weight: $\delta^{-2}$} &
$\infty$ & $1$\\\hline
\end{tabular}
\end{table}%

In this special example a consistent result is obtained in the Bayesian method
with uniform prior for the decay constant. This indicates how critical the
choice of the parameter is in the Bayesian approach. It is also clear that an
educated choice is also important for the pragmatic procedures. From the
discussion of Example 22 it should be clear that the decay constant is the
better parameter. Methods approximating the likelihood function provide
reasonable results unless the likelihood function is very asymmetric. The
conservative weighting procedure of the PDG is inferior to the extended scheme
(which may be dangerous in other cases). The simple classical average is the
least precise of all. As is well known, the combination of likelihood
functions is optimum.

\subsection{Bias}

Most of the text books on statistics and also the PDG claim that parameter
estimates should be unbiassed\footnote{The expectation value of an unbiassed
estimator of a parameter is equal to the parameter.}, and when they are not,
we are advised to correct for the biasses. One argument for preferring
unbiased estimators is the property of the maximum likelihood estimate to be
efficient (most precise) when unbiassed. Estimates from small samples often
are biassed.

Since both properties, efficiency and bias, are not invariant under change of
parameter, the requirement that estimated parameters should be unbiassed is
not so obvious. The situation is more tricky than it appears at first sight.
What should we conclude from the fact that the maximum likelihood estimate
$\hat{\gamma}$\ of the decay constant from an exponential decay distribution
is biassed but $\hat{\tau}=1/\hat{\gamma}$ is not? Is the result $\hat{\tau}$
more precise than $\hat{\gamma}$ and should we correct $\hat{\gamma}$ for its bias?

A bias in itself is not necessarily a bad property. What really matters is a
bias of the estimator weighted with its inverse error squared, the quantity
which is used for averaging data. When we compute an average of several
unbiassed measurements the result may very well be biassed and the average of
a large number of biassed estimates may be unbiased. \

An average of unbiassed measurements with errors proportional to the parameter
value will be systematically too small. We have demonstrated this in the
previous section (see Table 3). The unbiassed lifetime estimate produces a
strongly biassed average and the biassed decay parameter estimate gives an
unbiassed result.

The correct procedure is obviously to combine the log-likelihoods. A bias in
the maximum likelihood estimate present in the individual experiments
asymptotically disappears with increasing number of data. The PDG
prescription, when using the asymmetric likelihood ratio errors is an
approximation of the exact method, namely, adding the log-likelihoods.

Biasses usually are correlated with small event numbers and large errors which
require a cautious treatment also in the context of error propagation. A
non-linear function $y(x_{ub})$ with $x_{ub}$ unbiassed and large uncertainty
will give a biassed result $y$. The bias corrected estimators $x_{ub}$,
$y_{ub}$ would not follow the same function, i.e. $y_{ub}(x_{ub})\neq
y(x_{ub})$ and consequently confidence limits which are invariant under the
variable transformation $y(x)$ would not be invariant under $y_{ub}(x_{ub})$
and thus not be relevant for the estimate $y_{ub}$.

We will have to be very careful in the treatment of biasses.

Not much can be said about the bias observed in classical methods. There are
too many ways to define the limits and also a clear prescription of combining
measurements is missing.

\subsection{Subjectivity}

In the field of statistics the word \emph{subjective} usually is associated to
probabilities not definable through frequencies and also as a somewhat
negative label for the Bayesian school. Here, we will use it in its common
meaning. In experimental design and data analysis many decisions have to be
made which are not obvious and consequently depend on personal judgements.
However, these decisions should be concerned merely with the quality
(precision) of the result and avoid as much as possible to introduce biasses.
Where this is unavoidable, the procedure has to be documented and allow to the
reader of a publication to use the results without adopting the personal
judgements of the publishing group. To be more precise, we would like the
results to be as much as possible independent of event selection criteria,
stopping rules, Monte Carlo parameters, fitting procedures, etc..

Bayesian methods often are accused of being subjective and sometimes they
really are. However, the Bayesian confidence intervals as defined in this
article obviously contain no subjective elements except for the choice of the
parameter which is documented. Transformations to other parameters are
possible with the published results, of course with limited precision.

Likelihood ratio intervals are rather insensitive to subjective decisions.

On the other hand it is very difficult to avoid subjective elements in a
classical confidence level analysis: Consider two identical experiments
looking for a rare event to be identified in a mass plot. In the first
experiment within a range corresponding to 3 st. dev. of the mass resolution
no event is observed at the expected mass and a 90\% upper limit for the
occurrence is computed. In the second experiment there are a few events in the
region of interest but these events are compatible with background. Usually
the experimentalist would try to reduce the background by applying some clever
cuts and if this is not successful he would estimate the background
probability and include it in the computation of the limit. He also might
decide not to publish, because his colleague of the first experiment has
obtained a more restrictive limit. These subjective decisions completely ruin
the classical probability interpretation which is only valid if the scientist
decides when to stop an experiment, how to treat the data and whether to
publish before he looks at the data - not a very realistic scenario.

A blind analysis avoids some of these subjective elements and in some
situations it is quite sensible independent of the statistical procedure used.
Adopted as a general principle it would inhibit discoveries and delay the
progress in science.

\subsection{Universality}

As has been shown in our examples, the classical scheme has a limited range of
applications: There are problems with physical bounds and discrete samples
(digital measurements). It produces useless intervals in other cases and so
far there is no general method to eliminate nuisance parameters. Especially in
Poisson processes with background and uncertainty on the background mean there
are unsolved problems. The unified approach in the version proposed by Feldman
and Cousins improves the situation of some problems but reduces the
applicability of the method to others. There remains the problem of
interpretation and application: Are the classical bounds useful measures of
the uncertainty of measurements and how should we use them in error handling
and testing of parameter predictions?

The use of simple likelihood ratio intervals is subject to similar
restrictions as the classical confidence intervals. When there is no smooth
likelihood function with a maximum in the physical region, the standard limits
do not indicate the same precision as in the standard case. For this reason,
upper limits deduced from the likelihood ratio should not be transformed into
confidence statements. Likelihood ratio intervals are more general than
classical limits in some respects: They are able to handle a discrete sample
space and there is a logical transition from continuous parameter problems to
discrete hypothesis evaluation. They have a clear and simple interpretation,
the interval is linked to the maximum likelihood estimate and combination of
errors is straight forward.

The Bayesian method is the most general approach. It is able to cope with all
the difficulties mentioned above.

\subsection{Simplicity}

Simplicity is an important aspect for the judgement not only for physical
theories but also for statistical methods.

While methods based on the likelihood function are fairly simple both in the
conception and their application, the classical approaches are complicated.
This is the main reason why the latter are not used for standard interval
estimation. Hardly any physicist knows the different ways to define the
probability intervals. Thus, for standard applications in physics, classical
methods are mainly of academic interest.

The newly proposed unified approaches for the computation of upper limits are
even more complicated. Even specialists have difficulties to understand the
details. I am convinced that only a fraction of the authors of the recent
papers on neutrino oscillations is familiar with the underlying calculation of
the limits and able to interpret them correctly. The variety of different
classical approaches to Poisson upper limits adds to the confusion. Practical
applications will in some situations require considerable computing time.
While combining results based on the likelihood function is transparent and
simple in most cases, the combination of classical limits, if possible at all,
is a quite tricky task.

\section{Summary and proposed conventions}

There is no common agreement on how to define the error of an estimated
parameter, but obviously we would like our error interval to include the true
value with high probability, to be well defined, and to be as small as
possible. Of course, as discussed above, there are other features which are
important when we compare different methods: A procedure should be simple,
robust, cover all standard cases and it should enable us to combine results
from different experiments. Parameter and interval estimation is used in very
different applications, such as simple few parameter estimates, upper limit
computation, unfolding and image reconstruction. It is not obvious that one
single method would be optimum for all these situations.

In the standard situation with high event numbers the likelihood can be
approximated by a Gaussian and all methods give the same results within an
adequate precision.

\subsection{Classical procedure}

The evaluation of classical approaches is problematic: There is no
prescription how to use the limits.

The essential principle of classical confidence bounds is \emph{coverage}.
Knowing that a certain fraction of published measurements cover within their
error limits the true parameter value is an attractive property at first sight
but unfortunately it is not very useful in practice. After all, experimental
results should be used to make decisions but even frequentists admit that
classical bounds are not well suited for this purpose.

The main difficulties of the standard classical approaches are.

\begin{itemize}
\item  Occasionally, they produce inconsistent results (see Sect.~6.2).

\item  Their range of applicability is restricted.

\begin{itemize}
\item  Discrete variates can only be treated by relaxing the requirement of
exact coverage. Digital measurements cannot be treated.

\item  Parameters defined in a restricted region may lead to empty or
unphysical intervals.
\end{itemize}

\item  The results in many cases do not represent the precision of the measurement.

\item  The requirement of pre-experimental fixing of the whole analysis method
is very unrealistic as a general scheme and would, if followed, in many cases
prevent the scientist from extracting the full information from the data and
inhibit discoveries.

\item  Error intervals depend on the choice of the sample variable. An
exception are one-dimensional central intervals and unified likelihood ratio
ordered intervals.

\item  A prescription how to combine frequentist intervals from different
experiments is missing. Biasses are not discussed.

\item  Nuisance parameters (background with uncertainty in the mean) often
cannot be handled satisfactorily. Estimating them out produces undercoverage
in some cases.
\end{itemize}

The unified classical approach provides the required coverage (approximate in
the Poisson case) but shares most of the other difficulties mentioned above
and adds new ones. It is applicable only under special conditions and it
introduces artificial correlations between independent parameters when the
parameters are near physical boundaries.

If I had to choose a frequentist method, I would select the likelihood ratio
ordering for standard problems. In situations with physical boundaries, I
would follow Bouchet \cite{bouc00} and ignore those. The corresponding limits
would represent a reasonable documentation of the experimental result and make
the combination of experimental results easier. I would relax the coverage
requirement which is anyhow violated at many occasions, not follow the
unification concept and treat Poisson limits as exposed in \cite{zech89}.

\subsection{Likelihood and Bayesian methods}

The log-likelihood function contains the full experimental information. It
measures the relative support of different hypothesis or parameter values by
the data. It fulfills transitivity, it is additive and it is invariant against
transformations of the parameters or the data. Thus it provides a sound basis
for all kind of parameter and interval inferences.

Likelihood ratio limits parametrize the experimental observations in a
sensible way. They allow the combination of data, error propagation and
provide a useful measure of the uncertainty. The use of likelihood ratio
errors (without integration) has some restrictions:

\begin{itemize}
\item  The likelihood function has to be smooth.

\item  Digital measurements cannot be treated.

\item  Likelihood ratio limits deduced from likelihood functions that are very
different from Gaussians (Upper Poisson limits) have to be interpreted with care.
\end{itemize}

A fully Bayesian treatment with a free choice of the prior parameter density
is not advisable and is not considered further. Fixing the prior density to be
uniform retains the mathematical simplicity and thus is very attractive but
also has disadvantages.

\begin{itemize}
\item  The choice of the parameter is rather arbitrary.

\item  It is less directly related to the data than the likelihood function.
\end{itemize}

Bayesian methods have been criticized \cite{cous00}, because they do not
provide goodness-of-fit tests.\ This is correct, but misleading in the present
context. Goodness-of-fit tests are irrelevant for parameter interval
estimation which relies on a correct description of the data by the
parametrization, whereas the tests question the parametrization. Also
frequentist intervals do not depend on the quality of the fit.

\subsection{Comparison}

Obviously all methods have advantages and disadvantages as summarized in
\ Table 4 where we compare the conventional classical, the unified classical,
the likelihood and the Bayesian methods with uniform and arbitrary prior
densities. The distribution of the plus and minus signs indicate the author's
personal conclusion. The attributes have to be weighted with the importance of
the evaluated property.%

\begin{table}[tbp]
\caption{Comparison of different approaches to define error
intervals, see text}%
\vskip7pt
\small
\centering
\begin{tabular}{|lccccc|}\hline
method: & classical & unified & likelihood & Bayesian & Bayesian\\
& & & & u.p. & a.p.\\
\hline
{consistency} & {- -} & {- -} & {++} & {+}%
& {+}\\
\hline
{precision} & {- -} & {-} & {+} & {+} &
{+}\\
\hline
{universality} & {- -} & {- -} & {-} & {+}%
& {++}\\
\hline
{simplicity} & {-} & {- -} & {++} & {+} &
{+}\\
\hline
{variable transform.} & {-} & {++} & {++} &
{- -} & {- -}\\
\hline
{nuisance parameter} & {-} & {-} & {-} &
{+} & {+}\\
\hline
{error propagation} & {-} & {-} & {+} & {+}%
& {+}\\
\hline
{combining data} & {-} & {-} & {++} & {+} &
{-}\\
\hline
{coverage} & {+} & {++} & {- -} & {- -} &
{- -}\\
\hline
{objectivity} & {-} & {-} & {++} & {+} &
{-}\\
\hline
{discrete hypothesis} & {-} & {-} & {+} &
{+} & {+}\\
\hline
\end{tabular}
\end{table}

\subsection{Proposed Conventions}

There are two correlated principles which I am not ready to give up:

\begin{enumerate}
\item \textbf{Methods have to be consistent}. (see Sect.~6.2)

\item \textbf{Since all relevant information for parameter inference is
contained in the likelihood function, error intervals and limits should be
based on the likelihood function only. }
\end{enumerate}

These principles exclude the use of classical limits.

All further conventions are debatable. I propose the following guide lines
which present my personal view and may serve as a basis for a wider discussion.

\begin{enumerate}
\item  Whenever possible the full likelihood function should be published. It
contains the full experimental information and permits to combine the results
of different experiments in an optimum way. This is especially important when
the likelihood is strongly non-Gaussian (strongly asymmetric, cut by external
bounds, has several maxima etc.) A sensible way to present the likelihood
function is to plot it in log-scale and normalized to its maximum. A similar
proposal has been made by D' Agostini \cite{dago00cr}. This facilitates the
reading of likelihood ratios. Also in two dimensions the likelihood contour
plots give an instructive presentation of the data. Isn't the two-dimensional
likelihood representation of Fig.~9 more illuminating than drawing
confidence contours supplemented by sensitivity curves?

\item  Data are combined by adding the log-likelihoods. When not known,
para\-metri\-za\-tion are used to approximate it.

\item  If the likelihood is smooth and has a single maximum the likelihood
ratio limits should be used to define the error interval. These limits are
invariant under parameter transformation. For the measurement of the parameter
the value maximizing the likelihood function is chosen. No correction for
biassed likelihood estimators is applied. The errors are usually asymmetric.
These limits can also be interpreted as Bayesian one standard deviation errors
for the specific choice of the parameter variable where the likelihood of the
parameter has a Gaussian shape.

\item  Nuisance parameters are eliminated by integrating them out using a
uniform prior. Correlation coefficients should be computed. Alternatively, the
nuisance parameters are estimated out. Then it is mandatory to document their
values, errors and correlation coefficients.

\item  For digital measurements the Bayesian mean and r.m.s. should be used.

\item  In cases where the likelihood function is restricted by physical or
mathematical bounds and where there are no good reasons to reject a uniform
prior the measurement and its errors defined as the mean and r.m.s should be
computed in the Bayesian way.

\item  Upper and lower limits are computed from the tails of the Bayesian
probability distributions. An even better alternative are the likelihood ratio
limits. They have the disadvantage of being less popular.

\item  Non-uniform prior densities should not be used.

\item  It is the scientist's choice whether to present an error interval or a
one-sided limit.

\item  If measurements depending significantly on systematic errors are used
to establish new physics, conservative error estimates should be used.

\item  In any case the applied procedure has to be documented.
\end{enumerate}

These recipes contain no subjective element and more or less reflect our
everyday practice. An exception are Poisson limits where for strange reasons
the coverage principle - though only approximately realized - has gained preference.

Frequentists will object to these propositions. They are invited in turn to
present their solutions to the many examples presented in this article, to
define precisely error limits and to explain how to use them for error
propagation, combination of results and hypothesis exclusion. If they accept
the LP, they will have to explain why the likelihood function or its
parametrization is not sufficient to compute confidence limits.

It is obvious from the discussion of the treatment of errors in the preceding
section that statistical framework is lacking. Even in the pure Bayesian
approach it is not clear whether to choose moments or probability intervals to
characterize the errors. In any case, compared to frequentist methods,
likelihood based methods are better suited for error calculations. This does
not mean that classical methods in general are useless and that only Bayesian
recipes should be adopted.

For checking statistical methods, frequentist arguments may be useful.
Bay\-esian confidence intervals with low coverage could be suspicious but they
are not necessarily unrealistic.

An example of a bad use of the Bayesian freedom are some unfolding methods. In
some cases the regularization reduces the error below the purely statistical
one and results are obtained which are more precise than they would have been
in an experiment with no smearing where unfolding would not have been necessary.

\subsection{Summary of the summary}

Frequentist intervals depends on the probability of observations which are not
there and on the choice of the probability interval in the sample space. The
Bayesian approach requires a prior density or a choice of the variable space.
Only the likelihood function is free of arbitrariness. It is the natural link
between the observation and the parameter. We have to parameterize it in a
sensible way such that the resulting parameters represent the accuracy of the
measurement. If this is not possible, we should provide the full function.

\subsubsection*{Acknowledgment.}
I am grateful to Fred James for many critical comments to an early
version of this report, to Bob Cousins and Giulio D'Agostini for
numerous interesting oral and email transmitted discussions. I
thank Dieter Haidt for a careful reading of the manuscript and for
many comments which have contributed to clarify some parts of the
manuscript. I would like to acknowledge the numerous competent and
helpful comments and suggestions of the second EJP referee.

\section*{Appendix A: A short remark on probability}

In Kendall and Buckland's \emph{Dictionary of Statistical Terms} \cite{kend82}
we find: ``\textbf{proba\-bil\-ity} A basic concept which may be taken either as
undefinable, expressing `degree of belief', or as the limiting frequency in an
infinite random series. Both approaches have their difficulties and the most
convenient axiomatization of probability theory is a matter of personal taste.
Fortunately both lead to much the same calculus of probabilities.''

This statement is a bit too short to be of much help in concrete statistical
applications and it is not entirely correct. Let us be more explicit.

Probability must satisfy Kolmogorov's axioms and these theorems define the
calculous of probability. This is not a matter of taste.

Probability statements are based on partial information. They should include
the whole of the available information. Otherwise contradictions may result
between procedures using different bits of information.

The essential problem is to relate probability to the real world. We need a
concept which allows us to handle and to document in a well defined way
partially uncertain results.

Most people accept the frequency concept of probability. Assuming that an
experiment can be repeated under identical situations, the frequency of a
certain outcome in the limit of an infinite number of trials is the
probability to be associated to the outcome. One may question both the idea of
infinite repeatability and the idea of identical conditions since in reality
the conditions will never be exactly the same in two experiments. We should
ignore this kind of destructive objections. We have to use idealizations. It
would be stupid to reject the theory of special relativity because it is
impossible to realize constant velocity. I see no difficulty with
probabilities based on the frequency concept.

In our every day life we have to base decisions on probabilities which we
cannot relate easily to well defined repeatable situations: Will it be raining
today? Will my daughter get sick? Will the stock values raise?

Also in science we find difficulties to incorporate uncertainties in our
knowledge into a strict frequency concept. Are we allowed to talk about the
probability that the Higgs mass is located between 100 and 200 GeV? How likely
is it that the calibration constant of a certain device was within certain
limits? Can we associate probabilities to something that has happened in the
past but is not known to us?

In the mathematical description of a deterministic world there is no
substantial difference between future and past. It is quite natural to
attribute probabilities to something fixed but where we have only partial
information. Probability here refers to our knowledge and not to the facts.
The probability that a fair coin thrown yesterday gave head is one half.

What about constants like particle masses? Can we incorporate all
probabilities in a frequency scheme?

Well, we can imagine a huge number of worlds, all with different Higgs masses
distributed according to a certain probability distribution. Similarly, we can
easily imagine that similar wether, health or economic conditions occur
repeatedly. Considering different alternative states, one of which is realized
but unknown to us, with some partial knowledge we can attribute probabilities
to them. In our life we will encounter many such situations, and we hope that
the probabilities on average will correspond to the frequencies with which we
are right. I believe that probability always can be understood in terms of
limiting frequency. It does not matter that situations often cannot be
reproduced in reality, it is enough that it can be done in our imagination.

Is then the debate between frequentists and Bayesians about probability just a
useless academic game? The answer is No!

Bayesians associate prior probabilities to parameters which are to be measured
in an experiment and they produce updated probabilities for the parameters
after the measurement\footnote{Of course, we cannot produce a distribution of
a parameter like a particle mass but a probability distribution of the
parameter. The parameter has a fixed value but our knowledge is incomplete
\cite{jayn84}.}. There, the real problem is not that we cannot define the
notion of prior probability but that we do not know its pdf. In fact, we know
the prior to some extent, we have a crude idea of it, otherwise it would be
impossible to design the experiment. But we cannot, prior to an experiment,
associate to the Higgs mass a well known pdf like we do for the lifetime of an
unstable particle.

In physics, usually there is a quite clean distinction between exactly know
pdfs (sometimes up to parameters) and hardly known pdfs. In sciences like
biology or economics often empirical pdfs have to be used, hence there is a
continuous transition between the two extremes. Also in our discipline, we are
sometimes forced to use empirical, badly known pdfs, for example when we
discuss systematic errors or when we apply a least square fit to observations
without knowing the distribution of the fluctuations.

The inclusion of partially known pdfs considerably widens the range of
applications of statistical inference, and we should profit from the
corresponding tools, but it also introduces problems. The modern Bayesian
school has tried to solve some of them. A scheme has been developed by De
Finetti and Savage to associate pdfs to the ``personal probabilities'' as they
are called by Savage but their method is not relevant for typical physics situations.

When we use prior probabilities or empirical pdfs, we should state them and
the results of the analysis partially depend on the validity of our
assumptions. In many cases the dependence is weak.

\section*{Appendix B: Shortest classical confidence interval for a lifetime measurement}

In some cases we can find a \emph{pivotal quantity} $Q(X,\theta)$ of a variate
$X$ and a parameter $\theta$. A pivot has a probability density which is
independent of the parameter $\theta$ (like $X-\theta$ for a Gaussian with
mean\textbf{\ }$\theta$) and consequently its probability limits $Q_{1},Q_{2}$
of $Q$ are also independent of $\theta:$
\begin{equation}
P(Q_{1}<Q(X,\theta)<Q_{2})=\alpha
\end{equation}
By solving the inequality for $\theta$ we find:
\begin{equation}
P\left(  \theta_{1}(Q_{1},X)<\theta<\theta_{2}(Q_{2},X)\right)  =\alpha
\end{equation}

Equation (21) relates $Q_{1}$ and $Q_{2}(Q_{1},\alpha)$. The remaining free
parameter $Q_{1}$ is used to minimize $|\theta_{2}-\theta_{1}|$.

We look at the example of a lifetime observation $x$. The width of a
confidence interval $[\gamma_{1},\gamma_{2}]$ of the decay constant $\gamma$
is minimized. Here $Q=\gamma X$ is a pivotal quantity.
\begin{align*}
f(X)  & =\gamma e^{-\gamma X}\\
Q  & =\gamma X\\
g(Q)  & =e^{-Q}%
\end{align*}

\begin{sloppypar}
For a given confidence level $\alpha$ the two limits depend on each other:
$Q_{2}(\alpha,Q_{1})$. The limits transform to limits on the parameter
$\theta$: $[\theta(Q_{1}),\theta(Q_{2})]$ Then
$\vert$%
$\theta(Q_{2}(\alpha,Q_{1}))-\theta(Q_{1})|$ is minimized:
\begin{align*}
e^{-Q_{1}}-e^{-Q_{2}}  & =\alpha\\
Q_{2}  & =-\ln(-\alpha+e^{-Q_{1}})\\
Q_{1}  & <\gamma X<Q_{2}=-\ln(-\alpha+e^{-Q_{1}})\\
\frac{Q_{1}}{X}  & <\gamma<\frac{-\ln(-\alpha+e^{-Q_{1}})}{X}%
\end{align*}
Minimizing the interval length is equivalent to minimizing
\[
\ln\frac{e^{-Q_{1}}}{-\alpha+e^{-Q_{1}}}
\]
Taking into account the boundary condition $Q_{1}>0$, we find $Q_{1}%
=0,Q_{2}=-\ln(1-\alpha)$ and%
\begin{align*}
\gamma_{1}  & =0\\
\gamma_{2}  & =-\frac{1}{X}\ln(1-\alpha)
\end{align*}
\end{sloppypar}

For $\alpha=0.6826,$ and an observation $x=1$ we find $\gamma_{1}=0,\gamma
_{2}=1.15$. (The likelihood ratio limits are $\gamma_{1}=0.30,\gamma_{2}=2.36$)

When we play the same game with the parameter $\tau=1/\gamma$ and the same
pivot we obtain
\begin{align*}
Q_{1}^{\prime}  & <\frac{X}{\tau}<Q_{2}^{\prime}=-\ln(-\alpha+e^{-Q_{1}%
^{\prime}})\\
\frac{x}{-\ln(-\alpha+e^{-Q_{1}^{\prime}})}  & <\tau<\frac{x}{Q_{1}^{\prime}}%
\end{align*}
Minimizing
\[
\frac{1}{Q_{1}^{\prime}}+\frac{1}{\ln(-\alpha+e^{-Q_{1}^{\prime}})}
\]
we find $\tau_{1}=0.17$ $,\tau_{2}=2.65$ (The likelihood ratio limits are
invariant $\tau_{1}=0.42,\tau_{2}=3.31$)

The pivotal limits obviously depend, as expected, on the choice of the parameter.

\section*{Appendix C: Objective prior density for an exponential decay}

In the literature we find arguments for \emph{objective priors}. A nice
example which illustrates the methods used, is the derivation of the decay
constant for particle decays.

Two different particles with exponential decays have the probability densities
of the decay time $T$ and decay constant $\lambda$%
\begin{align*}
f(T,\lambda)  & =f(T|\lambda)\pi(\lambda)\\
f^{\prime}(T^{\prime},\lambda^{\prime})  & =f(T^{\prime}|\lambda^{\prime}%
)\pi(\lambda^{\prime})
\end{align*}
where $\pi(\lambda)$ is the prior density and $f(T|\lambda)=\lambda
\exp(-\lambda T)$. The decay constants $\lambda$ and $\lambda^{\prime}$ are
related by $\lambda^{\prime}=\alpha\lambda$. This fixes $\alpha$. Then for
times $t^{\prime}=t/\alpha$ we have the transformation
\[
f(T,\lambda)dTd\lambda=f^{\prime}(T^{\prime},\lambda^{\prime})dT^{\prime
}d\lambda^{\prime}
\]
from where we obtain
\[
\alpha\pi(\alpha\lambda)=\pi(\lambda)
\]
This relation is satisfied by
\[
\pi(\lambda)=\frac{const.}{\lambda}
\]

The prior density is inversely proportional to the decay constant. The flaw of
the argument is easily identified: Why should there be a universal prior
density $\pi(\lambda)$ for all particles? There is no reason for this
assumption which was used when we tacitly substituted $\pi^{\prime}%
(\lambda^{\prime})$ with $\pi(\lambda^{\prime})$.

We can also argue in a slightly different way: Two physicists measure the
lifetime of the same particle in different time units, $T$ in seconds and
$T^{\prime}$ in minutes. The numerical values are then related by $T^{\prime
}=T/\alpha$ and $\lambda^{\prime}=\alpha\lambda$ as above with $\alpha=60$.
The prior densities have to fulfill
\[
\pi(\lambda)d\lambda=\pi^{\prime}(\lambda^{\prime})d\lambda^{\prime}
\]
Now the two physicists choose the same functional form of the prior density.
Thus we get
\begin{align}
\pi(\lambda)d\lambda & =\pi(\lambda^{\prime})d\lambda^{\prime}\nonumber\\
\pi(\lambda)d\lambda & =\pi(\alpha\lambda)\alpha d\lambda\nonumber\\
\pi(\lambda)  & =\frac{const.}{\lambda}%
\end{align}

The resulting prior densities are not normalizable, a fact which already
indicates that the result cannot be correct. Again there is no compelling
reason why $\pi$ should be equal to $\pi^{\prime}$. It is also easily seen
that prior densities different from (22) do not lead to contradictions. For
example a uniform prior in the interval $0<T<1\sec$ would produce $\pi(T)=1$
and $\pi^{\prime}(T^{\prime})=1/60$.

Similar plausibility arguments as used above are applied for the derivation of
the a prior density for the Poisson parameter. They are in no way convincing.

It is impossible to determine prior densities from scaling laws or symmetries
alone. We can use non-uniform prior densities when they are explicitly given
as in Example 19.

\end{document}